\documentclass{report}
\usepackage[a4paper,left=3cm,right=2cm,top=2.5cm,bottom=2.5cm]{geometry}
\usepackage{amsmath}
\usepackage{booktabs} 
\usepackage{multirow} 
\usepackage{multicol}
\usepackage{amssymb} 
\usepackage{textcomp, gensymb}
\usepackage{graphicx} 
\usepackage{pdfpages}

\usepackage[labelfont=bf,textfont=md]{caption}  

\usepackage{chngcntr}   
\counterwithout{table}{chapter} 

\usepackage{lscape} 

\usepackage{xcolor, soul}
\sethlcolor{red}

\usepackage{titlesec}   
\usepackage{fancyhdr}   
\usepackage{color}
\usepackage{xcolor}
\usepackage{hyperref}
\hypersetup{
    colorlinks=true, 
    linktoc=all,     
    linkcolor=black,  
    citecolor=black 
}

\title{Supplementary Information for: \\
\textbf{Exact particle-enhanced point-spread function unlocks 
3D super-resolution localization microscopy on nanoparticles
}}

\author{Teun A.P.M. Huijben, \textit{et al.}}
\date{2023}

\begin{document}
\pagenumbering{arabic}

\renewcommand{\thetable}{S\arabic{table}} 



# Exact particle-enhanced point-spread function unlocks 3D super-resolution localization microscopy on nanoparticles

**Teun A.P.M. Huijben[1†], Sarojini Mahajan[2†], Peter Zijlstra[2], Rodolphe Marie[1], & Kim I. Mortensen[1]**

**†** *These authors contributed equally*

[1] Department of Health Technology, Technical University of Denmark (DTU), Denmark

[2] Department of Applied Physics and Science Education, Eindhoven University of Technology (TU/e), The Netherlands

Correspondence to: p.zijlstra@tue.nl, rcwm@dtu.dk, mortensen.kim@gmail.com

## Abstract

Nanoparticles (NPs) have proven their applicability in biosensing, drug delivery, and photo-thermal therapy, but their performance depends critically on the distribution and number of functional groups on their surface. When studying surface functionalization using super-resolution microscopy, the NP modifies the fluorophore's point-spread function (PSF). This leads to systematic mislocalizations in conventional analyses employing Gaussian PSFs. Here, we address this shortcoming by deriving the first-ever analytical PSF model for a fluorophore near a spherical NP. Its calculation is four orders of magnitude faster than numerical approaches and thus feasible for direct use in localization algorithms. We fit this model to individual 2D images from DNA-PAINT experiments on DNA-coated gold NPs and demonstrate extraction of the 3D positions of functional groups with <5 nm precision, revealing inhomogeneous surface coverage. Our method is exact, fast, accessible, and poised to become the standard in super-resolution imaging of NPs for biosensing and drug delivery applications.

## Introduction

Functionalized nanoparticles (NP) have made a rapid and substantial entrée in the modern healthcare industry. The large surface-to-volume ratio and versatile chemical functionalization leading to biocompatibility and functionality make these colloidal particles invaluable for targeted nanomedicine and diagnostics. These nano- to micrometer-sized particles of various materials, sizes, shapes, and compositions have great potential in the biomedical fields of prevention, diagnosis, and treatment of diseases. For example, lipid NPs are used as carriers of mRNA in vaccines against Covid-19. Dielectric NPs (e.g., hydrogel, polymer, or silica), are promising candidates

for vehicles of targeted drug delivery[1-3]. Metallic (e.g., gold) NPs, have optical and plasmonic properties which make them suited as biosensors in diagnostics[4-6] and photothermal agents in cancer therapy[7-9].

However, the functionalization of the NPs, i.e., the chemical conjugation of molecules such as DNA, peptides, aptamers, drugs, or proteins to the NP surface is crucial for their function. The density and distribution of the functional groups on their surface influence the performance of these biofunctionalized NPs[10-13]. Hence, it is crucial to control both parameters to enable rational design of the NPs for the desired application. Conventionally, ensemble-based methods such as spectroscopy or electrophoresis are used to quantify the density of ligands. However, such ensemble averaging does not reveal particle-to-particle heterogeneity let alone the distribution of ligands on each NP.

Single-molecule localization microscopy (SMLM), on the other hand, is ideally suited to study the surface functionalization of individual NPs and to characterize inter- and intra-particle functionalization heterogeneity of the NP in their native, typically aqueous, environment[14-16]. To circumvent the diffraction limit, one repeatedly images sparse (random) subsets of fluorophore-labeled functional sites in a sample, as schematically depicted in Figure 1a. Various SMLM techniques exist, with different mechanisms to ensure a sparse subset of fluorophores, e.g., blinking of the fluorophores tuned by the imaging buffer (*d*STORM[17]), the photophysics of the fluorophore itself (PALM[18]) or sequential binding and unbinding of the fluorophore to the target molecule (PAINT[19]). The sparsity of emitting fluorophores at each time point ensures that each fluorophore images as a diffraction-limited spot. The position of every fluorophore can be determined by fitting a model for the spatial distribution of photons in the image plane, i.e. the point-spread function (PSF), to the spot in an image. When all such positions of fluorophores on the NP surface are combined, one obtains a super-resolved mapping of the functional groups on the NP[14-16] (Fig. 1a).

In conventional SMLM, a 2D Gaussian PSF model is used for analysis of images. For most practical applications the 2D Gaussian PSF well-approximates the distribution of photons in a spot[20]. However, when employing SMLM close to NPs, the presence of the NP in the vicinity of the fluorophore influences its radiation pattern[21-23], which results in a complex distortion of the PSF (Fig. 1a,b). Furthermore, the PSF shape is not

distorted in a simple manner but depends strongly on the position of the fluorophore relative to the NP (Fig. 1b). Consequently, fitting such spots with a 2D Gaussian PSF leads to systematic localization biases, commonly referred to as mislocalizations (Fig. 1a). Especially the use of large pixel sizes conceals the PSF deformations in images, and hence do not warn the user about their presence.

Although present for any NP, mislocalizations occur especially in the case of plasmonic NPs, such as gold and silver, where the emission of the fluorophore couples to the NP plasmon thereby heavily influencing the emission pattern. In fact, there is a great interest in using metal structures in SMLM due to the fluorescence enhancement that results in an increase of fluorescent signal of multiple orders of magnitude[24–30]. This enhancement provides an increased localization precision, which scales as $1/\sqrt{N}$, where $N$ is the number of detected photons[20]. Unfortunately, the increase in precision is rendered useless since it is accompanied by a dramatic decrease in accuracy due to PSF deformation and displacement.

Multiple studies have previously investigated the physical mechanism and severity of mislocalizations when performing localization microscopy close to nanophotonic structures using a Gaussian PSF model[21,22,31–47]. The degree of mislocalization was determined either directly, based on a known position of the emitter relative to the NP[43,44], or indirectly, based on the reconstructed shape of the NP compared to the known shape[21,33–42]. It was found that the degree of mislocalization is affected by the spectral overlap between the emitter and the NP[31,33,48] emission, and by the physical distance between the emitter and the NP[33,35].

These studies employ numerical approaches[21,22,32,35,37–43,46] (Fig. 2a) to compute the photon distributions in each spot which in turn enables the study of mislocalizations. The computational time of each numerically calculated PSF (in the order of minutes) is prohibitive for use in conjunction with localization microscopy image analysis, since a single PSF fit requires many evaluations of the PSF and a typical SMLM experiment consists of several thousand localizations.

Furthermore, the focus has mostly been on the magnitude of the mislocalization, but never on the origin of the problem, the deformation of the PSF. Previously, mislocalizations induced by the NP have been considered a disadvantage and has

been sought minimized by tuning experimental parameters like the NP material, fluorophore emission wavelength, or refractive index of the medium. Instead, we here take advantage of the fact that the particle-emitter coupling encodes the fluorophore's position relative to the NP in images to unlock 3D super-resolution microscopy on NPs from 2D images without explicit optical engineering of the PSF. In order to do so, we derive the first-ever analytical PSF model for a fluorophore near a spherical NP, based on Mie[49–51] and vector diffraction theory[52,53]. Our analytical model is exact, consistent with numerical calculations, yet up to four orders of magnitude faster than numerical approaches, enabling direct fitting to localization microscopy datasets.

We employ this new PSF model in a fitting procedure to study the surface distribution of single-stranded DNA on spherical gold NPs in 3D. For a typical image containing 25,000 photons, we obtain a localization precision on the surface of <5 nm. Such rapid and precise localization enables mapping of the full surface functionalization in 3D of individual nanometric particles hence making it possible to visualize particle-to-particle heterogeneity. Our analysis reveals an inhomogeneous surface coverage of the DNA docking strands over the surface of single particles and high heterogeneity between particles, which we attribute to local variations in density and accessibility of DNA. Moving forward, this will enable researchers to relate inhomogeneous surface coverage to the employed chemical functionalization protocols, in order to better understand and improve desired nanoparticle conjugates for a multitude of purposes.

## Results

*Exact PSF model for fluorescent molecules near spherical NPs*

The PSF is the probability density function for photons to arrive on the camera. In order to derive it, we calculate the propagation of the emitted light through the microscope. Figure S1a depicts the microscopy system we consider. The NP and a fluorophore modelled as dipole emitter are present in water, just above a glass coverslip, and imaged using an objective lens.

The process of deriving the PSF comprises multiple steps (Fig. 2b, S1c, and S2). We start by expressing the electro-magnetic (EM) field radiated by the fluorophore, modelled as a dipole emitter, in a linear combination of vector-spherical harmonics (VSHs). Expressing the field in spherical basis functions is required to account for interactions with a spherical object, the NP, in contrast to previous PSF models for

emitters in water[20,54] for which a basis of plane waves was sufficient. Next, we use Mie scattering theory[49–51] to define how the spherical NP scatters the light emitted by the dipole. Both contributions, the initial dipole field and the NP scattered field, are subsequently decomposed into plane waves[55,56], refracted at the water-glass interface, projected into the far-field (i.e., back-focal plane) and focused onto the camera by the objective. Note that for simplicity we do not consider reflection of light at the water-glass interface back into the water, that can interact again with the NP, since it appears to have a negligible effect on the PSF (Fig. S3). The PSF is represented by the intensity of the light at the camera plane, given by the $z$-component of the Poynting[57] vector $\vec{E} \times \vec{B}^*$. Concatenating all these steps results in a single analytical expression for the PSF at any position on the camera, which we will refer to as our PSF model (see Methods for more details, Suppl. Note 2 for the full mathematical derivation, and the GitHub repository for the Python implementation). We evaluate the required integrals in this expression numerically since no closed form solution exists[20]. Implemented in Python, and computed on a regular single-core CPU, the evaluation of a single PSF takes approximately 100 ms for $10 \times 10$ pixels in the PSF and 10 truncated Mie orders (for dependency of the computational time and the PSF accuracy on the number of pixels in the PSF and the number of considered truncated Mie orders, see Fig. S4 and S5).

*PSF model is consistent with numerical approaches, but efficiency permits direct use in image analysis*

The accuracy and computational efficiency of our analytical PSF model are quantified by comparing it to numerical calculations of the same NP-dipole system. In order to validate the analytical PSF and highlight the shortcomings of numerical calculations, we additionally developed and optimized a numerical procedure for the calculation of the PSF based on an existing implementation[21]. The numerical approach is by construction an approximation, due to the discretization of the system (Fig. 2a). Therefore, the results depend on numerical parameters like the size of the simulated domain and the resolution of the discretization, referred to as mesh size. We investigated the influence of these two parameters on the numerical PSF (Fig. S6) and selected parameters (Methods) that result in high precision, i.e., 1% difference with the analytical PSF, with a reasonable computational time in the order of tens of minutes. We quantitatively compare the difference between the analytical and numerical PSF for multiple NP materials, sizes and compositions, starting with the

arbitrary system of a horizontally oriented dipole located 10 nm above the surface of a 200 nm gold spherical NP (Fig. 2c). The numerical calculation obviously gives more accurate results when the mesh size is smaller, since then the discretized system better resembles reality. By decreasing the mesh size in the numerical approach, the numerical PSF converges towards the analytical PSF (Fig. 2c, left $y$-axis), indicating that the analytical PSF is indeed the exact solution. The associated drastic increase in the numerical computational time (Fig. 2c, right $y$-axis) demonstrates the impracticality of the numerical method for direct use in PSF fitting in localization microscopy. The evaluation of the analytical PSF is four orders of magnitude faster compared to the near-exact numerical PSF (we consider two PSFs identical when the maximum pixel difference is within 1%).

For completeness, we compare both the electric far-field and the PSF between the analytical and numerical approach for NPs with other materials, sizes, and compositions such as gold spheres, silica spheres and gold shells (Fig. 2d). Advantageously, Mie scattering theory allows us to extend our model to other spherical particle geometries such as core-shell particles consisting of a spherical core with concentric shells with arbitrary thicknesses and materials. In particular, we calculate PSFs for particles with a silica core and a gold shell (Fig. 2d) that have their plasmon resonance in the near-infrared, making them interesting candidates for targeted photothermal cancer therapy[58]. In all examples, we chose fluorophore positions and orientations that lead to exotic PSFs to exemplify concrete cases that would challenge the applicability of conventional SMLM. We can see that the complex electric vector field measured on an arc of the far-field hemisphere is identical between the analytical and numerical approaches, for different NP compositions, and dipole orientations and positions (Fig. 2d). Consequently, the numerical PSFs are almost identical to the analytical PSFs with maximal pixel differences around only 1%, explained by the limited discretization resolution of the numerical approach. The full comparison, including all electric field components in the far-field and differences with the analytical PSF, can be found in Fig. S7.

In summary, our analytical PSF is correct, compared to numerical approaches that suffer from meshing. Additionally, we have demonstrated a drastic improvement in computational speed of up to four orders of magnitude, compared to the numerical approach using the finite-difference time domain (FDTD) method [21] (Fig. 2c). This

makes our PSF model reliable and practical as an analysis tool for super-resolution imaging of functional groups on NPs.

*Distribution of function sites on gold NPs is patchy*

We showcase this analytical approach by applying it for 3D super-resolution localization microscopy of emitters on a spherical gold NP. Standard SMLM approaches would fail in this case, since the PSF for an emitter on a spherical NP resembles the PSF for an isolated fixed emitter, due to the strong plasmonic coupling between the emitter and the NP (Fig. S21). This results in biased Gaussian fitting[20,54,59]. To demonstrate our approach, we perform SMLM on DNA-coated 100 nm gold nanospheres (GNS) by using DNA-based point-accumulation for imaging in nanoscale topography (DNA-PAINT[60]). We conjugate single-stranded 30-nucleotide (nt) DNA docking strands to the NP surface with a thiol bond (see Methods) and image the sample on an inverted total internal reflection fluorescence (TIRF) microscope. Freely diffusing 9-nt complementary single-stranded DNA imager strands, labelled with ATTO655, transiently bind to the NP-bound docking strands (Fig. 3a). Each binding event creates a fluorescent burst on the camera (Fig. 3b), whose PSF is fitted using our analytical PSF model. Acquisitions of 30 minutes accumulate hundreds of binding events per individual particle (Fig. S8). We perform necessary controls with non-complementary imager strands (Fig. S9) to verify that the observed binding events are indeed specific (i.e., due to DNA hybridization rather than non-specific sticking) and we perform DNA-PAINT on the glass substrate without NPs to demonstrate that the fluorophores are freely rotating in orientation and that the presence of exotically-shaped PSFs (Fig. S8) are solely explained by presence of the NP (Fig. S10).

We detect binding events by thresholding of the intensity time trace of each individual GNS, and subsequently average the frames within each individual event to obtain the PSF with a higher signal-to-noise ratio (Fig. 3b,c). The measured intensity not only contains fluorescence originating from the bound imager strand, but also background (BG) photons and one-photon photoluminescence (PL) from the gold. Note that these contributions do not modify the shape of the PSF due to a lack of coherence with the emission of the NP-dye conjugate. To remove these two components, the true fluorescent PSF is obtained by subtracting the average of a set of frames around each event (Methods and Fig. S11). The added value of using gold, besides the high signal-

to-noise ratio due to plasmon-enhanced fluorescence (Fig. S12), is that the PL is very photostable and can be used to localize the GNS. This provides us with the GNS position at the time of every measured PSF, which (i) aids the fitting, since only the relative position of the emitter to the NP needs to be determined, and (ii) substitutes the need for in-plane sample drift correction (Fig S13).

We fit all the average PSFs obtained from a single GNS with our analytical PSF model, resulting in the positions of the binding events on the GNS surface (Fig. 3d). The analytical PSF model is corrected for field-dependent aberrations quantified on unspecific binding events away from the particles (Methods). We know the position of the GNS from a Gaussian fit of the PL, the only remaining free fit parameters are: the position of the emitter on the GNS surface, expressed in spherical coordinates $\theta$ and $\phi$, the number of photons in the PSF, and the background generated by camera offset and background photons. These four parameters we extract by a least-squares fitting approach of our analytical PSF model to the experimental PSF (see Methods for details). We project the obtained emitter positions onto a spherical surface to visualize the location of the binding event on the GNS surface (Fig. 3d). In principle, every emitter position on the NP surface results in a unique PSF, and with enough photons the fit would be able to discriminate between them. However, for practical signal-to-noise levels, this is not possible due to the spherical symmetries of the GNS that make multiple emitter positions on the surface result in PSFs with a similar shape. We solve this ambiguity by using additionally the position of the GNS from the PL signal (Fig. S14).

The fit assumes a spherical NP with known diameter. To confirm that the analyzed GNSs are indeed single gold NPs, we performed two additional correlated measurements on the same sample, namely single particle spectroscopy and scanning electron microscopy (SEM). The parameters of the Lorentzian fit to the single particle scattering spectrum in addition to the visual information from the high-resolution SEM image confirm that the GNSs are not clustered and have a size and shape as expected (Fig. S15). Interestingly, we note that the PSFs originating from clusters of GNSs (dimers, trimers, heptamers, etc.) show a different set of intriguing shapes (Fig. S16), but we exclude them from analysis here.

We collect hundreds of PSFs per GNS, providing us with good statistics on the surface functionalization. Figure 4a shows the fitting result of all 828 localizations on a single GNS (see Figure S15 for the scattering and SEM results). To better visualize the entire GNS, we perform a 2D unfolding of the spherical surface (Fig. 4b), which visualizes all sides of the GNS. We also compute a localization density map (Fig. 4c) by rendering each localization as a normalized 2D asymmetric Gaussian, where the two Gaussian axes represent the localization uncertainty (Fig. 4d) in the polar- ($\Delta_\theta$) and azimuthal direction ($\Delta_\phi$), respectively. The angular localization uncertainties in the polar- and azimuthal directions are obtained directly from the PSF fitting procedure, and we convert them to distances on the surface by multiplying with the radius of the NP. The histogram of the combined uncertainty $\sqrt{\Delta_\theta^2 + \Delta_\phi^2}$ shows that we can localize the binding events with an uncertainty of 4.9 $\pm$ 2.1 nm on the surface (Fig. 4d). The dependency of the uncertainty with respect to the fluorophore positions and the number of photons can be found in Fig. S17.

We clearly see that the localizations are not homogeneously distributed over the GNS. Due to the statistical nature of DNA-PAINT, we are not necessarily seeing every available docking strand, and every localization does not necessarily represent a unique docking strand. The number of localizations per accessible strand follow the Poisson distribution. To draw solid conclusions about the inhomogeneous surface coverage of accessible docking strands, we apply the fitting analysis to multiple GNSs, perform qPAINT analysis, and assess the localization uncertainty. First of all, applying the same fitting procedure to multiple GNSs, results in surface reconstructions that all show an inhomogeneous distribution of binding sites (Fig. 4e). Secondly, we employ quantitative PAINT (qPAINT, Fig. S18), a technique to determine the actual number of available docking strands participating in PAINT by looking at the binding statistics, knowing the binding kinetics of the considered DNA sequence[61]. By performing qPAINT analysis on multiple analyzed GNSs we show that the number of available docking strands is 7 $\pm$ 0.1 times lower than the average number of PSFs we detect per GNS (Fig. S18). From this we conclude that 1) every accessible docking strand creates on average 7 localizations, referred to as overcounting, and that 2) it is very unlikely for an accessible docking strand not to have bound an imager strand during the course of the measurement (30 minutes). The latter point indicates that a patch of surface without localizations does not contain any

accessible docking strands, and that this is not the effect of undercounting. And thirdly, the localization uncertainty directly provided by the PSF fit is 4.9 nm (Fig. 4d), which is a small compared to the GNS circumference. From this we conclude that consistent inhomogeneous surface coverage is not a side effect of having a large localization uncertainty.

## Conclusions and Outlook

In conclusion, we have introduced the first-ever analytical PSF model for a fluorophore in close proximity to a spherical NP. The model is up to four orders of magnitude faster compared to state-of-the-art numerical approaches, it is exact, and it is open source: It should become an industry standard for analysis of SMLM data obtained from spherical particles of any composition and size.

Additionally, the computational speed of our model paves the way to identify parameter regimes in which PSF distortion is negligible, enabling the use of conventional localization algorithms that assume a 2D Gaussian PSF. Secondly, we show an effective way to exploit PSF deformations for 3D localization microscopy that works for any combination of spherical particle and fluorophore irrespective of wavelength and material. In this way, we can map out the complete surface functionalization on NPs, which we demonstrate by applying our approach to experimental data obtained through DNA-PAINT experiments on DNA-coated spherical 100 nm gold NPs. Surprisingly, we find that the functionalization with DNA via thiol chemistry is inhomogeneous over the NP surface, and that this is observed for many NPs.

The patchy nature of nanoparticle functionalization has never been visualized before but is not unexpected: the presence of facets and an inhomogeneous surface charge[62] could lead to an inhomogeneous surface coverage of DNA. It is also noteworthy that areas with a higher number of localizations do not necessarily contain a higher density of DNA strands. The frequency of observed binding events does not only depend on the number of docking strands $N$, but also on the accessibility of the strands, since a high density of DNA may cause steric hinderance and electrostatic repulsion. Indeed, the accessibility of the docking strands can modulate the binding kinetics of incoming imager strands and is known to depend strongly on DNA-density[63]. Our method

thereby uniquely reveals the presence of accessible docking strands (rather than total docking strands), which is valuable information in the context of biomedical applications where functionality arises from accessible surface groups[64].

The current model describes the PSF for fluorophores close to spherical NPs. The spherical symmetry of the NP is crucial for being able to employ Mie scattering theory. Even though the analytical solution for other NP shapes is not (yet) derived, having an exact solution for the spherical case allows finetuning of the parameters for the numerical calculations. In this way, PSFs for non-spherical NPs can be computed reliably. For subsequent fitting to experimental data, one will have to resort to data-driven PSF fitting approaches like decomposition into a set of basis functions[22] or convolutional deep neural networks[65].

In the current work, we validated our PSF model by comparison to numerical approaches and achieved consistency with experimental data. Though fitting we obtain localization precision down to 4.9 nm for 25,000 photons. Still, exciting future work is necessary to assess the localization accuracy by obtaining the ground-truth position of the fluorophore and comparing this to the inferred position by the PSF fit. To acquire the ground-truth location, one can think of approaches to deterministically place the fluorophore on a designed position relative the NP, for example with DNA-origami[44], or by correlating with SEM measurements after attaching quantum dots to the DNA docking strands[32].

Our approach can visualize functional groups on the molecular level. We anticipate that our method will be instrumental for the future characterisation of NP heterogeneity and that the combination with functional assays will provide much needed information on the correlation between nanoparticle coating (both density and distribution) and particle behavior in biomedical applications.

# Methods

### Analytical PSF model

In order to calculate the point-spread function (PSF) for a dipole emitter fixed in orientation near a spherical nanoparticle (NP), we start from the light emitted by the excited dipole emitter and trace these electro-magnetic (EM) fields down the microscope onto the camera. The calculation consists of six steps, which are schematically visualized in Fig. S2. Each of the six steps is explained in Suppl. Notes 1 and 2, where Note 1 gives a concise textual overview and Note 2 the full in-depth mathematical derivations.

The PSF for a freely rotating dipole is obtained by averaging the PSF for three orthogonal dipole orientations (Fig. S19). We mathematically prove that the average over three orthogonal orientations is identical to integrating over all possible orientations (Suppl. Note 3).

### Numerical PSF calculations

The numerical approach for calculating the PSFs is based on an existing implementation[21] and comprises three steps: 1) numerically calculating the EM near-field and recording them just below the coverslip, 2) far-field projection, and 3) focussing (Fig. S1b). In the first step, the EM fields are numerically calculated using finite-difference time domain (FDTD[66,67]) calculations performed in Ansys Lumerical FDTD software (Lumerical Inc., 3D Electromagnetic Simulator, version 2021 release 1.3) on the high-performance cluster of the Technical University of Denmark[68]. In the FDTD algorithm, the electric and magnetic components of the EM-field are propagated over a three-dimensional cartesian grid in a time-coordinated fashion. As depicted in Fig. S1b, the simulated domain encloses the water medium (containing NP and dipole emitter) and the top part of the glass coverslip. Discretizing the entire microscopy setup, from sample to camera, would result in impractical computational times and memory requirements, since these factors scale drastically with the size of the simulation domain. Since the immersion oil and the coverslip have the same refractive index as glass, they are treated as one medium. The complete simulation domain has dimensions 8 μm × 8 μm × 220 nm, with a mesh accuracy parameter of 6 (which Lumerical converts in a mesh size of 17 nm in case of a 200 nm NP). These parameters are determined by a parameter sweep and comparison with the analytical PSF (Fig. S6). The mesh is locally refined to a mesh size of 3 nm in a region of

200 nm × 200 nm × 180 nm around the NP, based on the rule of thumb to have a mesh size at least 10 times smaller than the radius of the NP. The simulation time is set to 40 fs, based on the speed of light and the dimensions of the simulation domain. Since the time it takes the light to travel from the dipole to the edge of the detector is approximately: $t = s * n/c = 8\,\mu m * 1.33 / 3e^8 = 35$ fs, with $s$ the distance, $n$ the refractive index and $c$ the speed of light. The EM fields are detected with a near-field detector that is positioned 20 nm below the water-glass boundary with a width extending the entire simulation domain. After running the numerical simulation, the detected EM near-fields are projected into the far-field[67] using the build-in function *farfieldexact*. Note that at this stage, the far-field represents the angular distribution of the collected light and is equivalent to the information found in the back-focal plane of the objective. The far-field is evaluated on a hemisphere with a radius of 1 meter, centered around the NP center. The far-field hemisphere is discretized in 100 × 100 points in, respectively, the polar and azimuthal direction. The calculated electric and magnetic complex vector fields on the hemisphere are saved and imported in MATLAB for further analysis. In the third step, the fields are focussed onto the camera plane forming the PSF using the Debye-Wolf diffraction integral, in the same way as for the analytical approach (Suppl. Note 2 chapter 6). As a unique addition to the field, the code of our validated numerical approach, presented here, including instructions for the FDTD simulations in Ansys Lumerical FDTD are both publicly available (see Methods).

All necessary scripts to construct the simulation elements, perform the far-field transformation and calculate the focussing are available on the GitHub repository connected to this publication.

**Sample preparation**

Glass coverslips (22 mm x 22 mm; thickness #1.5) are sonicated in a methanol bath for 20 mins and dried with nitrogen gas. The coverslips are then hydrophilized by ozone cleaning for 60 mins. For thiolation, the coverslips are dip-coated in a solution of MPTMS (5% v/v) in ethanol (purity >99.9%) for 3 mins followed by ethanol rinsing and nitrogen drying.

For immobilizing gold NPs (CTAB-coated 100 nm nanospheres obtained from Nanoseedz, product NS-100-50), 500 μL of stock solution was centrifuged at 1500 rcf

for 3 mins and the pellet was re-dispersed in 250 µL of 1mM CTAB in milliQ water to obtain 2x concentrated solution. Right before spin-coating the particles on the thiolated coverslips, the nanospheres are heated to 30 degrees Celsius for 1 min and sonicated for 30 seconds to minimize agglomeration. The nanospheres are then immobilized by spin-coating (at 2000 rpm for 1 min) on the thiolated coverslip. The coverslips with GNS are generously rinsed with methanol, phosphate buffered saline (PBS), and water, to get rid of the residual CTAB and weakly-bound particles.

The GNS were then incubated with a mixture of 5 µM of single-stranded DNA docking strands (5' SH-CAT CAT CAT ACG CTT CCA ATA ATA CAT CTA-3', from Integrated DNA Technologies) and 1mM tris(2-carboxyethyl)phosphine hydrochloride in citrate buffer (100 mM, pH 3, 1 M NaCl) for surface functionalization. After two hours, slides were rinsed with PBS and buffer B (5 mM Tris-HCl, 10 mM MgCl2, 1 mM EDTA, pH 8.0) to wash away free DNA strands. The slides were stored in buffer B in a humidity chamber, until the optical measurement.

A negative control experiment was performed to confirm that the observed fluorescent events represent specific binding of the imager strands to the docking strands. The sample with complementary imager strands was washed by buffer B two times. Then 50 µL of 200 pM (in buffer B) non-complementary imager strands (5'- TTC TTC TCC C- ATTO655-3', from Integrated DNA Technologies) were added to the sample slide, and imaging was performed.

**Fluorescence microscopy**

An inverted total internal reflection fluorescence (TIRF) microscope with a 100x/NA 1.49 objective (Nikon) was used for the DNA-PAINT experiment. Once the glass slide with GNS was mounted on the sample stage, buffer B was carefully removed, and then 50 µL of 200 pM of complementary imager strands (3' TAT GTA GAT C-5', from Integrated DNA Technologies) with ATTO 655 diluted in Buffer B was drop-casted with a pipette onto the slide. The sample was illuminated with a 637 nm laser (Coherent® OBIS™ FP 1193837 | 637nm LX 100mW) with a power density of 3.98 kW/cm$^2$, the intensity time traces were recorded with a CMOS camera (Prime BSI Scientific CMOS, physical pixel size 6.5 µm). The dichroic mirror (ZT640rdc-UF2, Chroma Technology) and several filters such as a notch filter (ZET635), a long-pass filter of 650 nm (Thorlabs) and a bandpass filter (center wavelength 700 nm,

bandwidth of 75 nm, Thorlabs) are used in the detection path to get high signal-to-noise ratio. Hundreds of PSFs were obtained from single gold nanospheres from 18000 frames recorded at 10 fps.

**Scattering spectroscopy and SEM**

Two additional measurements were performed to confirm that the NPs analyzed in the DNA-PAINT experiment were single, spherical GNS with a size of around 100 nm. After the fluorescent measurements, the same setup was used to execute single particle scattering spectroscopy by replacing the red laser with a supercontinuum white light source (SuperK Compact from NKT photonics) and all the wavelength filters in the detection path were removed. The microscope was still aligned in TIRF and a beam block was placed right below the objective to block the direct source light. The scattering signals were collected from 500 nm to 640 nm with a bandwidth of 10 nm for the same sample area as DNA-PAINT (Fig. S15a,b). The optical properties of plasmonic particles are dependent on the shape and size of the NP. Thus, the resonance peak and linewidth obtained from the single particle scattering spectroscopy by fitting a Lorentzian curve gives an insight into whether the spectrum is collected from a single particle or a cluster of particles. Particles 1 and 2 are highlighted in the images (Fig. S15a-c) to show the correlation. The single particle spectra from gold nanospheres 1 and 2 are shown in Figures S15e and S15h, respectively. The single particle spectra also depend on the environment of the particle, so deducing the size of the nanosphere is complex. Many bright spots were observed in the fluorescent measurements (Fig. S15a) and the particles which we chose for further PSF analysis were dimmer, mainly because the filter in the detection path was filtering out the PL signal from a single GNS. The brighter spots were thought to be clusters as the absorption cross-section scales with the size of the particle.

To be certain, SEM imaging was performed on the sample and the particles of interest were imaged at a higher magnification (Fig. S15c,f,i). For correlative SEM imaging, the glass slides were marked with an SEM pen marker (AgarScientific) before functionalizing the GNS with docking strands. After the DNA-PAINT and scattering measurements at the optical microscope, the sample slide was dried in air and treated with ozone for 10 mins to get rid of organic layers from the particle surface. The glass slides were then sputtered with a thin (approximately 8 nm) layer of gold to make the

substrate conducting for SEM imaging. The imaging was performed with a 10 kV electron beam at 250k magnification in SE mode (SEM Quanta 3D-FEG). The SEM images of the same field of view as optical microscope show that the brighter spots in the fluorescence measurements were indeed from dimers, trimers or clusters of nanospheres while the dimmer spots were single GNS of 100 nm on average (Fig. S15a,c). The particles which have been used for further analysis and fitting were confirmed as single nanospheres from scattering measurement and SEM images.

**Data processing and PSF fitting**

The data processing from a 3D (x-y-time) stack of raw microscopy data to PSFs is implemented in Python and performed as follows (Fig. S11). Spots of interest in the FOV are found by employing an algorithm to find local maxima (github.com/dwaithe/MaximaFinder) to the time-averaged stack of data, and maxima near the edge are removed. Intensity traces are generated by averaging the 15x15 pixel values around the found maximum in every time frame. Manually inspected spots (stable intensity background, many events, single spot in 15x15 neighborhood) are selected. For every selected spot, the events are found by detecting the time frames where the intensity is 100 values above the median of the entire trace. Consecutive time frames above the threshold are grouped and every group is considered an event. Per considered event, the 15x15 images are averaged over time, creating the raw PSFs. The one-photon photoluminescence (PL) signal and camera background noise (BG) are averaged over 10 consecutive frames before or after each event (from now on referred to as the PL+BG image), and subtracted from the raw PSF, resulting in the final experimental fluorescence PSF, which will be used for PSF fitting.

We fit all the PSFs obtained from a single GNS with our analytical PSF model, to localize the binding events on the GNS surface. The PSF model has a number of microscope-, nanoparticle-, and fluorophore-specific parameters that we do not have to fit since they are known, like the magnification, pixel size, NA, NP radius, NP refractive index, focus level, etc. (see Table S1 for the full list). Assuming the wrong NP radius has a minor effect on the obtained localizations (Fig. S20). The six free parameters we are left with are: the position of the emitter on the GNS (expressed in spherical coordinates $\theta$ and $\phi$), the 2D position of the GNS in the field of view ($x$, $y$), the number of photons $N$, and the background intensity $b$. The position of the GNS in

the image ($x$, $y$) is determined by performing a Gaussian fit using MLE[20] to the PL+BG image (Fig. S11), which means we only have four fit parameters left to fit with our analytical PSF ($\phi$, $\theta$, $N$, and $b$). We find these four parameters by a multi-step least-squares fitting approach of our analytical PSF model to the experimental PSFs.

In order to obtain a good starting guess for the fit, in the first step, we fit the PSF model of a fixed dipole emitter without the presence of a NP[20]. This is done for two reasons: 1) this fit is significantly faster and therefore an easy way to get a good initial guess, and 2) the PSF for an emitter on a gold NP resembles the PSF for an isolated fixed emitter quite well, since the strong plasmonic coupling between the emitter and the NP basically makes that the collective emission resembles a fixed dipole emitter (Fig. S21). This first fit is given the freedom to fit all six parameters ($\phi$, $\theta$, $x$, $y$, $N$, and $b$). The fit is initiated with the starting guess of (0, $\pi/4$, 0, 0, $N^*$, and $b^*$), where the center of the image has coordinates (0,0) and $N^*$ and $b^*$ represent estimated values from the PSF. The background $b^*$ is estimated as the average intensity on the border of the image of the PSF, and the number of photons $N^*$ is estimated as the sum intensity of the image minus the background estimate $b^*$. In the second step, we fit our analytical PSF model where we only fit the four necessary parameters ($\phi$, $\theta$, $N$, and $b$). Fitting $x$ and $y$ is not necessary, since they are known from the Gaussian fit to the PL+BG. The fit is initialized with the parameters found by the first fit of Mortensen *et al.*[20]. Fitting our full analytical PSF model in the second step is computationally feasible, since the initial parameters are already estimated approximately in the first step. Due to the strong plasmonic coupling, it is sufficient to reduce the analytical PSF model to only the dipole orientation that is perpendicular to the NP surface (Fig. S21), even though the fluorophore is freely rotating (Fig. S10) and the PSF should theoretically be averaged over three orthogonal dipole orientations (Fig. S19 and SI Note 3).

Given the ambiguity problem (see main text and Fig. S14), there are multiple emitter locations on the NP surface that result in a PSF with the same shape. More precisely, the two locations on opposite sides of the NP (e.g. north vs. south, or left vs. right) result in PSFs with a near-identical shape. Although these two PSFs are similar in shape, they have a different position in the image, allowing the fit to differentiate between the two ambiguous emitter locations by knowing the position of the NP from the PL. However, the fit may be biased towards finding the emitter location closest to the initial guess and thus not finding it on the other side of the NP. As a solution, we

initialize the fit four times with four opposite initial guess, to force the fit to sample the four ambiguous emitter locations on all side of the NP. All initializations result in a set of four parameters, of which we choose the optimal one by evaluating the mean-squared error between the fitted and the experimental PSF. All fitting is performed using non-linear least-squares (*scipy.optimize.curve_fit* in Python), since the PSFs contain a sufficient amount of photons[20] (Fig. S12).

**Aberration correction**

We correct the analytical PSF model for field-dependent aberrations before fitting to experimental data. We quantify the spherical aberrations over the FOV by looking at unspecific binding events away from the particles (Fig. S22). We first crop an area of approximately $5 \times 7$ μm around the NP of interest. We detect all spots in all frames using the ThunderStorm[69] plugin in Fiji and remove the localizations that are closer than 150 nm to a NP. We fit a 2D Gaussian function to each spot using the maximum-likelihood fitting approach of Mortensen *et al*[20]. For the example given in Fig. S22, we obtain $\sigma$ values (widths of the fitted Gaussians) of around 138 nm, which is bigger than the 118 nm expected from theory (Fig. S22c). The $\sigma$ values vary slowly over the FOV (Fig. S22). We assume the optical aberrations can be approximated by Gaussian blurring. We calculate that a convolution of the theoretical PSF with a Gaussian kernel with a $\sigma$ of 73.5 nm gives the experimentally obtained PSF width. For each analyzed NP, we obtain the width of the Gaussian blurring kernel individually since the severity of the aberrations depends on where the NP is located in the FOV. In the PSF fit, we correct for these spherical aberrations by blurring the analytical PSF model with the obtained Gaussian kernel. Correcting the PSF model with the wrong aberrations gives a bias in the obtained localizations (Fig. S20). We investigate whether a multivariate Gaussian is a better model for the aberrations, since it allows for elliptical aberrations (Fig. S23). However, we find that the ellipticities of the unspecific binding events are negligible, and that fitting with a PSF model that is corrected for elliptical aberrations does not significantly change the obtained emitter locations.

**Localization uncertainty**

The localization uncertainty as presented in Fig. 4d of the Main text is defined as the uncertainty on the NP surface. The least-squares fitting procedure provides us with the uncertainty (standard deviation) on each of the estimated parameters. The uncertainty on the emitter position on the NP surface (expressed in spherical

coordinates $\theta$ and $\phi$), are angular uncertainties, which we convert to positional uncertainties by multiplying with the radius of the NP. The overall uncertainty on the position is calculated as $\sqrt{\Delta_\theta^2 + \Delta_\phi^2}$. For example, a reported localization uncertainty of 5 nm means that the found localization could be off by a distance of 5 nm moved over the spherical surface. The localization precision is consistent with fitting the PSF model to individual frames of a single binding event (Fig. S24). Note that since the experimental PSFs are obtained by averaging multiple frames belonging to the same binding event, the localization uncertainty is in principle dependent on the length of the event, since averaging will improve the signal-to-noise. However, in practice, the signal-to-noise of single frames is already high enough for single frame fitting.

## Acknowledgements

**Author contributions**

T.A.P.M.H. and K.I.M. conceptualised the idea. T.A.P.M.H. derived theory, developed methods, ran simulations, performed PSF fitting, wrote code, and analysed data. S.M. performed the experiments. T.A.P.M.H. made main figures, SI figures, and SI notes. R.M., and K.I.M. supervised the work at DTU and P.Z. supervised the work at TU/e. T.A.P.M.H. wrote the original draft with help from S.M. and all authors commented on the paper.

**Data availability**

The raw microscopy data is available for download via a repository server after publication.

**Code availability**

The software is available for download under the terms of the Apache2.0 licence from GitHub after publication.

**Competing interests**

The authors declare no competing interests.

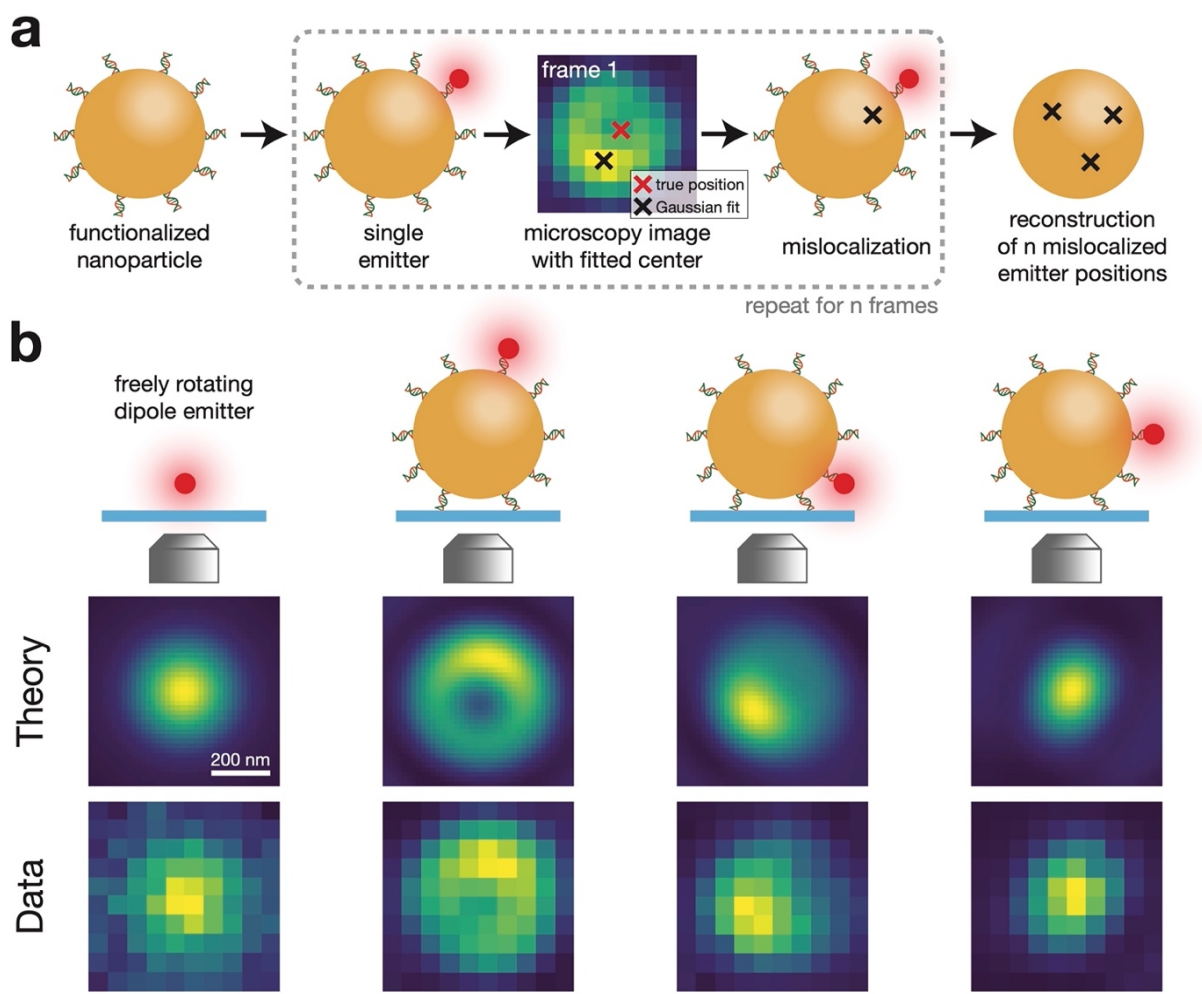

**Figure 1: Commonly employed methods for single-molecule localization microscopy on nanoparticles for mapping of functional groups are hampered by PSF distortion. a** Schematic of the process of localizing single emitters on a nanoparticle (NP) with localization microscopy. The surface of a functionalized NP is covered with chemical groups, here depicted as DNA strands. In single-molecule localization microscopy, light-emitting fluorescent dyes (red dots) bind transiently to functional groups of interest. Since only a small subset of available functional groups is labeled at each moment in time, each dye creates an isolated spot of light on the camera. The dye is localized (black cross) by fitting a point-spread function (PSF) model to the spot, to estimate the true position of the emitter (red cross) on the NP. Since the functional groups are transiently bound by a fluorophore, this process can be repeated sequentially for many frames in order to reconstruct a map of all functional groups on the NP. Due to PSF distortions, however, there is discrepancy, often substantial, between the true and fitted emitter position, referred to as a mislocalization. **b** Theoretical and experimental PSFs for a single freely-rotating fluorophore in the presence and absence of a NP, imaged on top of a glass coverslip. A single fluorophore (left column) creates a rotationally symmetric PSF, which may be modeled well by a 2D Gaussian PSF. When the fluorophore is located next to a NP (100 nm gold spherical NP, columns 2-4), the PSF does not resemble a 2D Gaussian, but is distorted in a manner that depends on the position of the fluorophore relative to the NP. Scale bar applies to all PSFs in **b**. Schematics are not to scale.

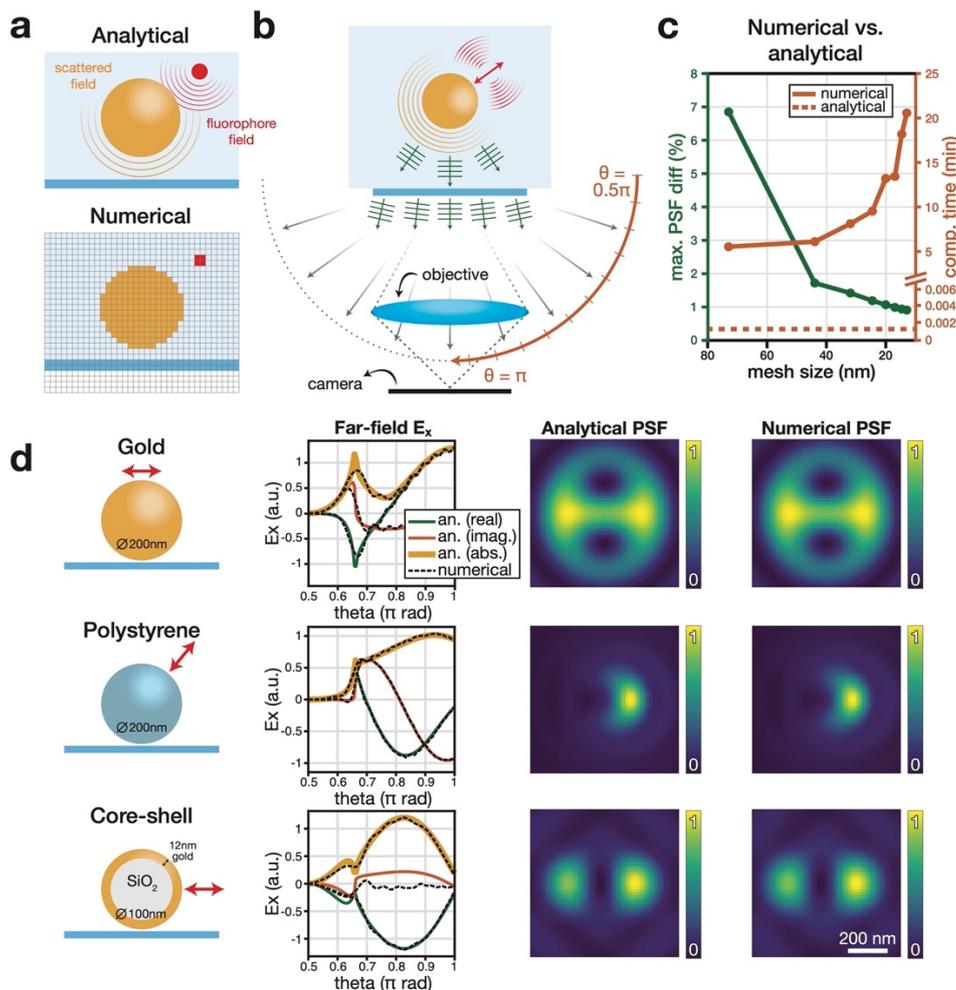

**Figure 2: Numerical validation of our analytical PSF model. a** Schematics of the analytical and numerical approaches for calculating the PSF for an emitter (red dot) in close proximity to a NP (gold sphere) in water on top of a glass coverslip (blue rectangle). In the analytical model, we directly and analytically derive the electromagnetic fields everywhere in the system. In the numerical approach, the fields are computed in a finite element manner on the discretized system. **b** Visual depiction of the steps of our analytical PSF model. Briefly, the dipole (red) and scattered (orange) fields are decomposed into plane waves (green), which are refracted at the water-glass interface, projected onto the far-field hemisphere (grey), collected by the objective (blue ellipse) and focused onto the camera. The far-field hemisphere is parametrized by a polar angle θ, which is further used in **d**. **c** The difference between analytically and numerically calculated PSFs for a horizontally oriented dipole emitter on top of a 200 nm gold spherical NP, as a function of mesh sizes in the numerical calculation. The numerical PSF converges towards the analytical PSF for decreasing mesh size (left green *y*-axis). However, the computational time (right orange *y*-axis) drastically increases for the numerical approach (solid line) for smaller mesh sizes. In comparison, the computational time for the analytical PSF (dashed line) is four orders of magnitude faster and not depending on discretization. **d** Comparison of the electric far-fields and the PSFs of the analytical and numerical approaches for a fixed dipole emitter positioned with various orientations close to different NPs. The left column shows schematics of the nanoparticle-dipole (red arrows) configurations. The second column shows the comparison of the absolute, real, and imaginary components of the *x*-component of the electric far-field calculated using the analytical (respectively yellow, green, and orange lines) and numerical (black dashed lines) methods (*y*- and *z*- components shown in Suppl. Fig. S7). The third and fourth columns show the normalized numerical and analytical PSFs. Scale bar applies to all PSFs. Parameter values used in the two approaches can be found in Suppl. Tables 1 and 2.

26

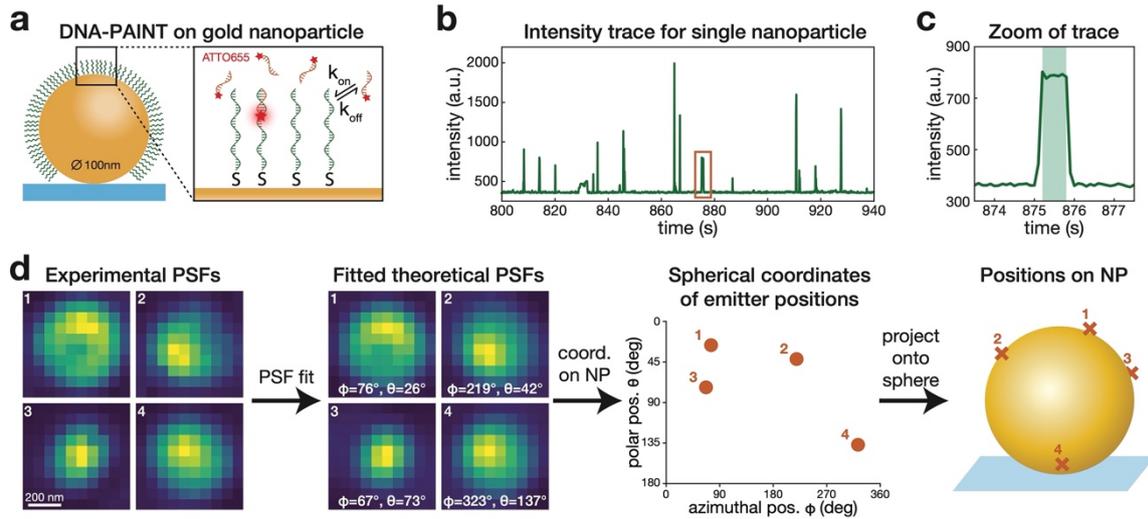

**Figure 3: Workflow for localizing single emitters on the surface of a nanoparticle using our analytical PSF model. a** DNA-PAINT experiments are performed on 100 nm gold spherical NPs. Single-stranded DNA docking strands (green, not to scale) are bound to the NP surface with a thiol bond, which mimic functional groups on the NPs. To these, freely diffusing single-stranded complementary DNA imager strands (red), conjugated to ATTO655 (red star), transiently bind, creating fluorescent bursts on the camera. **b** Example time trace of detected fluorescence intensity from a single NP shows multiple binding events. **c** Zoom of the orange-outlined event in **b**, of which the frames of the binding event itself (green shaded area) are averaged to obtain the PSF. **d** Each experimental PSF is fitted with our analytical model resulting in a fitted theoretical PSF. The high similarity between the experimental PSFs and the fitted theoretical PSFs shows that the analytical model perfectly describes the experimental data. The fit has four free parameters: the position of the emitter relative to the NP (expressed in spherical coordinates $\phi$ and $\theta$), the number of photons, and the background intensity. By PSF fitting, the position of each emitter on the 100 nm spherical NP surface is obtained, expressed in spherical coordinates. These positions are projected onto the spherical NP to map the binding sites in 3D. Scale bar applies to all PSFs.

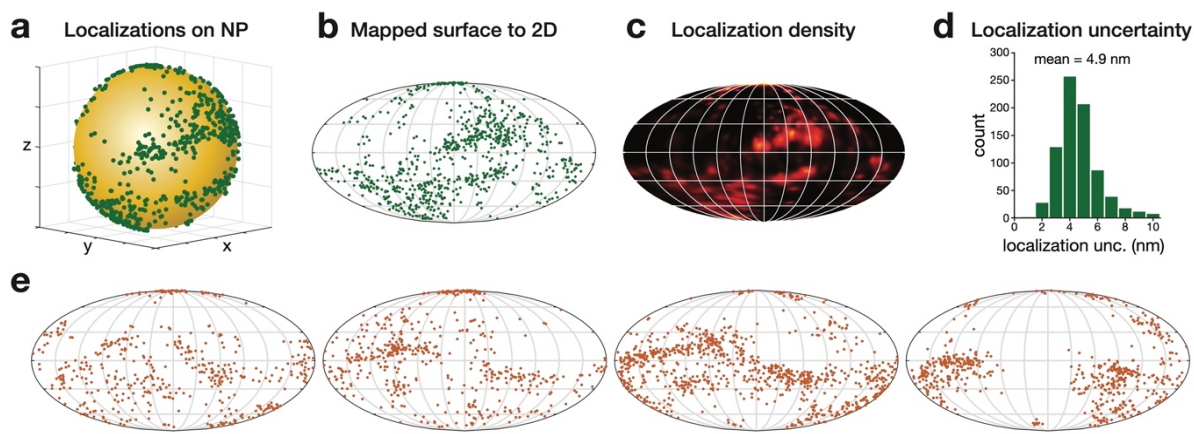

**Figure 4: Inhomogeneous surface coverage of functional groups. a** All localizations (green dots) of 828 PSFs from the same single NP. **b** Area-conserved 2D mapping of the NP surface to visualize the distribution of all localizations. **c** Localization density map resulting from rendering each localization as a normalized 2D asymmetric Gaussian, where the two Gaussian axes represent the localization uncertainty in the polar- and azimuthal direction, respectively. **d** Distribution of localization uncertainties of all 828 localizations, with an uncertainty mean ± std of 4.9 ± 2.1 nm. **e** Localization maps of 4 different NPs, containing respectively 477, 561, 533, and 882 localizations, highlighting the particle-to-particle variation in surface functionalization.



\renewcommand{\chaptermark}[1]{\markboth{#1}{#1}} 
\pagestyle{fancy}
\fancyhf{}
\fancyhead[L]{\leftmark} 
\fancyfoot[C]{\thepage} 

\counterwithout{section}{chapter}   

\maketitle

\setcounter{tocdepth}{1}

\tableofcontents

\chapter*{Supplementary Figures}
\addcontentsline{toc}{chapter}{Supplementary Figures}
\chaptermark{Supplementary Figures}

\phantomsection 
\begin{figure}[ht]
    \addcontentsline{toc}{section}{S1. Microscope model and its numerical/analytical implementations}
    \centering
    \includegraphics[width=1\linewidth]{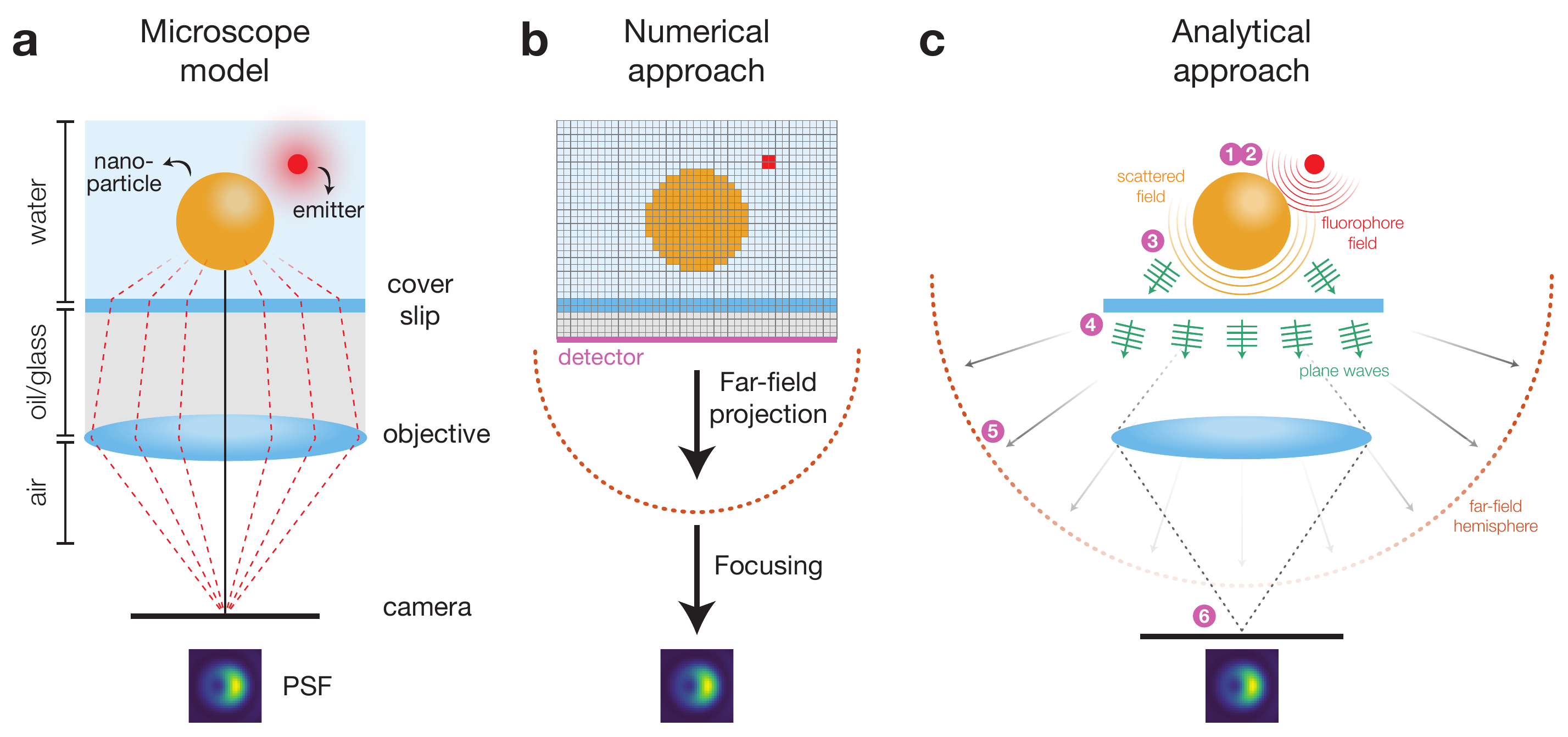}
    \caption{\textbf{Microscope model and its numerical/analytical implementations}. \textbf{a} The microscope system for the PSF calculations. The nanoparticle and dipole emitter are present in water, just above a glass coverslip. The coverslip is mounted above the objective with a drop of refractive index-matching immersion oil in between the coverslip and the objective. The imaging part of the microscope, consisting of the objective lens and multiple other lenses, which together focus the light onto a camera, is modeled as a single lens. \textbf{b} The numerical approach for calculating the PSF for the system in \textbf{a}, consists of three steps. In the first step, the electro-magnetic (EM) fields in the upper part (water, glass and oil) of the discretized system are numerically calculated using finite-difference time-domain (FDTD \cite{yee1966,taflove2005}) calculations implemented in the Ansys Lumerical FDTD software. Note that the glass and oil have the same refractive index, so they are considered as one medium in the numerical calculations. The EM fields are detected on the detector (pink) and, in the second step, propagated onto the far-field hemisphere (red) one meter away from the nanoparticle center, using the build-in function \texttt{farfieldexact} in the FDTD software. In the third step, the fields are focused onto the camera plane forming the PSF. \textbf{c} The analytical approach for calculating the PSF for the system in \textbf{a}, consists of 6 steps (see Figure \ref{fig:S2_overview} for the individual steps and Supplementary Note 2 for mathematical details). We start (1) by calculating the EM fields in the water by representing the dipolar emission of the fluorophore as a linear combination of vector-spherical harmonics (VSHs). Next, we use Mie scattering theory to define how the light emitted by the dipole is scattered by the spherical nanoparticle, still as a expansion in VSHs. In the next step (2) the VSH expansion is rotated around the nanoparticle center to place the dipole emitter on the desired position relative to the particle. The combination of dipole and scattered fields are subsequently: decomposed into plane waves (3), refracted in the water-glass interface (4), projected into the far-field (5) and focused by the objective onto the camera (6). The intensity of the light on the camera plane, given by the $z$-component of the Poynting vector, represents the PSF.}
    \label{fig:S1_microscope}
\end{figure}

\newpage
\begin{figure}[ht]
    \addcontentsline{toc}{section}{S2. Schematic overview of the analytical calculation of the PSF}
    \centering
    \includegraphics[width=1\linewidth]{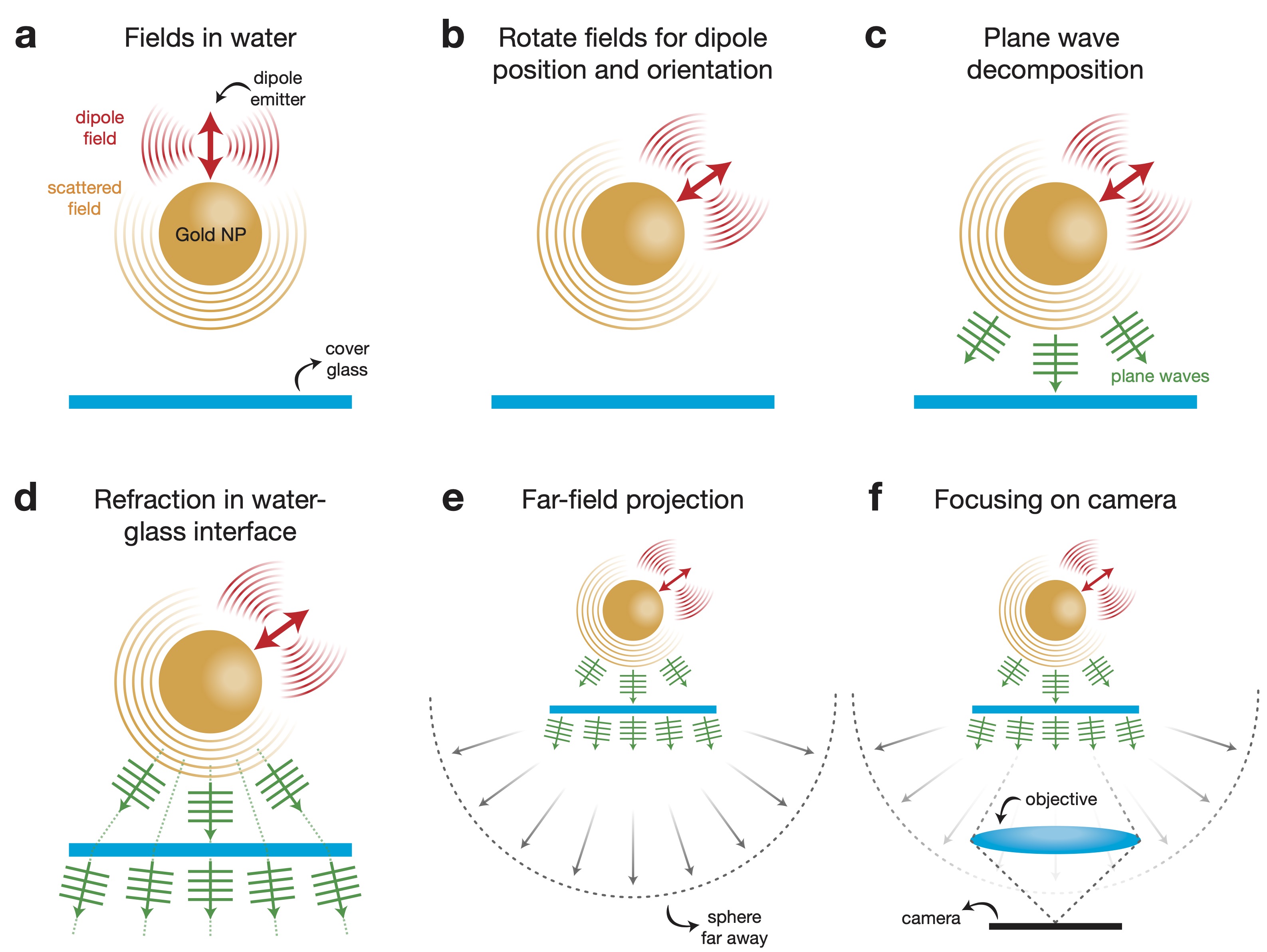}
    \caption{\textbf{Schematic overview of the analytical calculation of the PSF}. The process of calculating the PSF of a dipole emitter next to a spherical nanoparticle is composed of 6 steps: \textbf{a} calculation of the EM-fields in the water medium for a standard dipole, fixed in orientation and positioned on top of a NP, in terms of spherical waves, \textbf{b} rotation and interpolation of the fields to obtain the desired dipole position/orientation, \textbf{c} decomposition of the fields into plane waves, \textbf{d} refraction of the plane waves in the water-glass interface, \textbf{e} projection of the fields into the far-field, and \textbf{f} focusing the fields onto the camera. Schematics are not to scale.}
    \label{fig:S2_overview}
\end{figure}

\newpage
\begin{figure}[ht]
    \addcontentsline{toc}{section}{S3. Multiple scattering can be ignored}
    \centering
    \includegraphics[width=.64\linewidth]{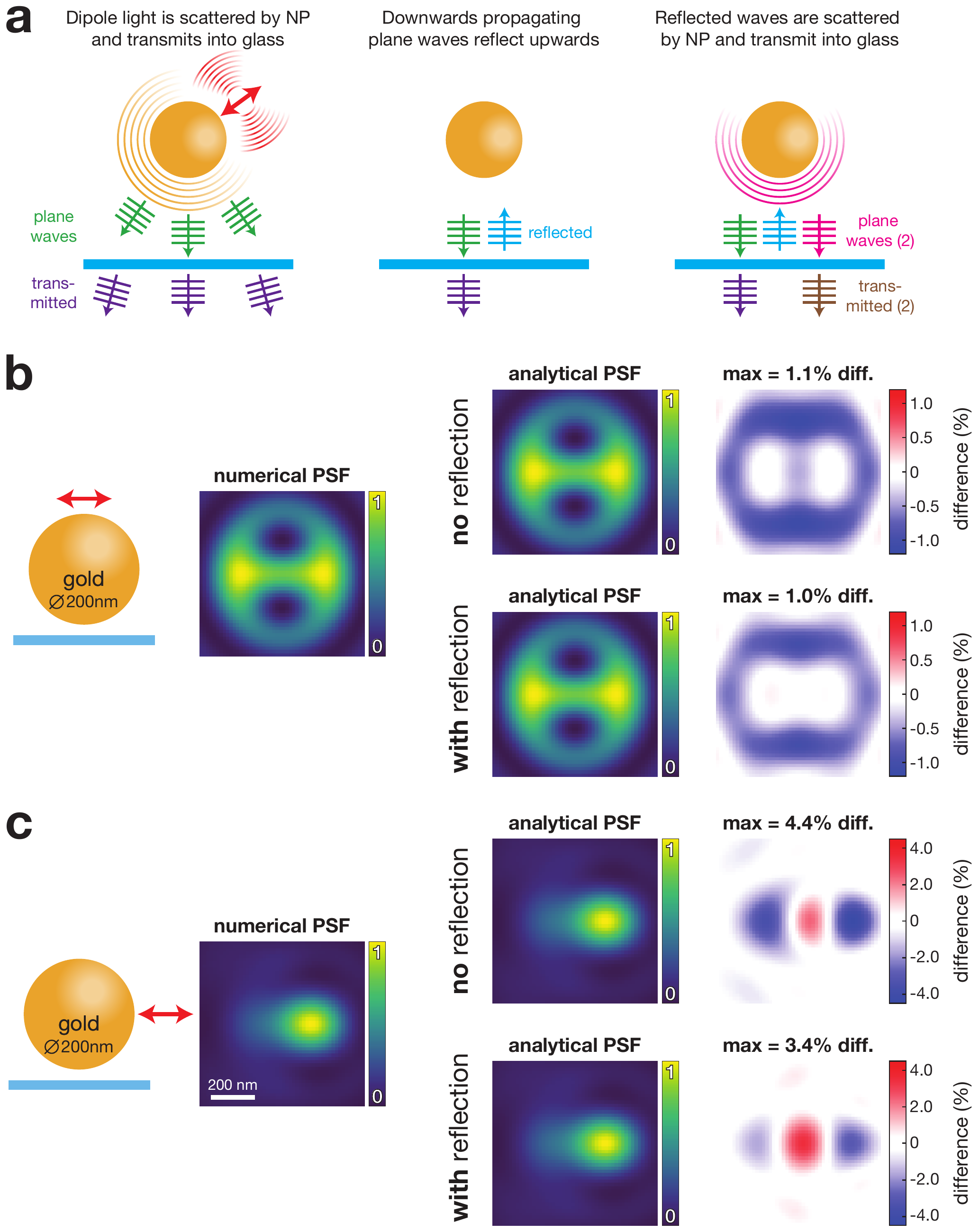}
    \caption{\textbf{Multiple scattering can be ignored}. \textbf{a} In the presented analytical model, the dipole field (red) and scattered field (orange) are jointly decomposed into plane waves (green) in order to refract them in the water-glass interface to obtain the plane waves in the glass (purple). In the model, we only consider the plane waves that transmit into the glass coverslip (purple) when calculating the far-field in the next step. \textbf{b} However, part of the downward propagating plane waves in the water (green) are reflected (blue) and will again interact with the NP. \textbf{c} The upward propagating plane waves (blue) are scattered by the NP for the second time, of which the downward propagating component (pink) again encounters the water-glass interface and undergoes refraction. This creates a second set of downward propagating plane waves in the glass (brown), that will travel through the microscope towards the camera.\\ \\
    In principle, this loop of reflection-scattering should be repeated an infinite amount of times. However, only a small fraction of light will be reflected back into the water, so every iteration the reflected component decays in intensity. To investigate whether it is necessary to include the reflected component (and further iterations) in the analytical pipeline, we calculate the PSFs for two different NP-dipole systems both without (only the purple contribution in \textbf{a}) and with (purple and brown) the reflected contribution. \textbf{b} Numerical and analytical normalized PSFs calculated for a horizontally oriented dipole on top of a 200 nm gold spherical NP. The analytical PSFs are calculated either without or with the reflection component. By calculating the difference between the numerical and analytical PSFs, we see that the biggest relative pixel differences are 1.1\% and 1.0\%, respectively. \textbf{c} Same as \textbf{b} for a horizontally oriented dipole positioned next to the NP, with biggest relative pixel differences of 4.4\% and 3.4\% (bigger differences explained by limitations in the numerical approach, not the analytical). \\ \\
    Both examples show that adding the reflection contribution is making the PSF more accurate, but only with a insignificant amount. From this we decide to not include the reflection component in the final model, since the extra computational time (due to an extra conversion from plane waves to spherical waves, and back to plane waves) is significant. Scale bar applies to all PSFs and full simulation parameters can be found in SI Tables 1 and 2.}
    \label{fig:S3_secondRefl}
\end{figure}

\newpage
\begin{figure}[ht]
    \addcontentsline{toc}{section}{S4. Infinite sum of Mie orders can be truncated without significant effect on the PSF} 
    \centering
    \includegraphics[width=.9\linewidth]{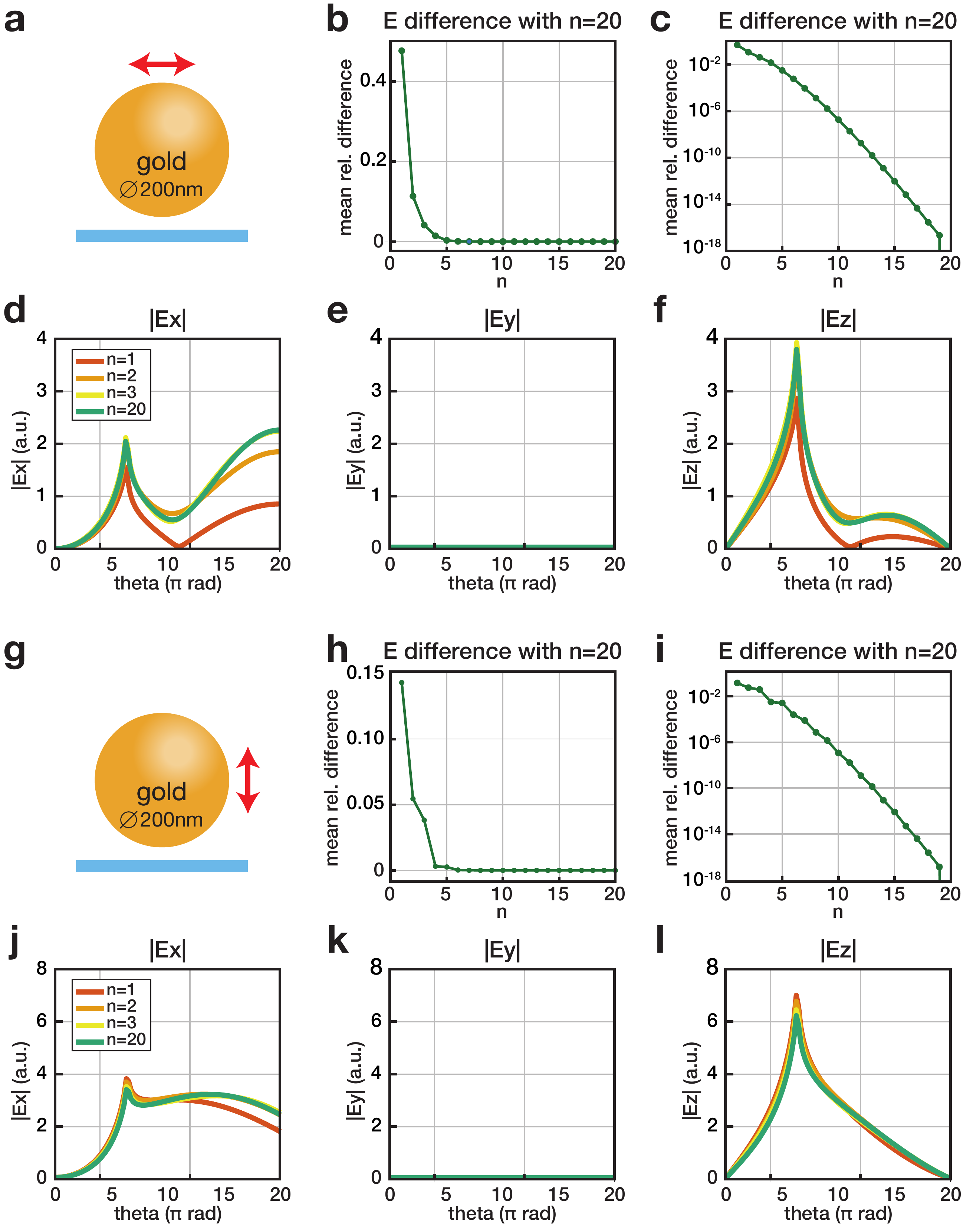}
    \caption{\textbf{Infinite sum of Mie orders can be truncated without significant effect on the PSF}. This figure shows that it is sufficient to truncate the Mie series at $n=10$, since the higher orders barely affect the $E$-fields. To investigate the effect of truncating the infinite series of Mie theory, we study how many orders are necessary to correctly describe the electric $E$-field in the far-field. Two different configurations are investigated: a fixed dipole positioned horizontally on top (\textbf{a-f}) or vertically on the side (\textbf{g-l}) of a 200 nm gold spherical NP. \textbf{a} Schematic representation of the system. \textbf{b} Mean relative difference in electric far-field for an increasing number of included Mie orders \textit{n}. The difference is calculated between the respective order and $n=20$. \textbf{c} Same as \textbf{b}, but on logarithmic scale. \textbf{d-f} Magnitude of $E$-field in the \textit{x}-, \textit{y}, and \textit{z}-direction, respectively, on an arc of the hemisphere parametrized by polar angle $\theta$, for multiple Mie orders (Note: green line in \textbf{d} is identical to the yellow line in the top panel of Fig. 2d of main text). \textbf{g-l} Same as \textbf{a-f}, but for vertically oriented dipole on the side of the NP. For all analytical parameters, see SI Table 1.} 
    \label{fig:S4_MieOrders}
\end{figure}

\newpage
\begin{figure}[ht]
    \addcontentsline{toc}{section}{S5. Computational time of analytical PSF depends on analytical parameters}
    \centering
    \includegraphics[width=1.0\linewidth]{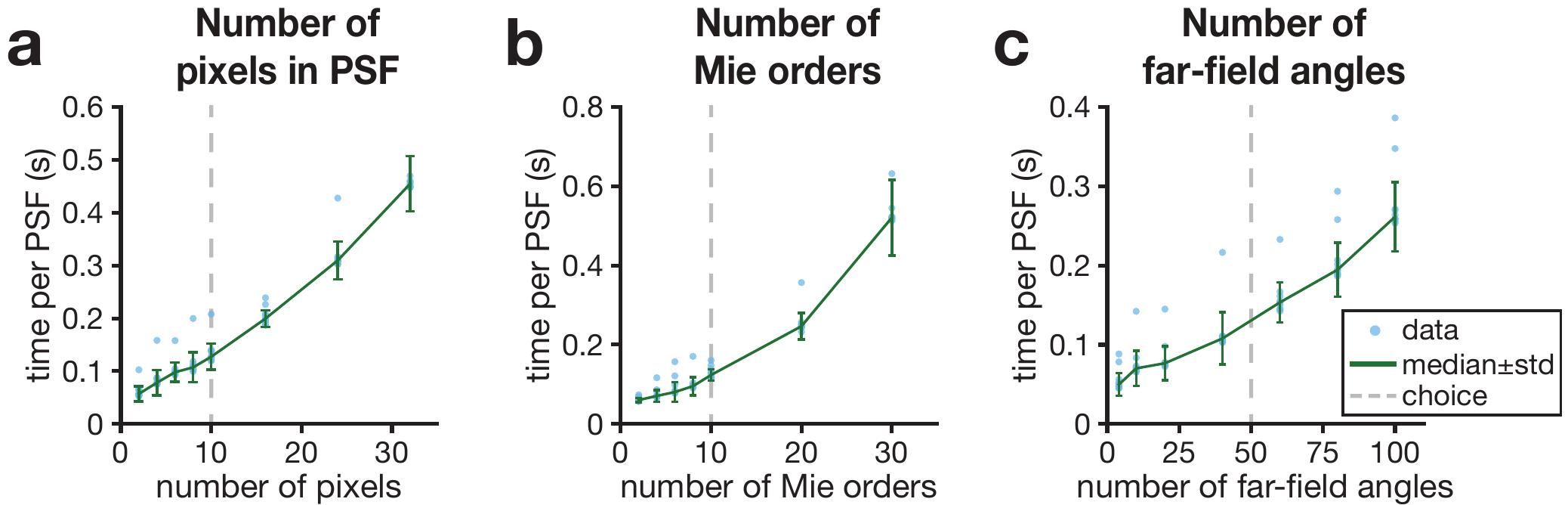}
    \caption{\textbf{Computational time of analytical PSF depends on analytical parameters}. The dependency of the computational time required for calculating a single analytical PSF on 3 parameters: \textbf{a} the number of pixels in the PSF, \textbf{b} the number of considered Mie orders, and \textbf{c} the number of discretized far-field angles. For clarity, with the number of pixels in the PSF we refer to the number of pixels per dimension ($x$ and $y$), so 10 pixels in plot \textbf{a} means a PSF of $10\times10$ pixels. For every condition in every plot, we repeat the PSF calculation 10 times (blue dots) and report the median (green line), where the errorbars represent the standard deviation. The vertical grey lines indicate the parameter values normally used in the PSF fiting procedure. The variation in computational times between the 10 repeats for the same parameter set is explained by the fact that the computational speed depends on other processes running on the computer.}
\end{figure}

\newpage
\begin{figure}[ht]
    \addcontentsline{toc}{section}{S6. Determining numerical parameters: simulation size and mesh size}
    \centering
    \includegraphics[width=1\linewidth]{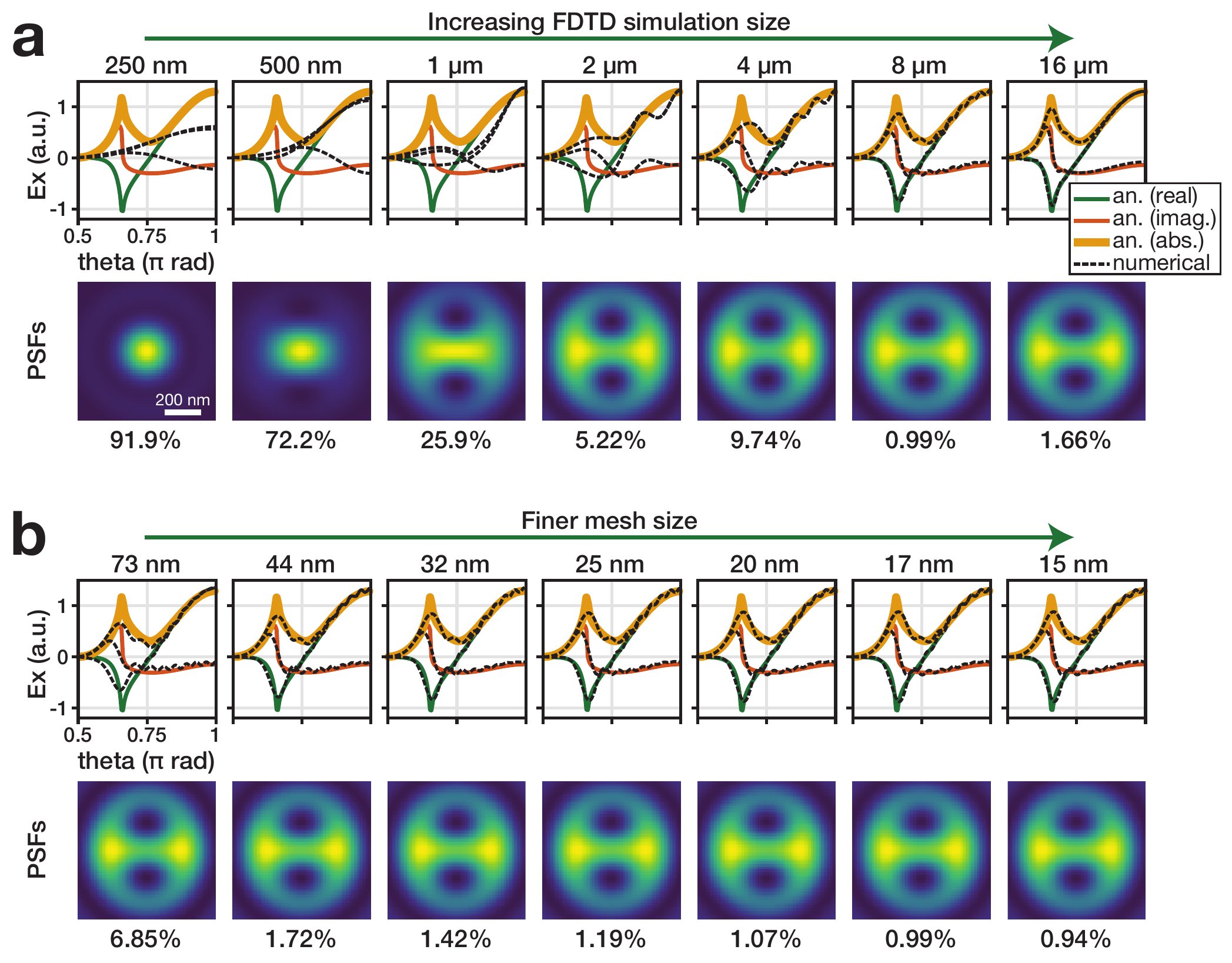}
    \caption{\textbf{Determining numerical parameters: simulation size and mesh size}. 
    When calculating the PSF numerically with FDTD calculations, we decide to use an FDTD simulation domain size of 8 $\mu$m and a mesh size of 17 nm. For the reasons that these values give accurate enough PSFs. A wider simulation domain and finer mesh would result in an impractical increase in computational time.
    In order to determine these optimal parameters for the numerical PSF calculations, we perform a sweep over a range of values. 
    \textbf{a} Multiple numerical calculations with different widths of the FDTD simulation domain (i.e., the width of the domain in FDTD that is discretized and simulated), ranging from 250 nm to 16 $\mu$m. The calculations are for a horizontally oriented dipole emitter positioned on top of a 200 nm gold spherical NP. Top row shows the comparison of the absolute, real, and imaginary components of the $x$-component of the electric field measured on an arc of the far-field, calculated using the analytical (respectively yellow, green, and orange lines) and numerical (black dashed lines) approach. Bottom row shows the numerical PSFs. The numbers below the PSFs indicate the maximum relative pixel differences between the numerical and analytical PSF, showing that a wider FDTD simulation size (direction of green arrow) gives a more accurate numerical PSF.
    \textbf{b} Same as \textbf{a} but varying the mesh size of the numerical calculation from 73 nm to 15 nm. The PSF differences indicate that a finer mesh (direction of green arrow) results in a more accurate numerical PSF. Scale bar applies to all PSFs and full simulation parameters can be found in SI Tables 1 and 2. }
    \label{fig:S6_numSweep}
\end{figure}

\newpage
\begin{figure}[ht]
    \addcontentsline{toc}{section}{S7. Full comparison of the analytical and the numerical PSF for multiple NP materials, sizes and composition}
    \centering
    \includegraphics[width=.75\linewidth]{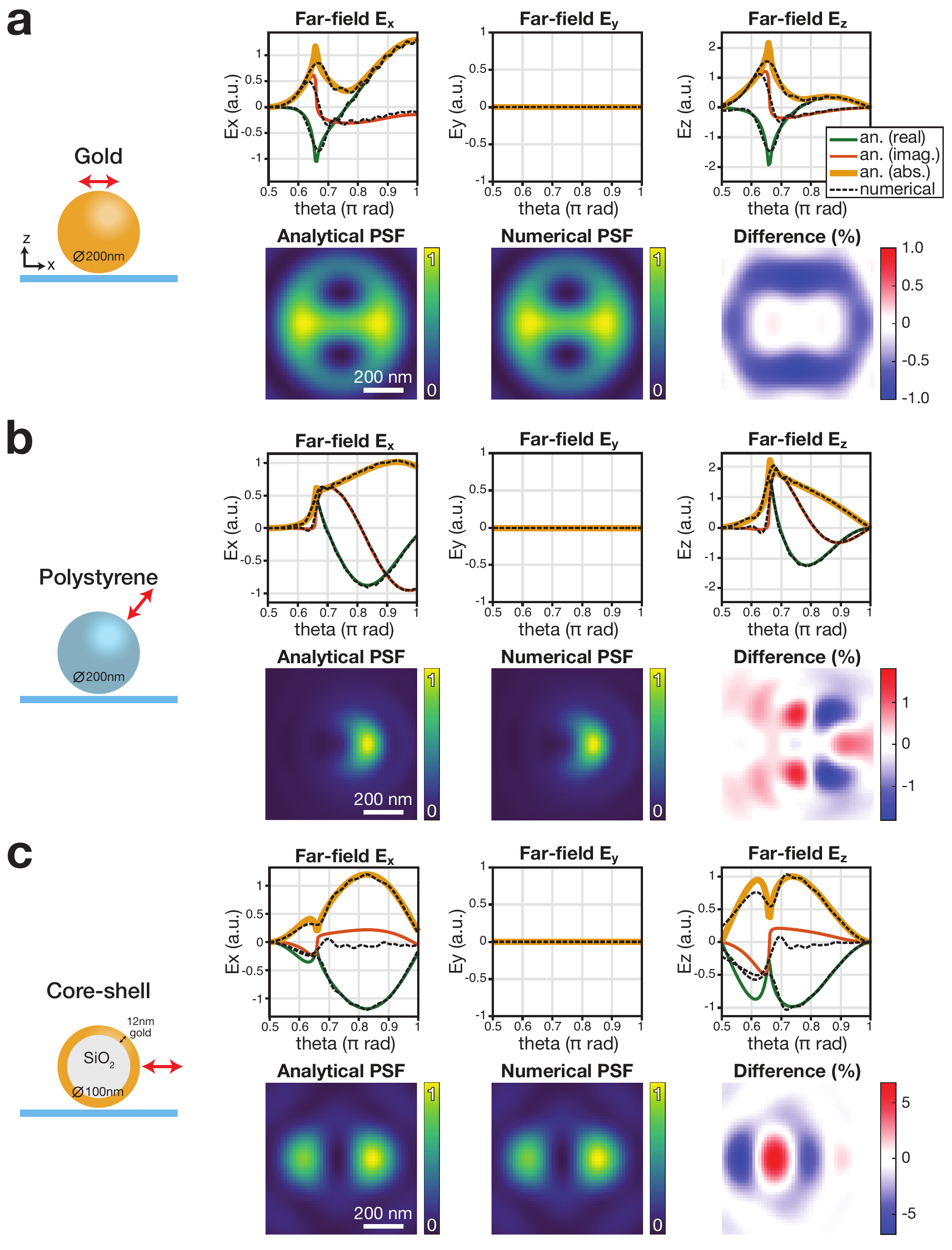}
    \caption{\textbf{Full comparison of the analytical and the numerical PSF for multiple NP materials, sizes and composition}. 
    Comparison of the electric far-fields and the PSFs between the analytical and numerical approach for a fixed dipole emitter positioned with various orientations and positions next to different NPs. Fluorophore positions and orientations are chosen to create exotically shaped PSFs. \textbf{a} The schematics on the left shows the nanoparticle-dipole configuration, a horizontally oriented fixed dipole emitter on top of a 200 nm gold spherical NP. Top row shows the comparison of the absolute, real, and imaginary components of the $x$-, $y$-, $z$-components of the electric far-field calculated using the analytical (respectively yellow, green, and red lines) and numerical (black dashed lines) method on an arc of the far-field hemisphere (see Fig. 2b in main text). The bottom row shows the normalized numerical and analytical PSFs, and their difference. \textbf{b}, same as \textbf{a}, but for a $45\deg$ tilted dipole emitter next to a 200 nm polystyrene spherical NP. \textbf{c}, same as \textbf{a}, but for a horizontally oriented dipole emitter next to a core-shell nanoparticle with a 88 nm silica-oxide core and a 12 nm gold shell. Scale bars apply to all PSFs, and details of the analytical and numerical parameters can be found in Suppl. Tables 1 and 2. \\ \\
    The results show that the analytical PSFs are almost identical to the numerical calculations, with maximal pixel differences around only 1\%. Keep in mind that the numerical calculations are limited by discretization resolution and computational time. The discrepancy between the two approaches is clearly visible for the core-shell calculations in \textbf{c}, because the gold shell of 12 nm is too thin relative to the numerical mesh size to result in a reliable far-field and PSF. A finer numerical mesh size would give better results, but would require impractically long numerical computational times.} 
    \label{fig:S7_fullComp}
\end{figure}

\newpage
\begin{figure}[ht]
    \addcontentsline{toc}{section}{S8. Library of distorted PSFs from a single NP}
    \centering
    \includegraphics[width=.95\linewidth]{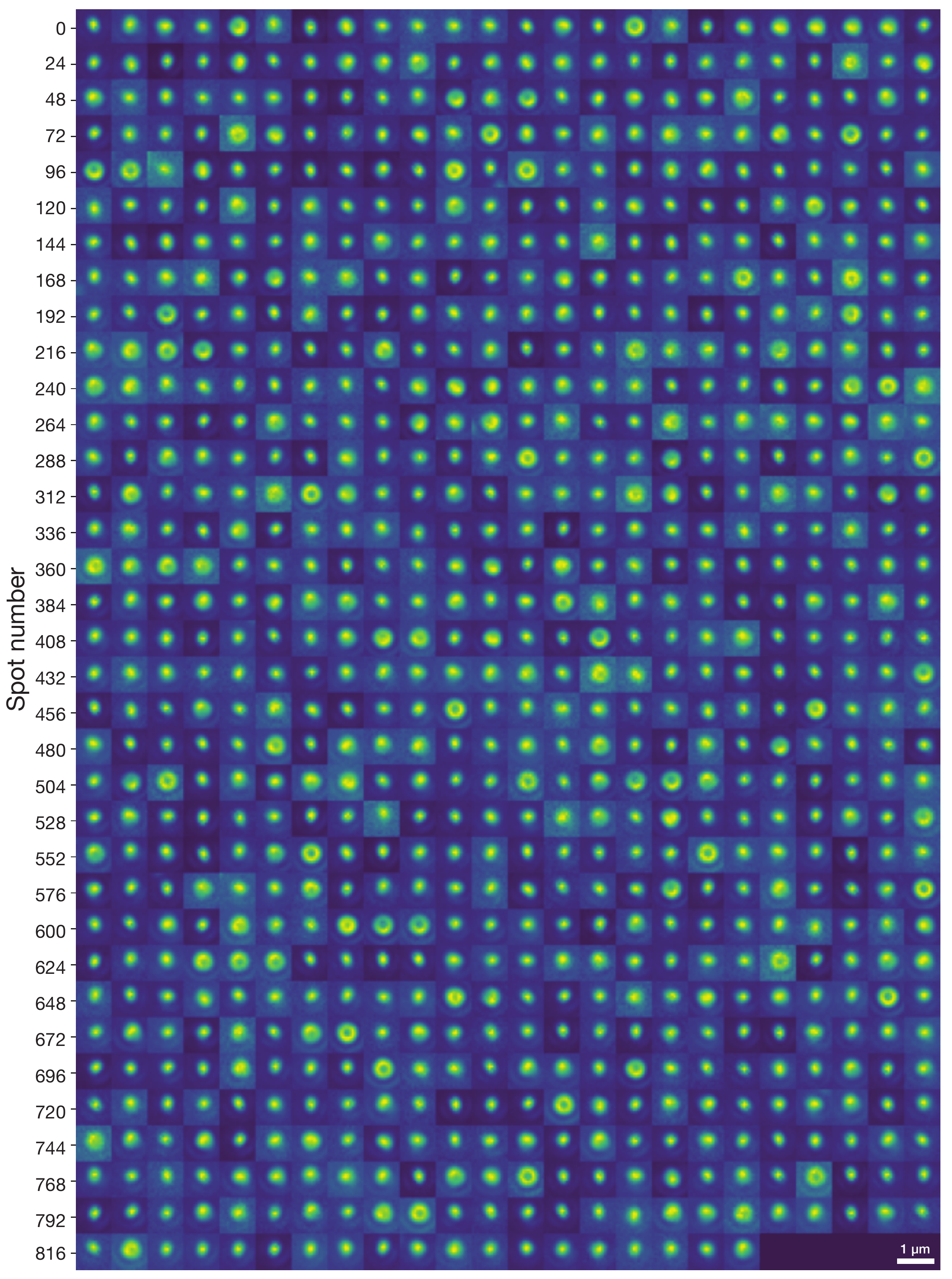}
    \caption{\textbf{Library of distorted PSFs from a single NP}. This figure shows all 835 detected PSFs from a single NP, obtained using the data analysis pipeline explained in the Methods. The PSFs are ordered based on when they happened during the 30 minute measurement, and every PSF is individually normalized to its maximum value.} 
    \label{fig:S8_libraryExotic}
\end{figure}

\newpage
\begin{figure}[ht]
    \addcontentsline{toc}{section}{S9. DNA binding events are specific, since non-complementary DNA-sequences result in no binding events.}
    \centering
    \includegraphics[width=.95\linewidth]{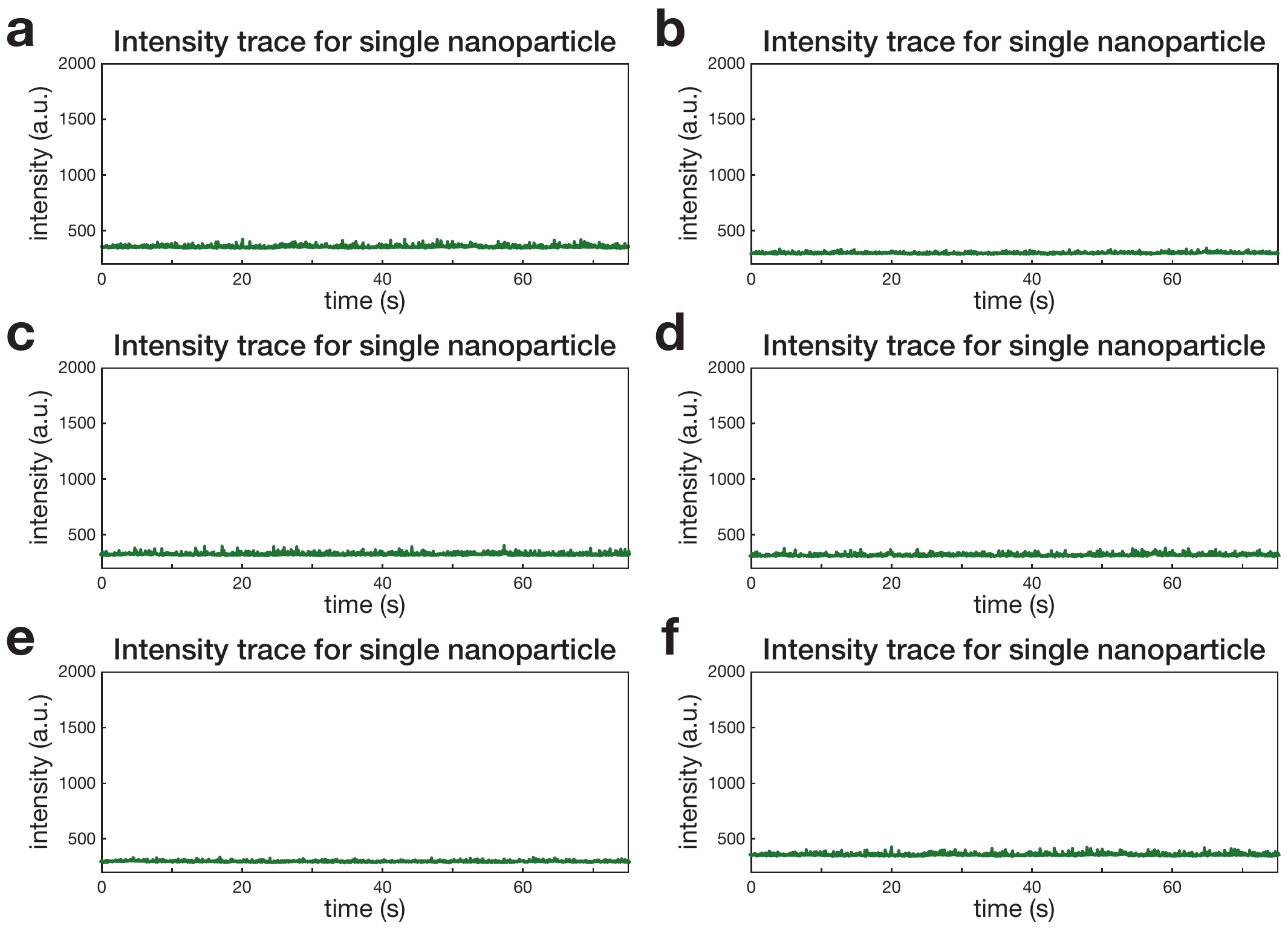}
    \caption{\textbf{DNA binding events are specific, since non-complementary DNA-sequences result in no binding events}. \textbf{a-f} Collection of 6 representative intensity time traces from single NPs from different positions in the field-of-view. In this control experiment, the DNA sequence of the imager strands is non-complementary to the sequence of the docking strands (see Methods), which results in no binding events on the NP (compare to Fig. 3b and S8a), seen by the absence of intensity peaks in the six time traces. From the absence of binding events in this control, we conclude that the binding events observed in the experiment with complementary sequences are specific. If the observed binding events resulted from unspecific binding to the NP or glass, we would also observe them in this control experiment, which is not the case. Note that the scale of the $y$-axis is chosen to be the same as in Fig. 3a, to show the absence of binding events. The slight difference in intensity level between the six traces shown here originates from the fact that the excitation intensity slightly varies over the field-of-view.} 
\end{figure}

\newpage
\begin{figure}[ht]
    \addcontentsline{toc}{section}{S10. Exotically-shaped PSFs are not visible for DNA-PAINT on glass in the absence of the nanoparticle.}
    \centering
    \includegraphics[width=.95\linewidth]{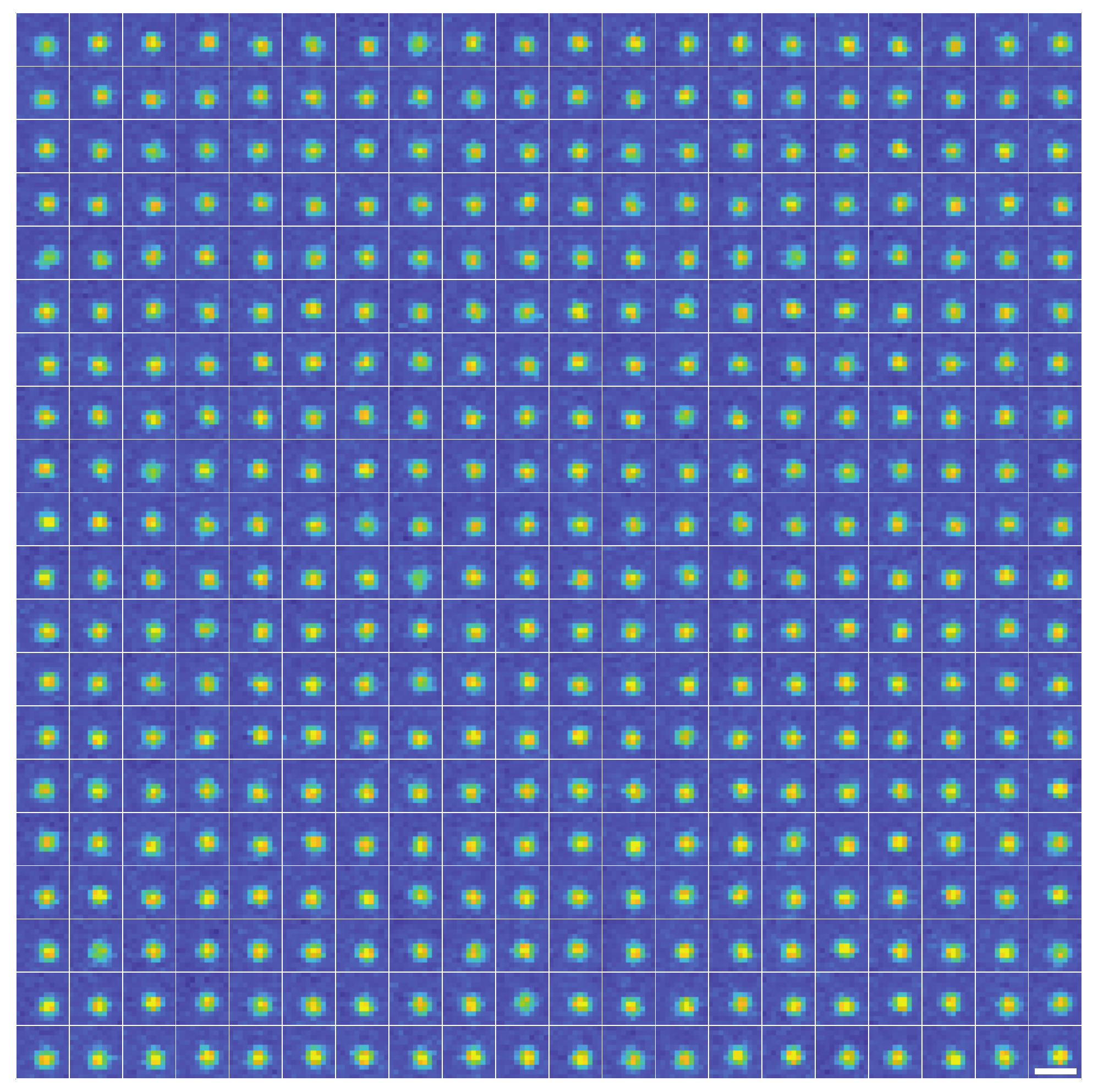}
    \caption{\textbf{Exotically-shaped PSFs are not visible for DNA-PAINT on glass in the absence of the nanoparticle}. Collection of representative PSFs obtained by performing DNA-PAINT on the glass in the absence of NPs. The docking strand are attached to the glass via biotin-streptavidin interactions. We see that all the PSFs have a Gaussian shape, from which we conclude that 1) in DNA-PAINT the fluorophores are freely rotating, otherwise we would have observed PSFs for orientationally-fixed fluorophores \cite{Mortensen2010}, and 2) the exotically-shaped PSFs for DNA-PAINT on NPs (see Fig. \ref{fig:S8_libraryExotic} for comparison) are caused by the presence of the NP. Scale bar 1 $\mu$m.} 
    \label{fig:S10_control2_glass}
\end{figure}

\newpage
\begin{figure}[ht]
    \addcontentsline{toc}{section}{S11. Data analysis: PSF extraction from raw data and background subtraction}
    \centering
    \includegraphics[width=1\linewidth]{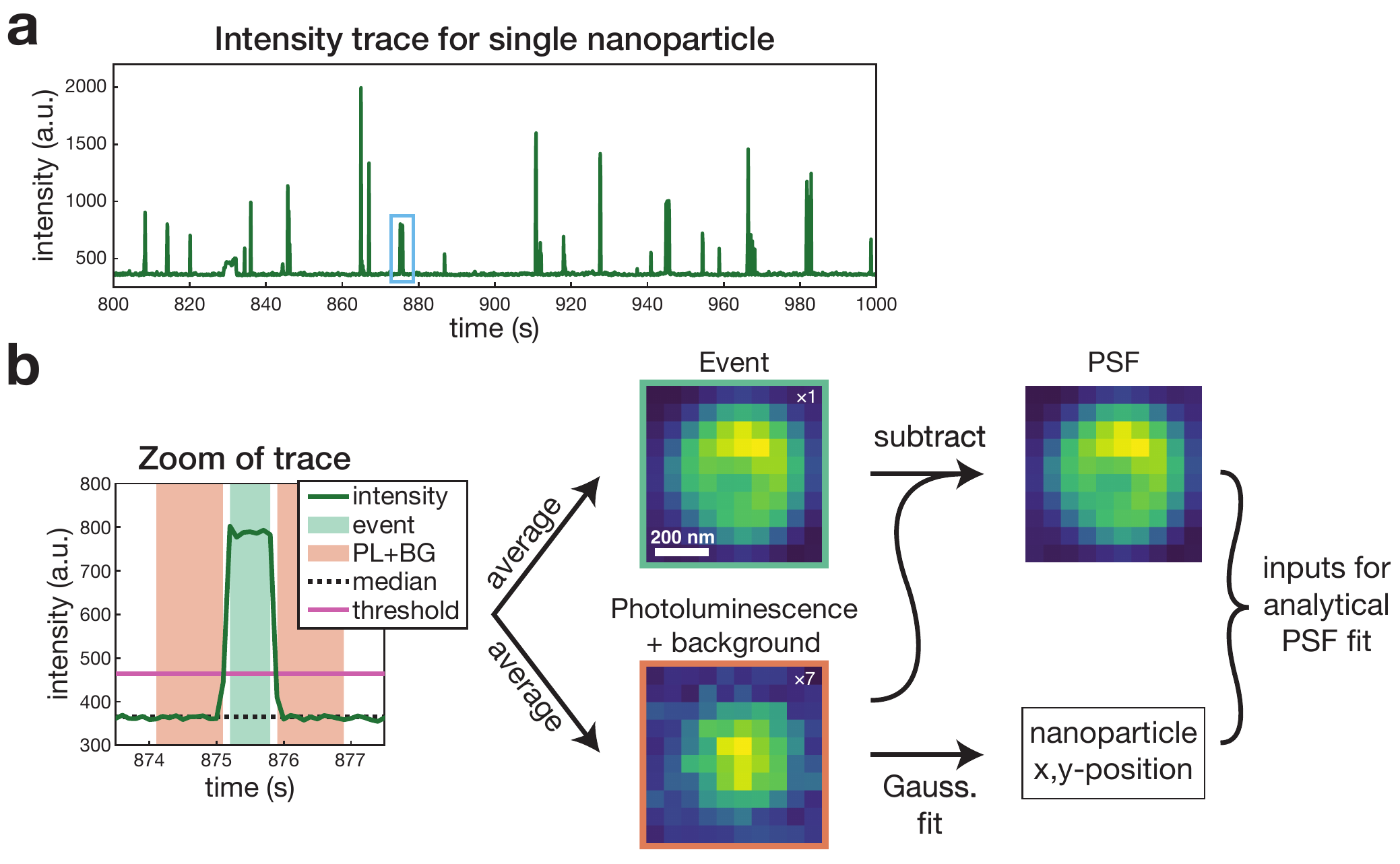}
    \caption{\textbf{Data analysis: PSF extraction from raw data and background subtraction}. 
    \textbf{a} An example time trace. For each spot of interest, the intensity trace is generated by averaging the 15x15 pixel values around the found maximum in every time frame. For every selected spot, the events are found by detecting the time frames where the intensity trace (green line) is 52 photons (100 intensity values for these plots, pink line) above the median (black dotted line) of the entire trace. 
    \textbf{b} Zoom of the blue-outlined event in \textbf{a}. Consecutive time frames above the threshold are grouped and every group is considered an event (green shaded area). Per considered event, the 15x15 images are averaged over time, creating the raw PSF (green outlined spot). The one-photon luminescence (PL) signal and background (BG, consisting of camera offset, background photons, etc.) are averaged over 10 consecutive frames below the threshold, before and after each event (orange shaded areas). The closest PL+BG block to the PSF is averaged over time (orange outlined spot) and subtracted from the raw PSF, resulting in the final experimental fluorescence PSF. Note that the intensity values in the PL+BG spot are multiplied by $7$ for visualization (hence the $\times7$ in the image), since the PL+BG signal is weaker in intensity compared to the fluorescent signal. The PL+BG signal (orange outlined spot) is fitted with a 2D gaussian to obtain the position of the NP (corrected by drift, see Fig. S10 for details), which serves, together with the fluorescent PSF, as the two inputs for the analytical PSF fit. Scale bar applies to all PSFs.}
    \label{fig:S11_dataAnalysis}
\end{figure}

\newpage
\begin{figure}[ht]
    \addcontentsline{toc}{section}{S12. Fluorescence enhancement of fluorophores on NPs}
    \includegraphics[width=1\linewidth]{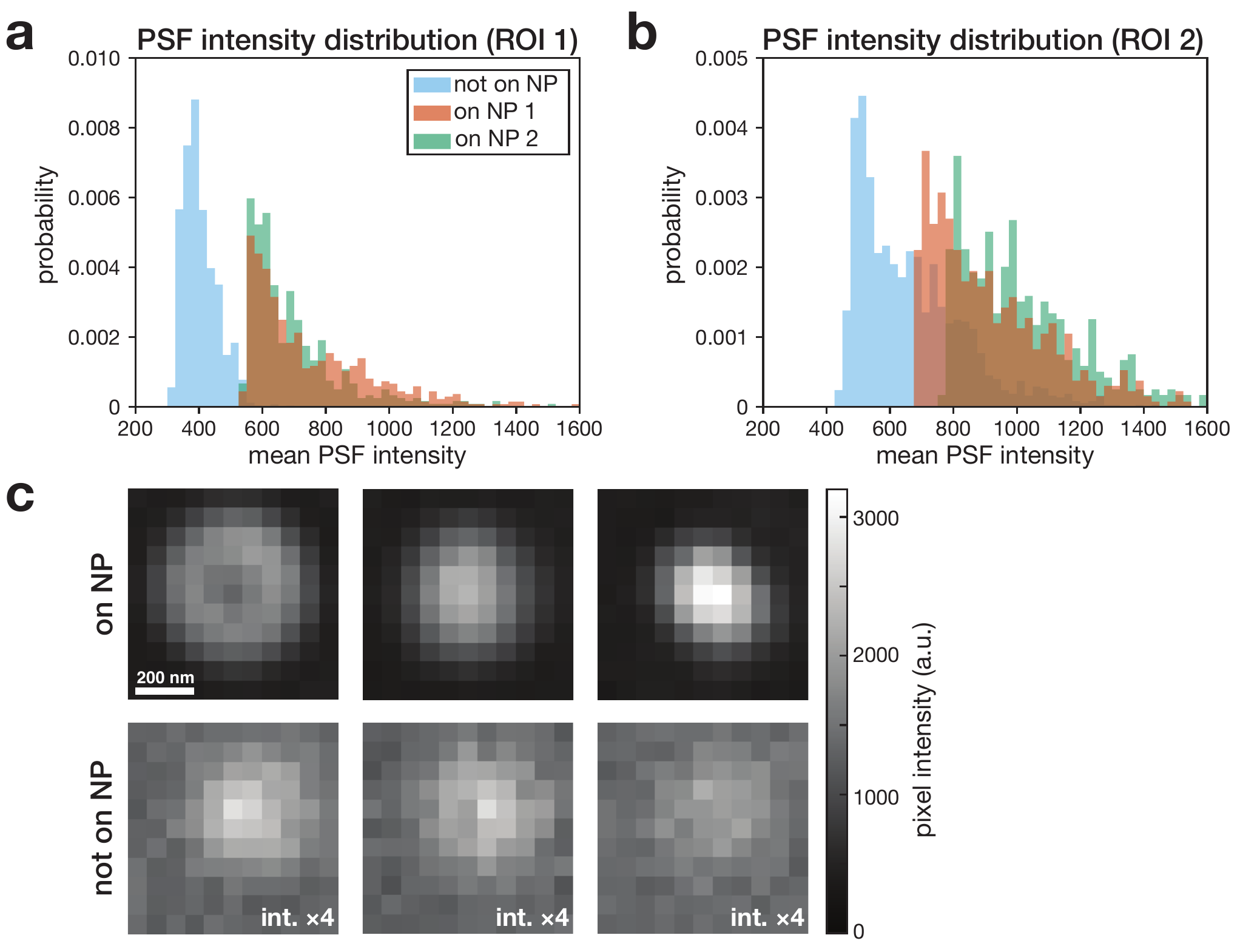}
    \caption{\textbf{Fluorescence enhancement of fluorophores on NPs}. Due to plasmonic coupling between the fluorophore and the metal nanoparticle, the fluorescent signal is enhanced in intensity. \textbf{a} Difference in PSF intensity between fluorophores located on a NP and not on a NP. For this analysis, we consider a small region of interest (ROI) of the FOV ($4 \times$ 6 $\mu$m) with 2 particles, and determine for each detected PSF whether it originates from a NP or not, based on its proximity to the NPs (of which we know the position due to the 1-photon photoluminescence signal). The three histograms show the mean PSF intensity and contains 2375, 546, and 482 PSFs, respectively. \textbf{b} Same as \textbf{a}, but for a different ROI in the FOV. The three histograms contain 1507, 534, and 478 PSFs, respectively. \textbf{c} Six representative PSFs (three on and three not on a NP) for NP 1 in \textbf{a}. Note that the three PSFs in the bottom row, located not on a NP, are each multiplied by a factor of 4 in intensity, in order to use the same color scale as for the top row of PSFs. Scale bar applies to all. \\ \\
    From this analysis, we can clearly see that the PSFs originating from fluorophores on a NP are significantly higher in intensity. As a consequence, the signal-to-noise ratio is higher (they share the same background signal).} 
    \label{fig:S12_PSFintensity}
\end{figure}

\newpage
\begin{figure}[ht]
    \addcontentsline{toc}{section}{S13. Global determination of NP position at all times using PL}
    \centering
    \includegraphics[width=.80\linewidth]{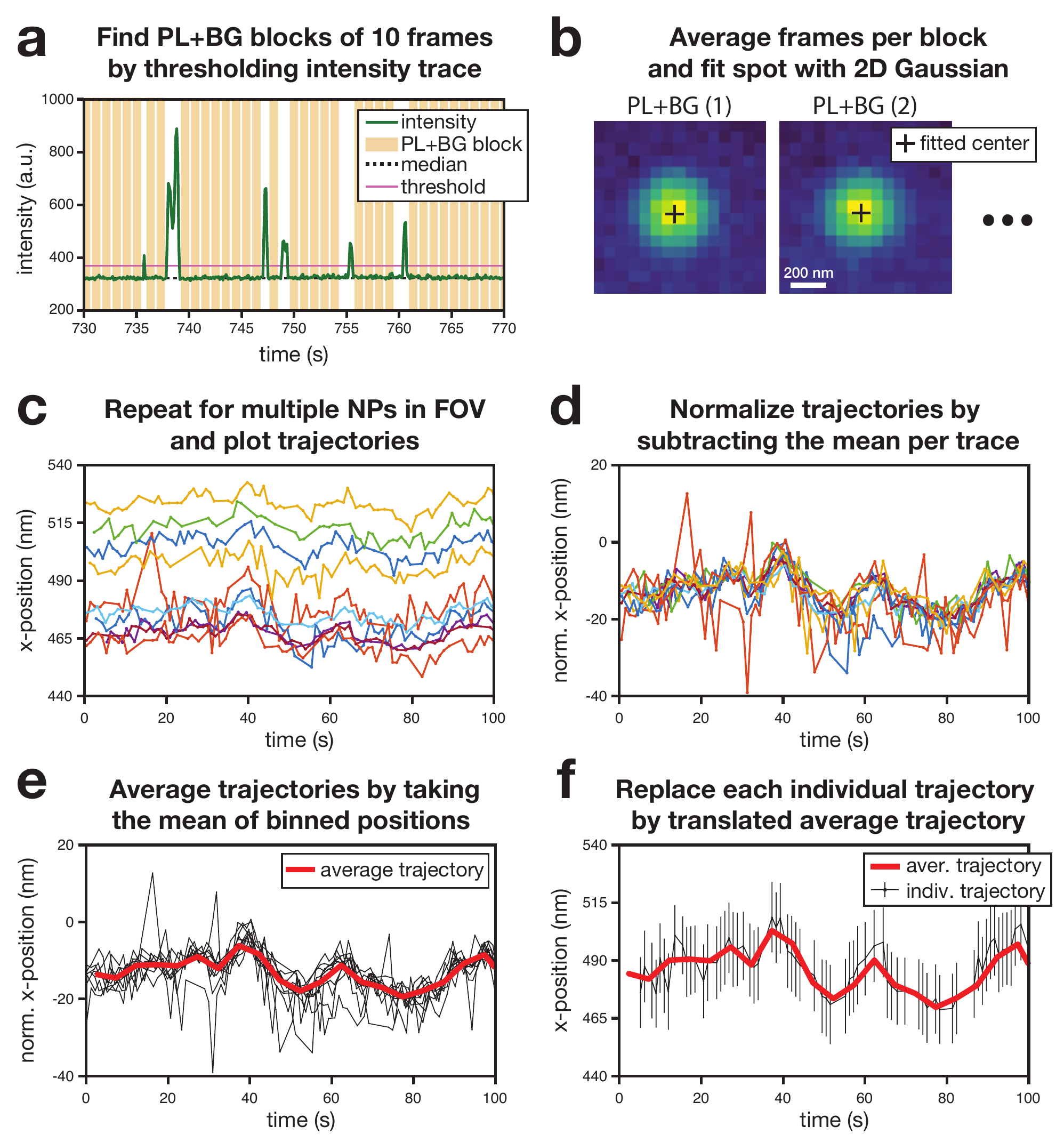}
    \caption{\textbf{Global determination of NP position at all times using PL}. The analytical fitting approach of our PSF model to experimentally obtained spots requires the position of the NP. We get the NP position by fitting a 2D Gaussian to the 1-photon photoluminescense (PL) and the background (BG) signal, which is detected when there is no binding event. We employ a strategy to refine the NP position by taking into account the drift of the ensemble of all particles, because of three reasons: 1) the NP position can not be determined at the exact time of the PSF, 2) the PL+BG image is sometimes affected by nearby unspecific fluorescent events hampering the PSF fit, and 3) the uncertainty of the Gaussian fit to the dim PL+BG signal is relatively large.\\ \\
    \textbf{a} Blocks of 10 consecutive frames where the intensity (green) is below a set threshold (pink), defined as 50 intensity values above the median (black) of the entire trace, are defined as 'PL+BG' blocks (yellow). Averaging of 10 frames is required, since the signal of a single frame is too low for a reliable Gaussian fit. \textbf{b}. We fit a 2D Gaussian to every averaged block of PL+BG signal with maximum-likelihood estimation \cite{Mortensen2010}. In order to obtain the global drift we repeat the blocking and fitting procedure for multiple NPs in the FOV (10 in this case), we subsequently: \textbf{c} plot the $x$-positions over time for all NPs and \textbf{d} subtract the mean per trajectory (note that subtracting the mean centers the entire trajectory around $x=0$, however, here we only show the first 100 $s$ of the trajectory). \textbf{e} The average trajectory (red) is obtained by binning the positions of all trajectories in bins of 0.5 s and taking the median position per bin. The binning is required because for every trajectory the positions are known for different points in time, since the PL+BG blocks have different positions per NP. Note that the outlier peaks visible in \textbf{d} and \textbf{e} are caused by biased Gaussian fits, probably caused by an unspecific binding event near the NP. \textbf{f} The NP position for each individual NP is obtained by using the translated average trajectory (red) instead of the original trajectory (black). Error bars represent the standard deviation of the 2D Gaussian fit. The average trajectory is linearly interpolated to acquire the position at the exact time of the PSF. This process is performed in same way for the $y$-coordinate.} 
    \label{fig:S13_PLdrift}
\end{figure}

\newpage
\begin{figure}[ht]
    \addcontentsline{toc}{section}{S14. Solving PSF shape ambiguity by knowing NP position}
    \centering
    \includegraphics[width=1.\linewidth]{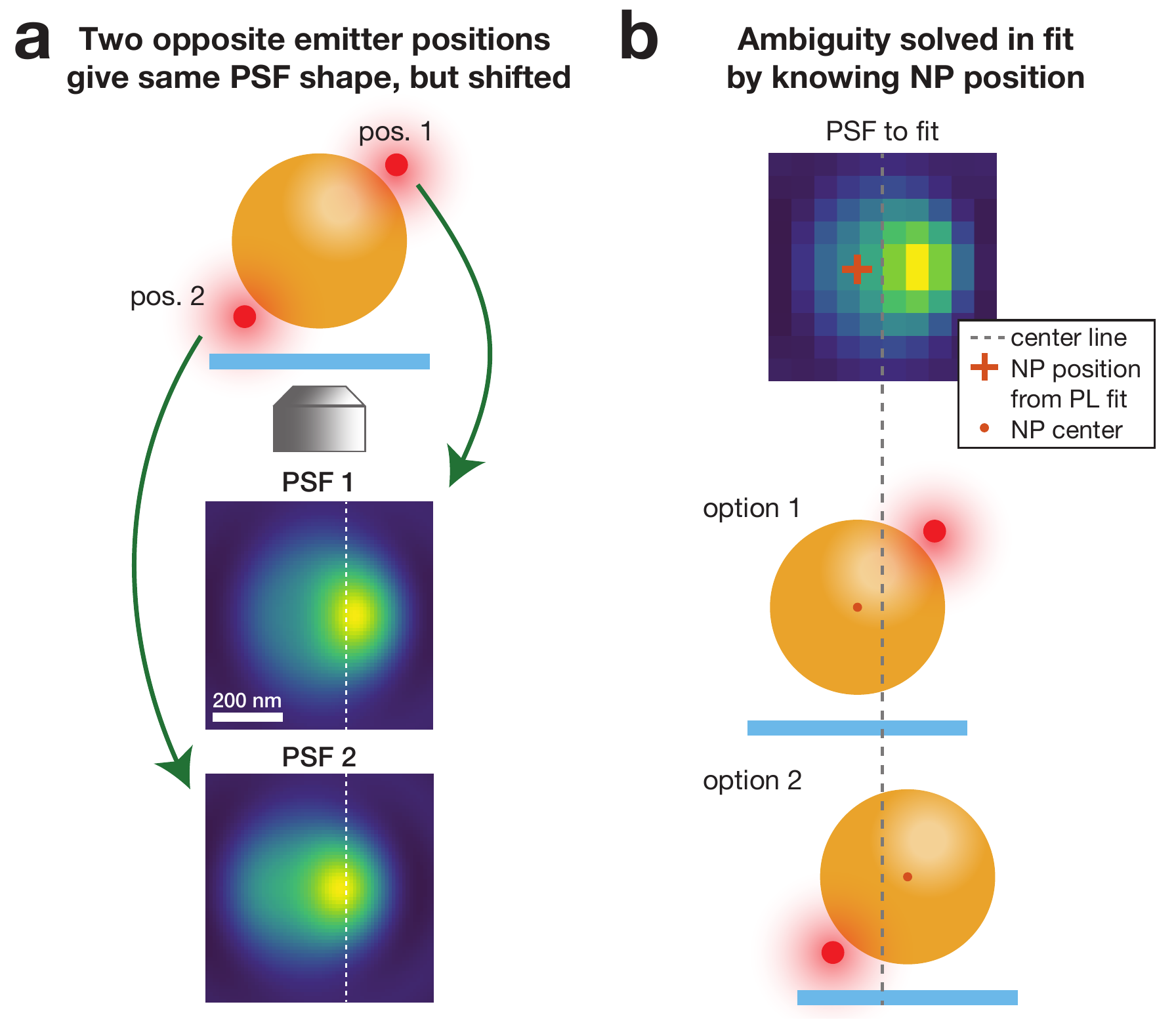}
    \caption{\textbf{Solving PSF shape ambiguity by knowing NP position}. \textbf{a} Due to the spherical symmetries of the spherical NP, multiple emitter positions on different locations on the NP surface result in similar PSF shapes. For example, opposite positions on the NP, result in PSFs that are different, but only marginally (PSF 1 and PSF 2, respectively). Even though every emitter position results in a unique PSF, with enough photons the fit would be able to discriminate between. However, for practical signal-to-noise levels, that is not possible for PSFs with similar shapes (like PSF 1 and PSF 2 shown here). Albeit the similar PSF shape, the PSFs are shifted from each other in position (compare the PSFs to the white dotted reference line). \textbf{b} We solve this ambiguity since we have access to, in addition to the PSF shape, the position of the NP from the 1-photon photoluminescense signal (PL). Purely from the shape of the PSF, there could be two configurations (option 1 and option 2) that would give a similar PSF shape. Which one the analytical PSF fit finds, depends on the initial guesses for the fit parameters. The difference between the two possible options is the position of the NP (orange dot) with respect to the PSF image (grey dotted line for guiding the eye). As a solution, we fix the NP position (orange cross) in the fit and we initialize the fit twice with two opposite initial guesses, forcing the fit to sample the two position on either side of the NP. We choose the optimal one (option 1, in this example) by evaluating the mean-squared error between the fitted and the experimental PSF. Scale bar applies to all PSFs. PSFs are calculated analytically and details about parameters can be found in Suppl. Tables 1 and 2.} 
    \label{fig:S14_ambig}
\end{figure}

\newpage
\begin{figure}[ht]
    \addcontentsline{toc}{section}{S15. Experimental characterization of single gold nanoparticles}
    \centering
    \includegraphics[width=1.\linewidth]{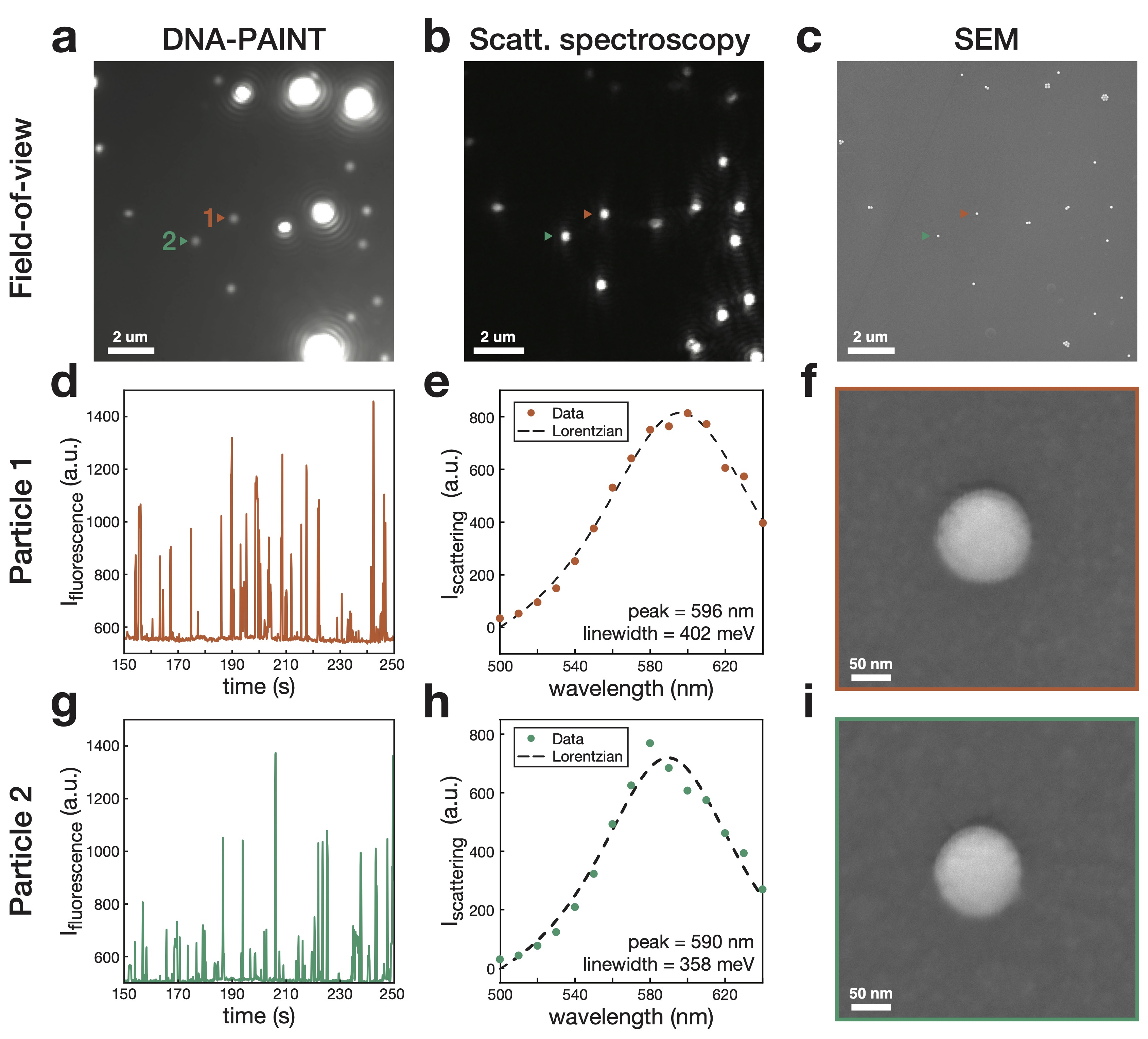}
    \caption{\textbf{Experimental characterization of single gold nanoparticles}. To confirm that the NPs studied with DNA-PAINT are indeed spherical gold NPs with the expected size of 100 nm, we performed two additional measurements. The same sample is sequentially: \textbf{a} illuminated with a red laser for performing DNA-PAINT measurements, \textbf{b} illuminated with white light to perform single-particle scattering spectroscopy, and \textbf{c}, imaged using scanning electron microscopy (SEM). The DNA-PAINT image in \textbf{a} is an average over 6000 frames and the scattering spectroscopy image in \textbf{b} is obtained at 580 nm excitation. \textbf{d-i} Detailed information from all three measurements for particle 1 and particle 2, marked in orange and green in \textbf{a}, respectively. \textbf{d,g} Fluorescent intensity over time, showing the DNA-PAINT intensity bursts. Note that the marked spots in the time-averaged image of \textbf{a} for these particles show mostly the photoluminescence of the gold. \textbf{e,h} Scattering intensity for different wavelengths (500-640 nm). Fitting the data with a Lorentzian reveals the location of the plasmon resonance peak and the spectral linewidth (values shown in lower-righthand corner), which match expected values from literature \cite{shafiqa2018} for single gold spherical particles with a diameter of 100 nm. \textbf{f,i} Zooms of the SEM images, demonstrating the that the particles of interest are single spherical NPs. Experimental details of all measurements can be found in Methods. Scattering and SEM data for clusters of gold NPs (as can be seen in \textbf{c}) are show in Fig. S13} 
    \label{fig:S15_HSMSEM}
\end{figure}

\newpage
\begin{figure}[ht]
    \addcontentsline{toc}{section}{S16. PSFs originating from clusters of gold 100 nm NPs}
    \centering
    \includegraphics[width=.7\linewidth]{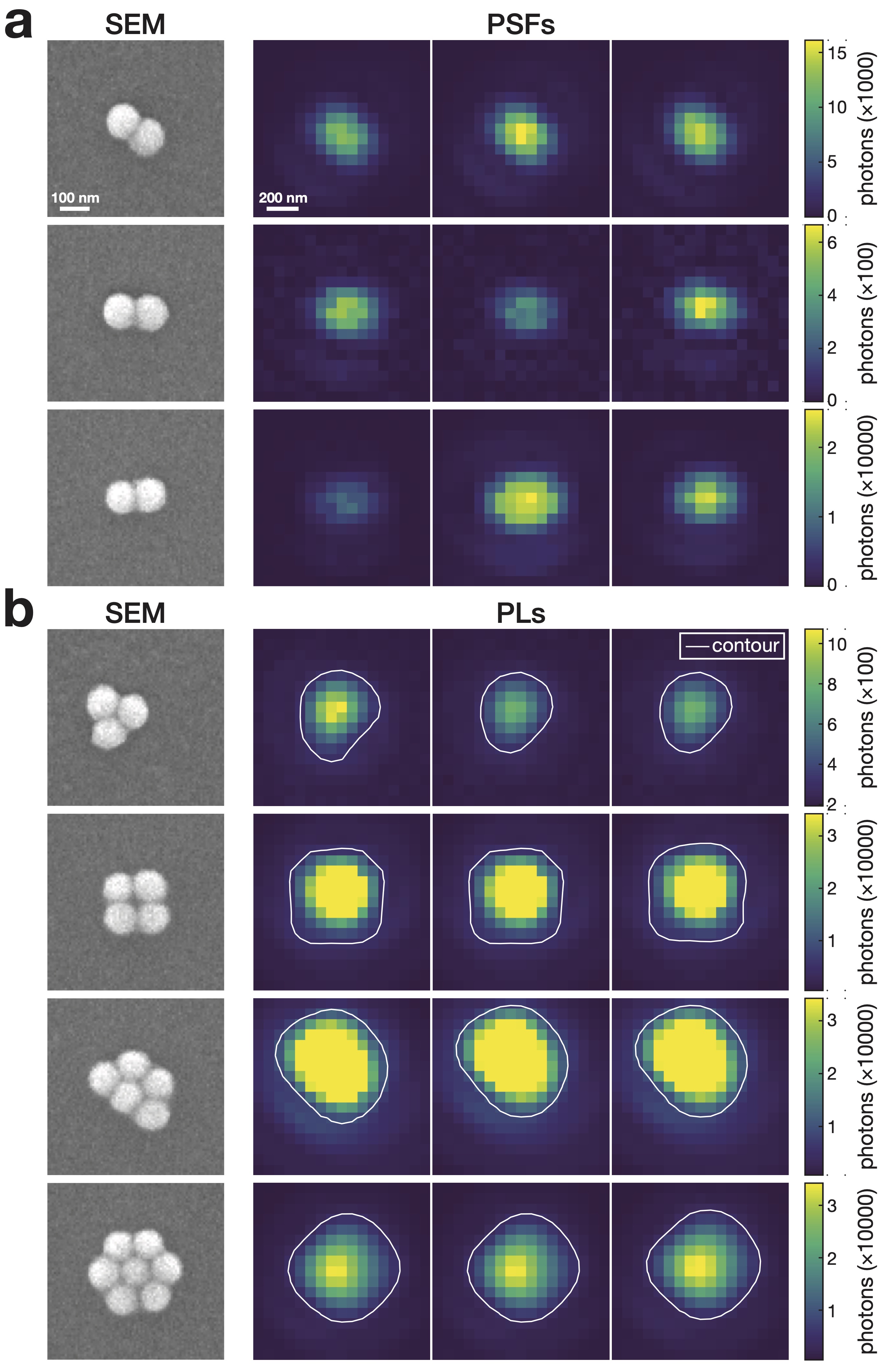}
    \caption{\textbf{PSFs originating from clusters of gold 100 nm NPs}. \textbf{a} Three examples of dimers of 100 nm gold NPs seen in SEM (left), together with three representative PSFs for each dimer (right). The 1-photon photoluminescence (PL) and background are subtracted from the shown PSFs. \textbf{b} Examples of clusters of 100 nm gold NPs containing 3, 4, 5, and 7 NPs (left), together with three representative PL images for each cluster (right). The white line represents the isocontour at 300, 5000, 9000, and 4000 photons, respectively. Note that the the PL images of the lowest three rows are saturated (hence the yellow center). The imaging conditions are optimized for the PSFs from single NPs, so the strongly enhanced PL of clusters saturates the camera. 
    Scale bars for the SEM and light microscopy images, respectively, apply to all. Color bars apply to all images in respective row.\\
    From the PSFs in \textbf{a}, we can clearly see that the PSFs have an elongated shape in the direction of the dimer. This is explained by the fact that a dimer has a strong longitudinal resonance axis (along the dimer direction), which strongly couples to the fluorophores emission. This creates a dipolar emission pattern, which results in an elongated PSF in the direction of the coupled dipole. For the clusters in \textbf{b} the PSFs could not be calculated, because of the saturation of the bit-depth of the camera. However, we can clearly see from the white contour lines that the PL has the shape of the cluster (at least up until the pentamer).}
    \label{fig:S16_multimeres}
\end{figure}

\newpage
\begin{figure}[ht]
    \addcontentsline{toc}{section}{S17. Localization uncertainty depends on position on NP}
    \centering
    \includegraphics[width=1.\linewidth]{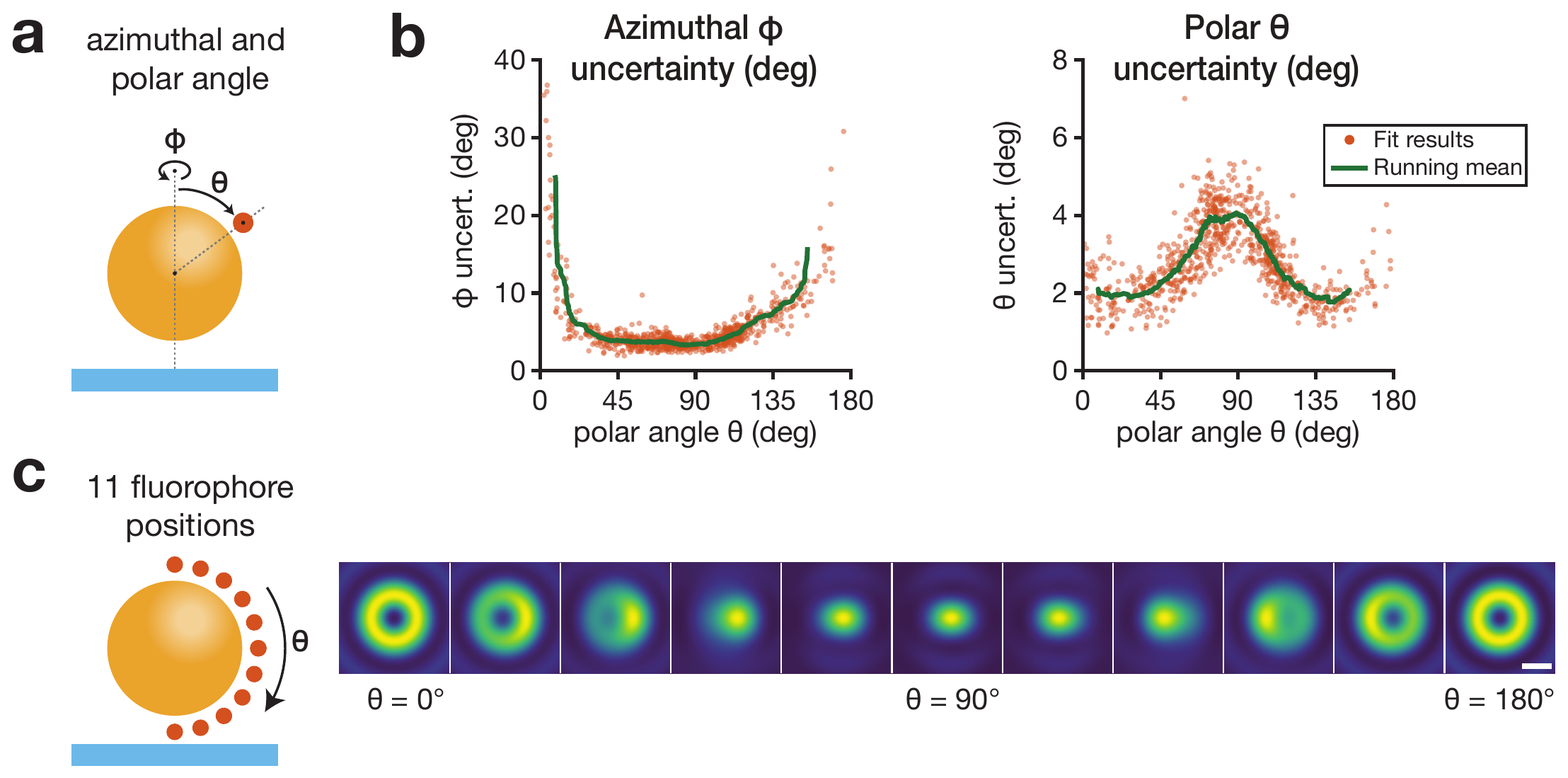}
    \caption{\textbf{Localization uncertainty depends on position on NP}. The reported localization uncertainty of $4.9 \pm 1.3$ in Fig. 4d is the average value over many localizations. Here we show how the localization uncertainty depends on the position of the fluorophore relative to the NP. 
    \textbf{a} The position of the fluorophore relative to the NP is expressed in spherical coordinates, with azimuthal angle $\phi$ and polar angle $\theta$. 
    \textbf{b} We fit our analytical PSF model to all 830 PSFs from a single NP. Each fit provides us with a value for the uncertainty of the obtained azimuthal and polar positions, respectively, which we report here expressed in degrees, as function of the obtained polar position on the NP. We see that the uncertainty in the azimuthal position $\phi$ (left plot) is lowest for positions at the equator of the NP ($\theta=90 \degree$) and increases for positions near the top and bottom of the NP ($\theta=0 \degree \text{ or } \theta=180 \degree$). 
    In contrast, the uncertainty on the polar position $\theta$ (right plot) is highest for positions at the equator, and decreases for positions near the top and bottom. \textbf{c} To qualitatively understand the dependency of the localization uncertainties in $\phi$ and $\theta$ on the polar position, we show the analytical PSFs for 11 position on the NP ($\phi=0 \degree$ and $\theta=0\degree:180\degree$). The fact that the uncertainty in the azimuthal position $\phi$ (left plot) is increasing for $\theta=0 \degree$ and $\theta=180 \degree$ is explained by the fact the for these positions the PSF becomes rotationally symmetric (the \textit{donut} shape), which means that $\phi$ is not defined, hence the high uncertainty.
    The fact that the uncertainty in the polar position $\theta$ (right plot) is highest for $\theta=90 \degree$ is explained by the fact the for this position the PSF does not vary significantly when $\theta$ is varied, resulting in a larger uncertainty. Which we see, since the middle 3 PSFs are almost identical. Scale bar of \textbf{c} is 200 nm and applies to all 11 PSFs, and the intensity of each PSF is individually normalized.} 
\end{figure}

\newpage
\begin{figure}[ht]
    \addcontentsline{toc}{section}{S18. qPAINT for counting number of available docking strands}
    \centering
    \includegraphics[width=1.\linewidth]{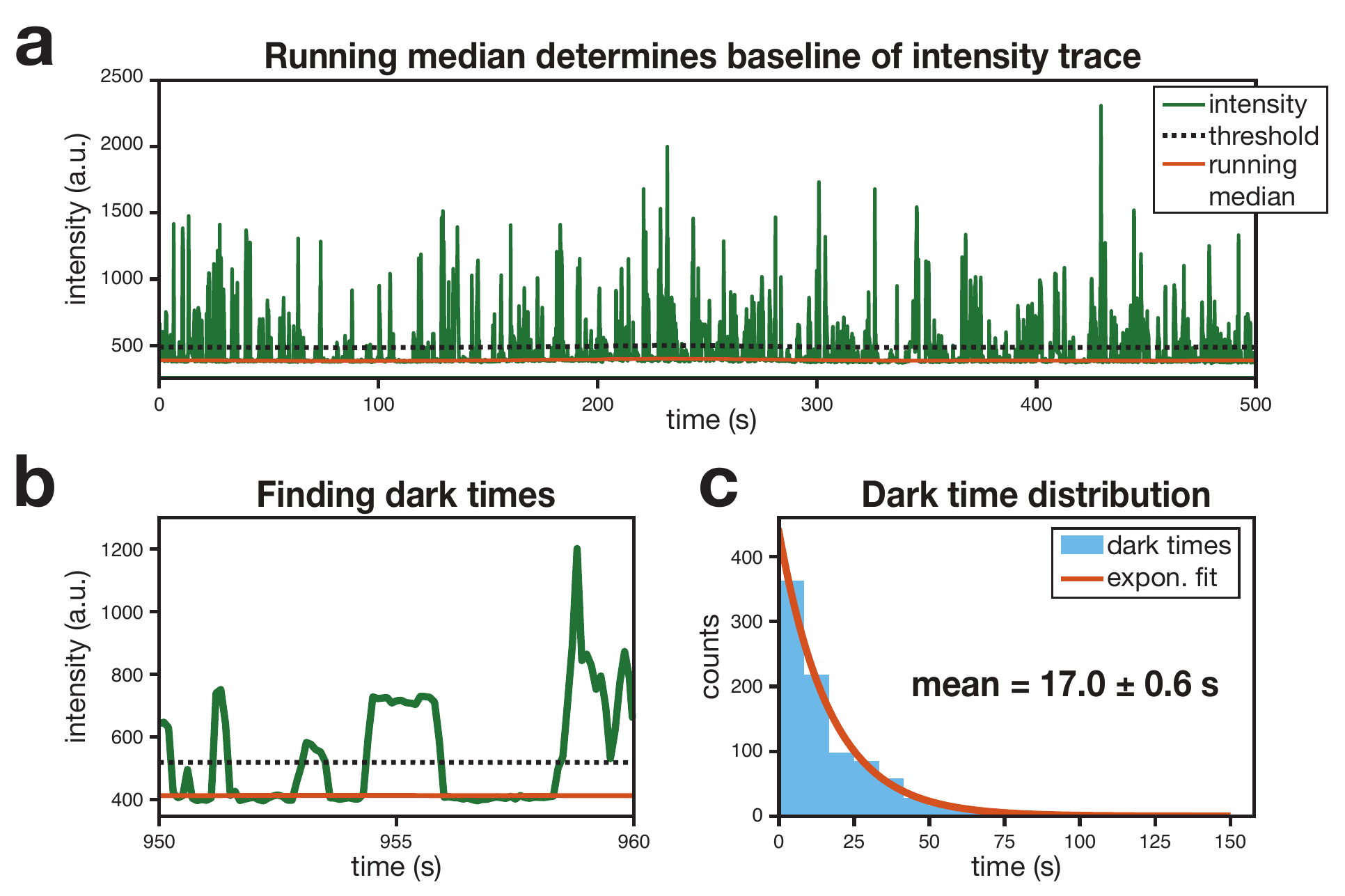}
    \caption{\textbf{qPAINT for counting number of available docking strands}. To get an idea about the number of available docking strands on the NP surface, we perform quantitative PAINT (qPAINT) analysis. In qPAINT, the actual number of docking strands available for PAINT are determined by looking at the binding statistics, knowing the binding kinetics of the considered DNA sequence. Assuming the binding events are governed by a single binding- ($k_{\textrm{on}}$) and unbinding-rate ($k_{\textrm{off}}$), the length of the periods without binding events (dark times) is exponentially distributed. The characteristic time of exponentially distributed dark times is $\frac{1}{k_{\textrm{on}} \cdot c \cdot N}$, with $c$ the concentration of imager strands and $N$ the number of emitters (in our case on the NP). \textbf{a} In order to calculated the dark times, we first find the baseline of the intensity trace (green) from a single NP by calculating the running median with a window length of 100 s (red). \textbf{b} The dark times are calculated by counting the number of consecutive frames the intensity trace is below the threshold (black dashed line, manually determined to be 100 intensity values about the running median line). \textbf{c} The distribution of dark times (blue) from a single NP follows clearly the exponential distribution with the calculated mean dark time as characteristic time. For this example, 883 dark times are found with a mean dark time of 17.0 $\pm$ 0.6 s. (mean $\pm$ sem). The mean dark time $\left \langle \tau \right \rangle$ is converted into the number of emitters $N$ using $N = \frac{1}{k_{\textrm{on}}\cdot c \cdot \left \langle \tau \right \rangle}$, We find $N=128$ by using $c=200$ pM and $k_{\textrm{on}}=2.3e^6$ $M^{-1}s^{-1}$ \cite{jungmann2010}. The qPAINT analysis is repeated for 10 NPs, and the found number of emitters $N$ is compared to the number of detected PSFs. On average, each particles has 7.0 $\pm$ 0.1 (mean $\pm$ sem) times more PSFs than available emitters $N$.} 
    \label{fig:S18_qPAINT}
\end{figure}

\newpage
\begin{figure}[ht]
    \addcontentsline{toc}{section}{S19. PSF of freely rotating dipole emitter only requires averaging over 3 mutually orthogonal dipole orientations} 
    \centering
    \includegraphics[width=1.\linewidth]{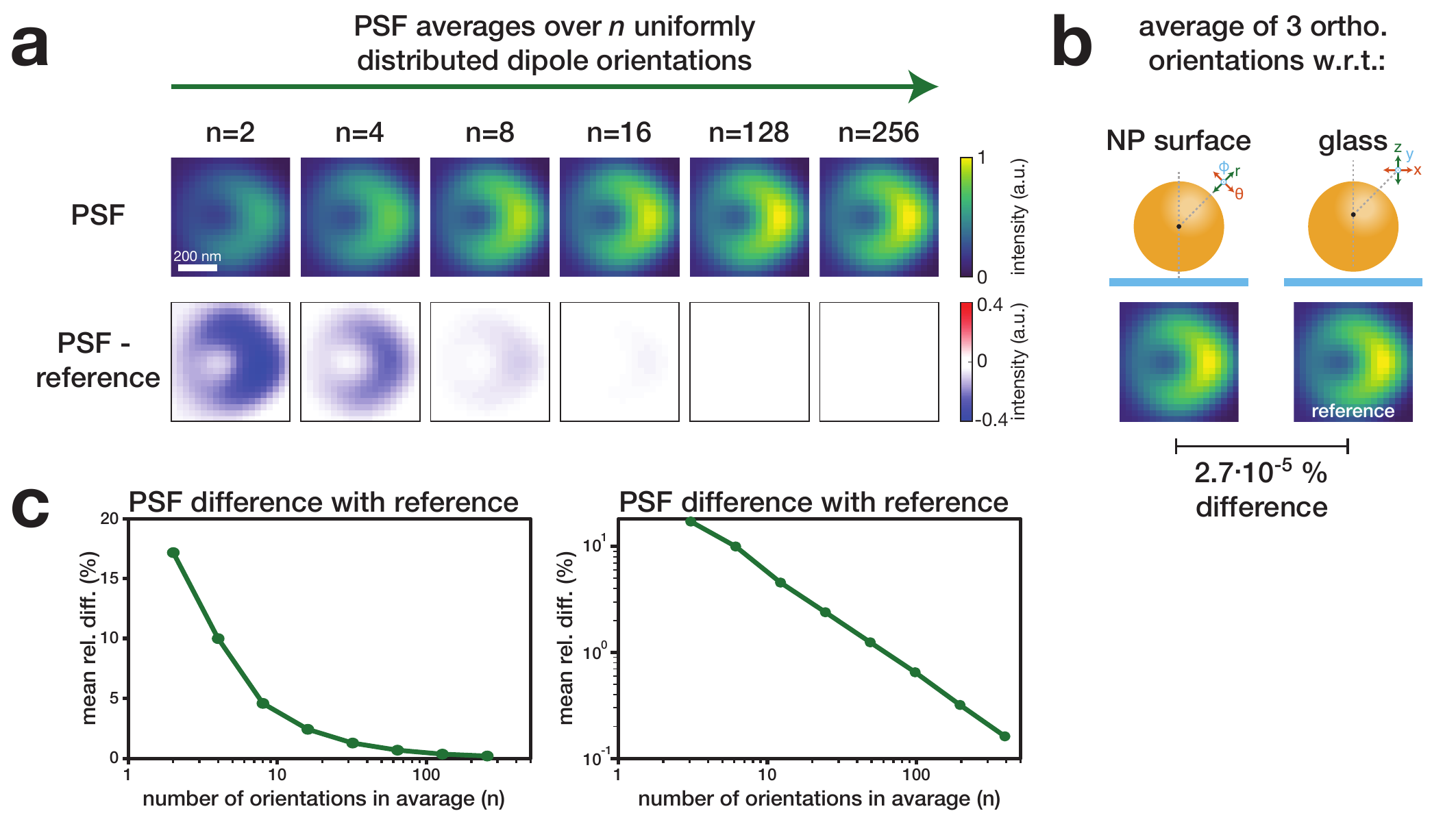}
    \caption{\textbf{PSF of freely rotating dipole emitter only requires averaging over 3 mutually orthogonal dipole orientations}. The developed PSF model directly gives the PSF for a dipole emitter that is fixed in orientation. However, in most experiments the fluorophore is able to rotate freely. When the rotational diffusion of the fluorophore is faster than the integration time of the camera, the fluorophore is assumed to sample all possible orientations during the camera integration time. The PSF of such a rotating fluorophore in principle is represented by averaging the PSFs for a fixed emitter over all possible orientations (which are infinitely many). In Supplementary Note 3 we show that averaging the PSF of all possible orientations is mathematically identical of averaging the PSFs for only three orthogonal orientations, which we will investigate here using our analytical PSF model. \\ \\
    \textbf{a} Top row shows the average PSFs as a result of averaging over $n$ uniformly distributed dipole orientation of the 3D rotation space, varying from $n=2$ to $n=256$ orientations. Bottom row shows the PSFs of the top row minus the reference PSF, which is the average of only 3 orthogonal orientations (see \textbf{b}). \textbf{b} The average of 3 mutually orthogonal dipole orientations, which are either orthogonal relative to the NP surface (left) or to the glass (right). The 3 individual dipole orientations are show as red, green, and blue arrows in the schematic. The difference between the two PSFs, which should theoretically be identical, is extremely small (0.000027$\%$), explained by numerical precision in the evaluation of the PSF model. This proves that the average PSF of three mutually orthogonal dipole orientations is not dependent on the individual three orientations. The right PSF (for $x$-, $y$- and $z$-oriented dipoles) is used as the reference in \textbf{a} and \textbf{c}. \textbf{c} Quantification of the difference in PSF between averaging over 3 orthogonal orientations and $n$ uniformly distributed orientations, in both linlog (left) and loglog scale (right). We can clearly see that the difference in PSF goes asymptotically to zero for an increasing number of orientations in the average. This supports our mathematical claim (SI Note 3) that averaging over all possible dipole orientations gives the same PSF as averaging over 3 mutually orthogonal orientations. The difference is defined as the mean relative difference with the reference image, calculated as $\textrm{mean}\big (|\textrm{PSF}-\textrm{ref}|/\textrm{max(ref)} \big )$. Scale bar applies to all PSFs and full analytical parameters can be found in SI Tables 1 and 2.} 
    \label{fig:S19_3ortho}
\end{figure}

\newpage
\begin{figure}[ht]
    \addcontentsline{toc}{section}{S20. Effect of NP size and blurring parameter on localizations.}
    \centering
    \includegraphics[width=1.0\linewidth]{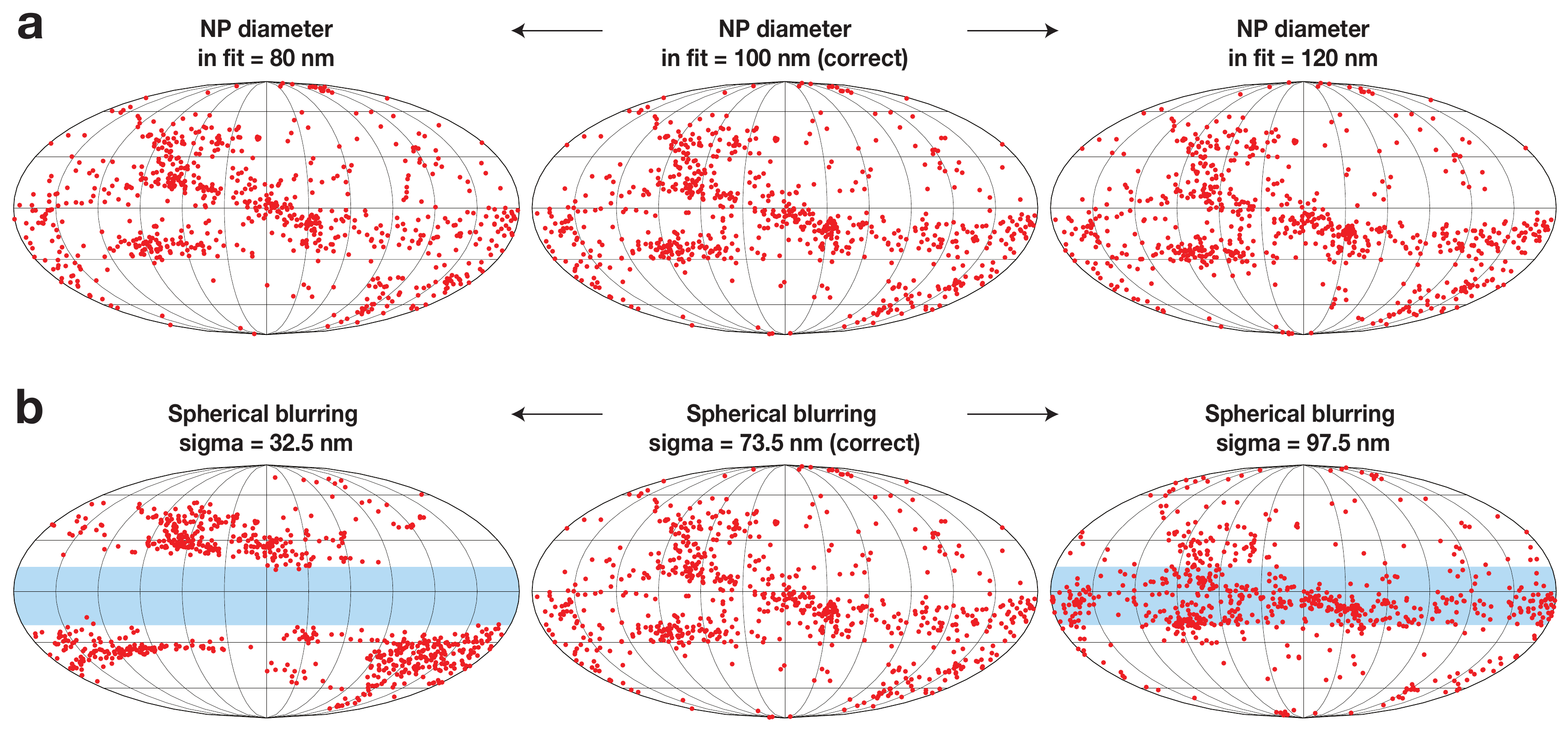}
    \caption{\textbf{Effect of NP size and blurring parameter on localizations}. 
    \textbf{a} The effect of the fixed NP diameter in the PSF fit on the localizations. We fitted all 734 PSFs of a single 100 nm NP three times, each with a different NP diameter fixed in the PSF fitting procedure, namely 80 nm (left), 100 nm (middle, the correct diamter), and 120 nm (right). Even though the fitting is performed with a wrong NP diameter, the localization distributions on the NP surface look very similar. From this we conclude (1) that the NP diameter is not a sensitive parameter towards the found distribution of localizations, and (2) that if the actual NP in the experiment has a slightly different diameter than 100 nm, the fitting procedure assuming a 100 nm NP will still perform well. 
    \textbf{b} The effect of the amount of Gaussian blurring ($\sigma$) applied to the analytical PSF model to correct for optical aberrations. We fitted all 734 PSFs of a single 100 nm NP three times, each with a different amount of Gaussian blurring, namely with a $\sigma$ of 32.5 nm (left), 73.5 nm (middle, representing the correct amount of blurring, see Fig. \ref{fig:S22_spherAber}), and 97.5 nm (right). We can clearly see that fitting with a wrong amount of Gaussian blurring has a significant effect on the distribution of localizations. When we blur the PSF model too little (left), the localizations have a bias away from the equator, highlighted by the blue shaded area without localizations. When we blur the PSF model too much (right), the localizations have a bias towards the equator, highlighted by the blue shaded area with more localizations than for the correct amount of Gaussian blurring (middle).
    } 
\end{figure}

\newpage
\begin{figure}[ht]
    \addcontentsline{toc}{section}{S21. Only one fluorophore orientation dominates the PSF for a rotationally free fluorophore next to a NP.}
    \centering
    \includegraphics[width=1.\linewidth]{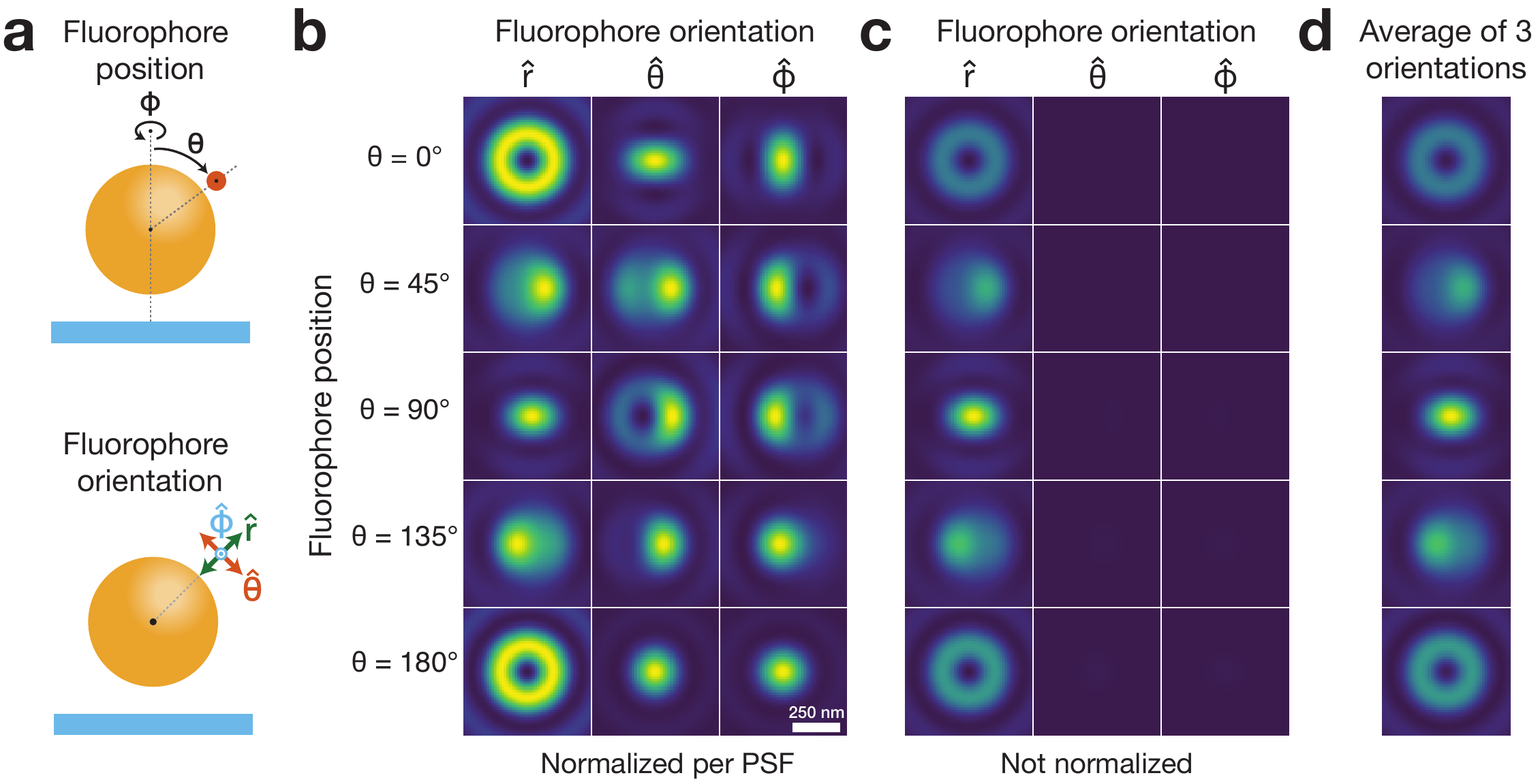}
    \caption{\textbf{Only one fluorophore orientation dominates the PSF for a rotationally free fluorophore next to a NP}. 
    \textbf{a} Schematic definitions of the fluorophore position and orientation relative to the NP. The fluorophore position (top schematic) is defined in spherical coordinates with $\theta$ the polar position measured from the top of the NP, and $\phi$ the azimuthal position measured from the positive $x$-axis. For the fluorophore orientation (bottom schematic), we consider three orientations here, where the fluorophore is either oriented in the radial ($\hat{r}$, green), polar ($\hat{\theta}$, orange), or azimuthal ($\hat{\phi}$, blue) direction. 
    \textbf{b} The normalized PSFs for 15 fluorophores next to a 100 nm gold NP (5 positions and 3 orientations). The PSFs are normalized per PSF, meaning that the intensity of each PSF is normalized to a value of 1, hence they all look equally bright. We see that for each position, the shape of the PSF is clearly dependent on the fluorophore orientation. \textbf{c} Same as \textbf{b}, but PSFs are not normalized. We see that the PSFs of radially-oriented fluorophores are significantly brighter then fluorophores oriented in the polar and azimuthal direction. 
    \textbf{d} The average PSF over three fluorophore orientations. The PSF of a rotationally free fluorophore can be calculated by averaging the PSFs of 3 fluorophores that are mutually orthogonal (See SI Note 3). Here, it means we can simply take the average of the 3 fluorophore orientations, since $\hat{r}$, $\hat{\theta}$, and $\hat{\phi}$ are mutually orthogonal. When comparing \textbf{d} with \textbf{c}, we see that the average PSF of \textbf{d} is identical so the $\hat{r}$-PSF of \textbf{c}, which makes sense since the PSF of the other two orientations have near-zero intensity compared to  $\hat{r}$. Considering computational time, in our PSF fitting approach we only fit the $\hat{r}$-PSF, instead of the average PSF, since calculating only the $\hat{r}$-PSF is three times faster and the difference is negligible. The reason why the intensity of the $\hat{r}$-PSF is significantly higher compared to other orientation, is that plasmonic coupling between the emission of the dipole emitter and the NP is highest for the radial orientation, they constructively interfere. The azimuthal and polar orientations destructively interfere, resulting in a dim PSF. The scale bar of \textbf{b} applies to all and the details for the analytical PSF calculations can be found in SI Table 1.} 
\end{figure}

\newpage
\begin{figure}[ht]
    \addcontentsline{toc}{section}{S22. Characterization and correction of spherical aberrations}
    \centering
    \includegraphics[width=1.\linewidth]{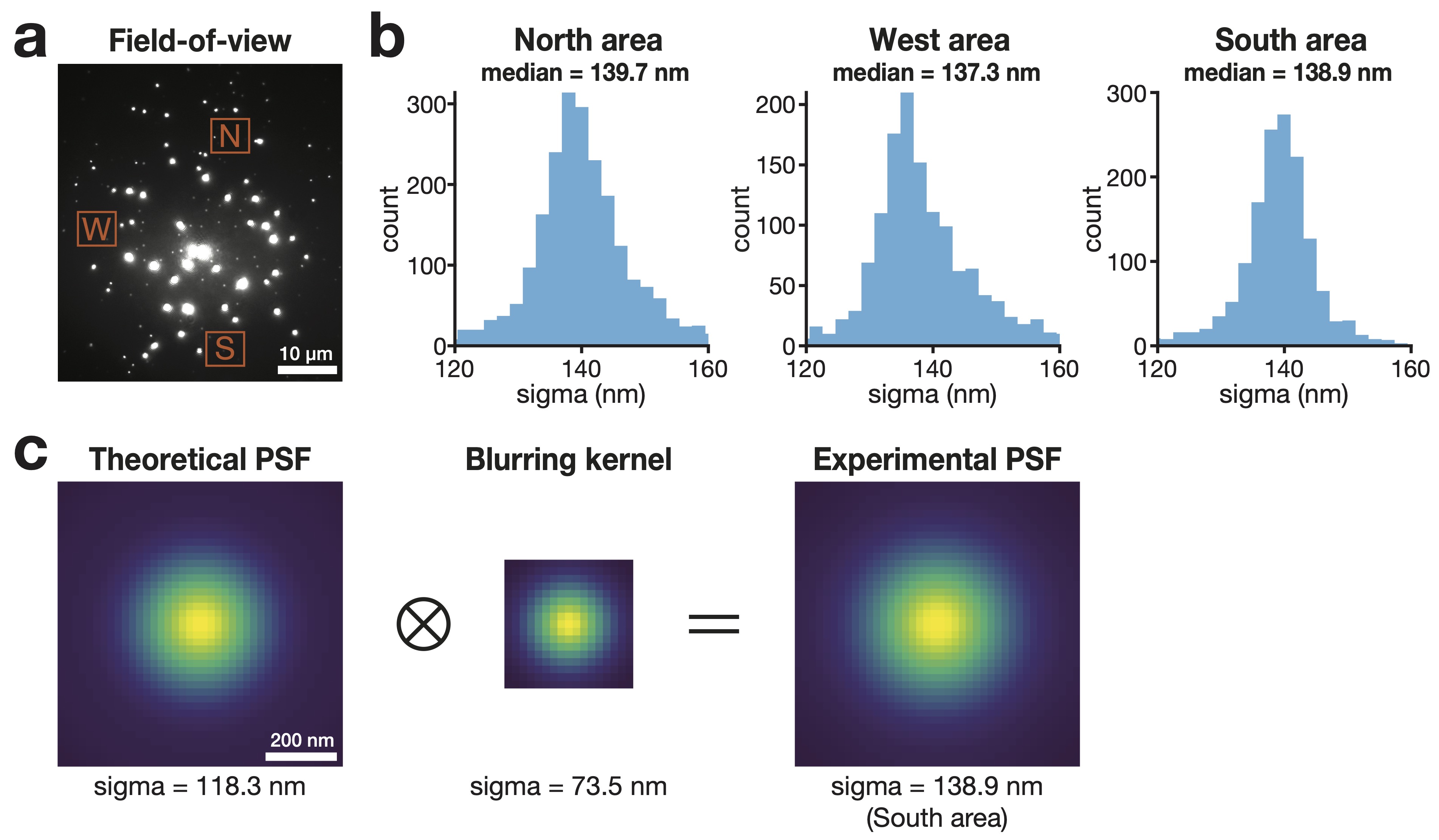}
    \caption{\textbf{Characterization and correction of spherical aberrations.}. \textbf{a} As example, we study the spherical aberrations in three areas in the field-of-view indicated by orange rectangles, referred to as North (N), West (W), and South (S). These areas are chosen, since they contain a number of nanoparticles that we have considered in our analysis. \textbf{b} For each area, we detect the spots of all unspecific binding events, which we define as spots not near a nanoparticle. We fit a 2D Gaussian function to each spot using maximum likelihood fitting \cite{Mortensen2010} which gives us the width of the spots expressed as sigma (the standard deviation of the fitted Gaussian). The three histograms show the sigma distribution of all 2322, 1442, and 1601 spots, respectively. The median of each distribution is listed above the respective distribution. We see that the median sigmas for different areas in the field-of-view vary slightly but are in the same range. \textbf{c} (left) The theoretical PSF for an rotationally free emitter located 5 nm from the coverslip, which we assume an unspecific binding event should give, has a sigma of 118.3 nm (calculated with a wavelength of 700 nm and a focous at 50 nm from the glass).   We assume that the deviation between the theoretical PSF width of 118.3 nm compared to the experimentally observed PSF width of around 138 nm (right) can be modeled by symmetric Gaussian blurring. We calculated that convoluting the theoretical PSF with a Gaussian kernel with a sigma of 73.5 nm (center) gives the experimentally obtained PSF width. See Methods for details of this analysis. Scale bar in \textbf{c}(left) applies to all images in \textbf{c}. }
    \label{fig:S22_spherAber}
\end{figure}

\newpage
\begin{figure}[ht]
    \addcontentsline{toc}{section}{S23. Characterization and correction of elliptical aberrations.}
    \centering
    \includegraphics[width=.95\linewidth]{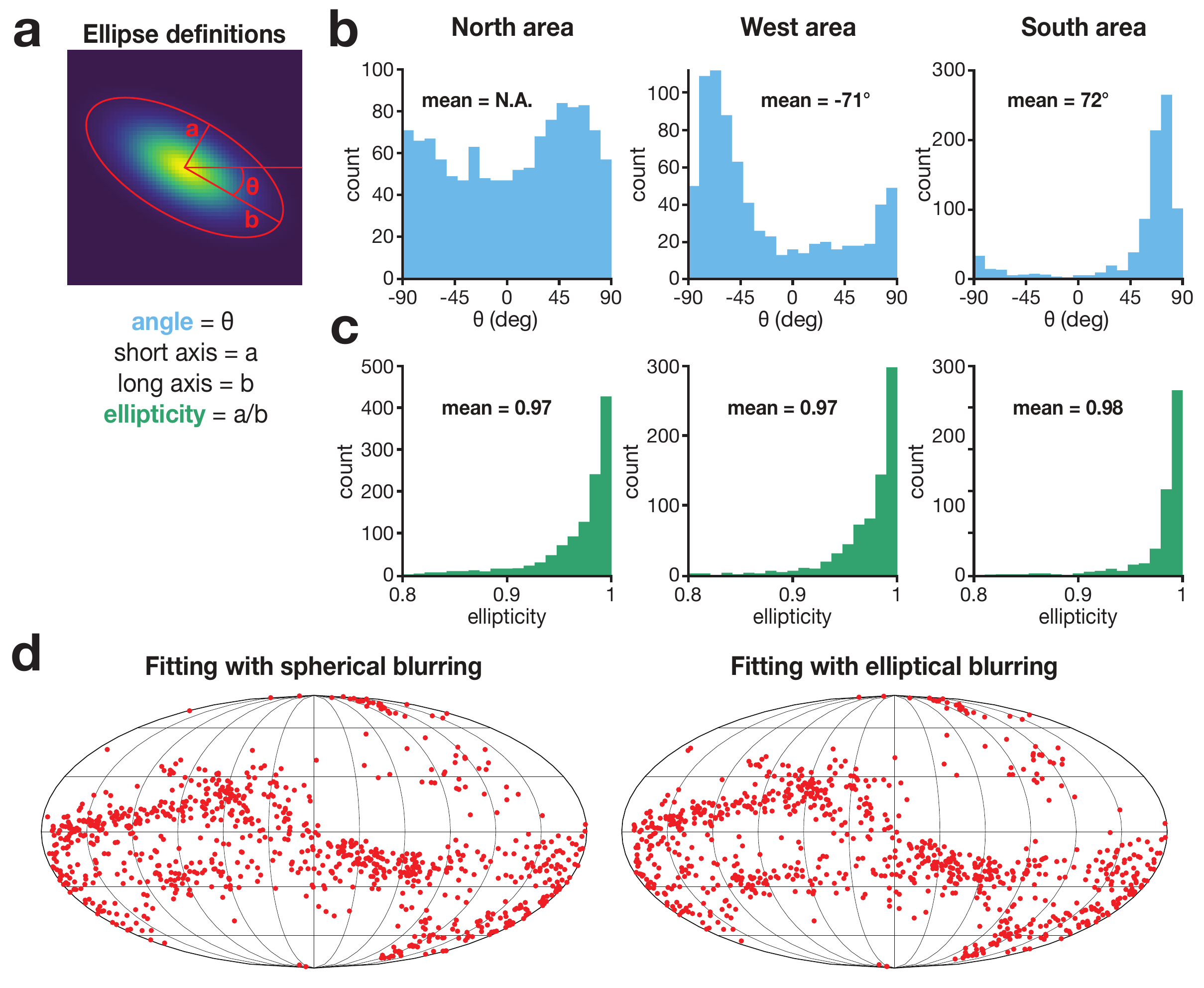}
    \caption{\textbf{Characterization and correction of elliptical aberrations.}. In our analysis, we assume that the correction of the analytical PSF model for aberrations can be modeled by symmetric 2D Gaussian blurring. In this figure, we characterize and investigate whether a multivariate 2D Gaussian is a better model for the aberrations. \textbf{a} A multivariate 2D Gaussian is a Gaussian with (possibly) two different standard deviations (a short axis $a$ and a long axis $b$), which can be rotated an angle $\theta$ around the center of the Gaussian. The ellipticity of the multivariate Gaussian is defined as the ratio of the short axis over the long axis, which gives that a circle has an ellipticity of 1 and a line an ellipticity of 0. \textbf{b} We characterize the orientation and ellipticity of the same unspecific binding events as in Fig. \ref{fig:S22_spherAber}, by fitting a multivariate 2D Gaussian to every detected spot. \textbf{b} and \textbf{c} show the angle $\theta$ and the ellipticity, respectively, for the unspecific binding events in three areas of the FOV, referred to as: North, West, and South. We see that spots in the South and West of the FOV have a distinctive orientation of the ellipse, in contrast to the spots in the North. The ellipticity values are on average around 0.97-0.98, disregarding the location in the FOV. The histograms in \textbf{b} and \textbf{c} contain data from 1188, 754, and 861 spots per area, respectively. \textbf{d} Since the ellipticity values are almost 1, we suspect that correcting the analytical PSF model with a multivariate 2D Gaussian instead of a symmetric 2D Gaussian will not have an effect. We first find the necessary elliptical kernel that, after convolving with the theoretical PSF, gives an blurred PSF with the same characteristics as the experimentally characterized PSFs (similar to \ref{fig:S22_spherAber}c). We calculate that (for a NP in the South area), we need an elliptical kernel with an ellipticity of 0.91, a long axis of 69 nm and an orientation of $72\deg$. We test the effect by blurring the analytical PSF model either with a symmetric Gaussian kernel (same as in \ref{fig:S22_spherAber}c) or with the mentioned multivariate Gaussian kernel. From the resulting surface reconstructions of 882 PSFs for a NP in the South area, we see that the two methods result in almost similar reconstructions. From this we conclude that fitting with a PSF model that is corrected for elliptical aberrations does not significantly change the obtained emitter locations, compared to correction with a symmetric Gaussian. }
\end{figure}

\newpage
\begin{figure}[ht]
    \addcontentsline{toc}{section}{S24. PSF fitting of individual frames of a binding event results does not affect found emitter positions.}
    \centering
    \includegraphics[width=.85\linewidth]{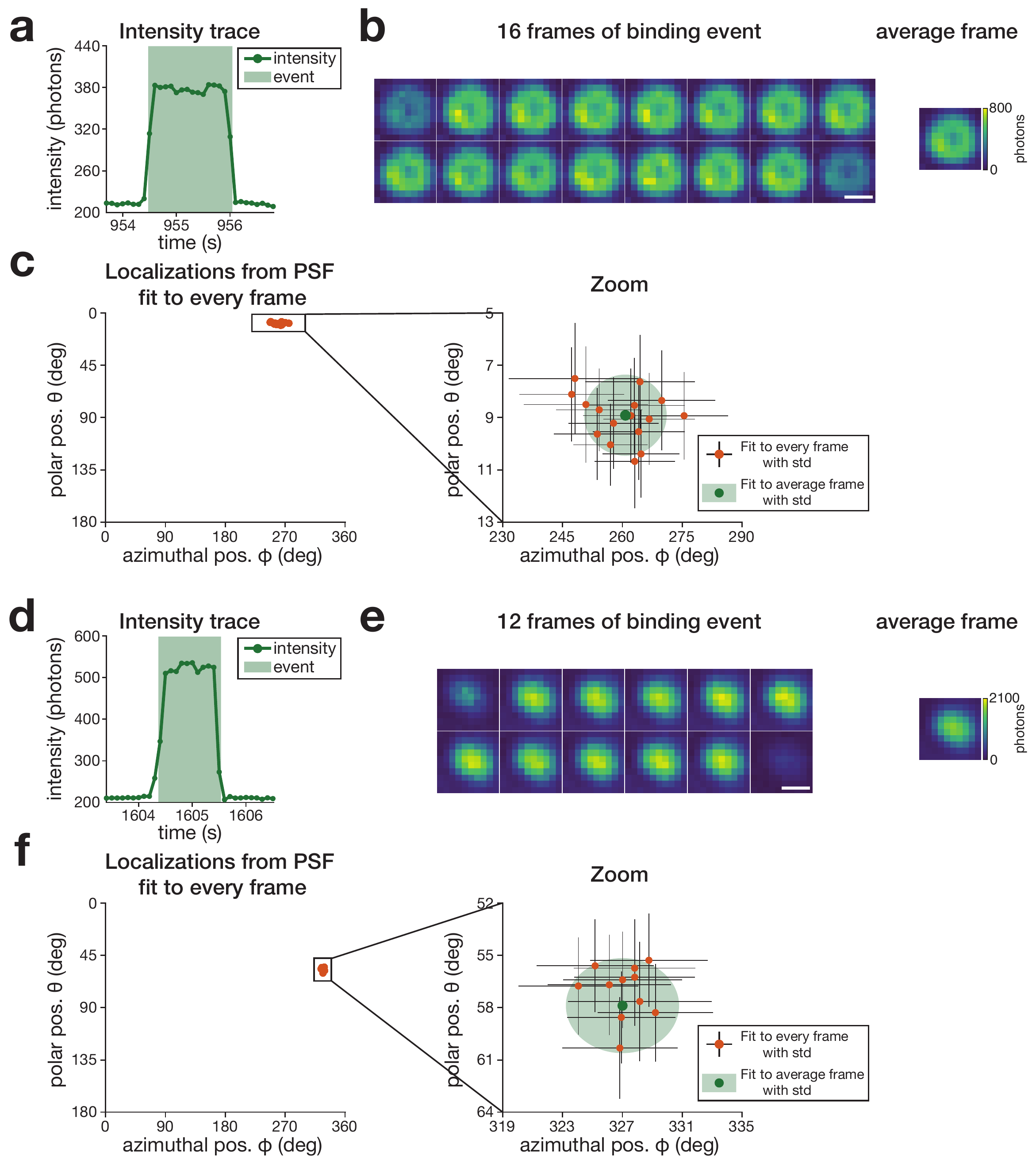}
    \caption{\textbf{PSF fitting of individual frames of a binding event results does not affect found emitter positions}. In our analysis, we detect a DNA binding event by thresholding the intensity time trace of a single NP. Then, we average all the image frames belonging to the same event and perform PSF fitting on the average frame. Here we investigate whether it makes a difference if we fit the average frame or every individual frame of the same binding event. 
    \textbf{a} Example partial intensity time trace (green line) for a single NP. The intensity is calculated as the average over $15\times15$ pixels at the location of the NP. The binding event (green shaded area) is found by thresholding the intensity trace. This event consists of 16 time frames (green dots). 
    \textbf{b} (left) The 16 image frames of the binding event in \textbf{a}, and (right) the average of these 16 frames. Color bar applies to all PSFs. The first and last image frames appear dimmer, since the binding/unbinding events happened during the integration time of these respective frames. 
    \textbf{c} By fitting our analytical PSF model to either every frame of the event or the average frame, we obtain the location of the emitter on the spherical NP surface expressed in spherical coordinates $\theta$ and $\phi$. The localizations resulting from the PSF fits to 16 individual frames (orange dots) are close to each other. In the inset on the right, we see that the spread on the localizations agrees well with the provided localization uncertainties by the fit (black crosses). Additionally, the 16 found localizations (orange) very well agree with the position and uncertainty interval of the PSF fit to the average frame (green dot and shaded area). From this we conclude that the provided localization uncertainty (green shaded area) of a single PSF fit to the average frame is correct, since 16 PSF fits to individual frames of the same event fall well within this interval. \textbf{d-f} same as \textbf{a-c}, but for another binding event on the same NP containing 12 frames. Scale bars in \textbf{b} and \textbf{e} are 300 nm and apply to all PSFs.} 
\end{figure}

\newpage
\begin{figure}[ht]
    \addcontentsline{toc}{section}{S25. Collection of figures for visual explanation of various topics covered in SI Note 2}
    \centering
    \includegraphics[width=1\linewidth]{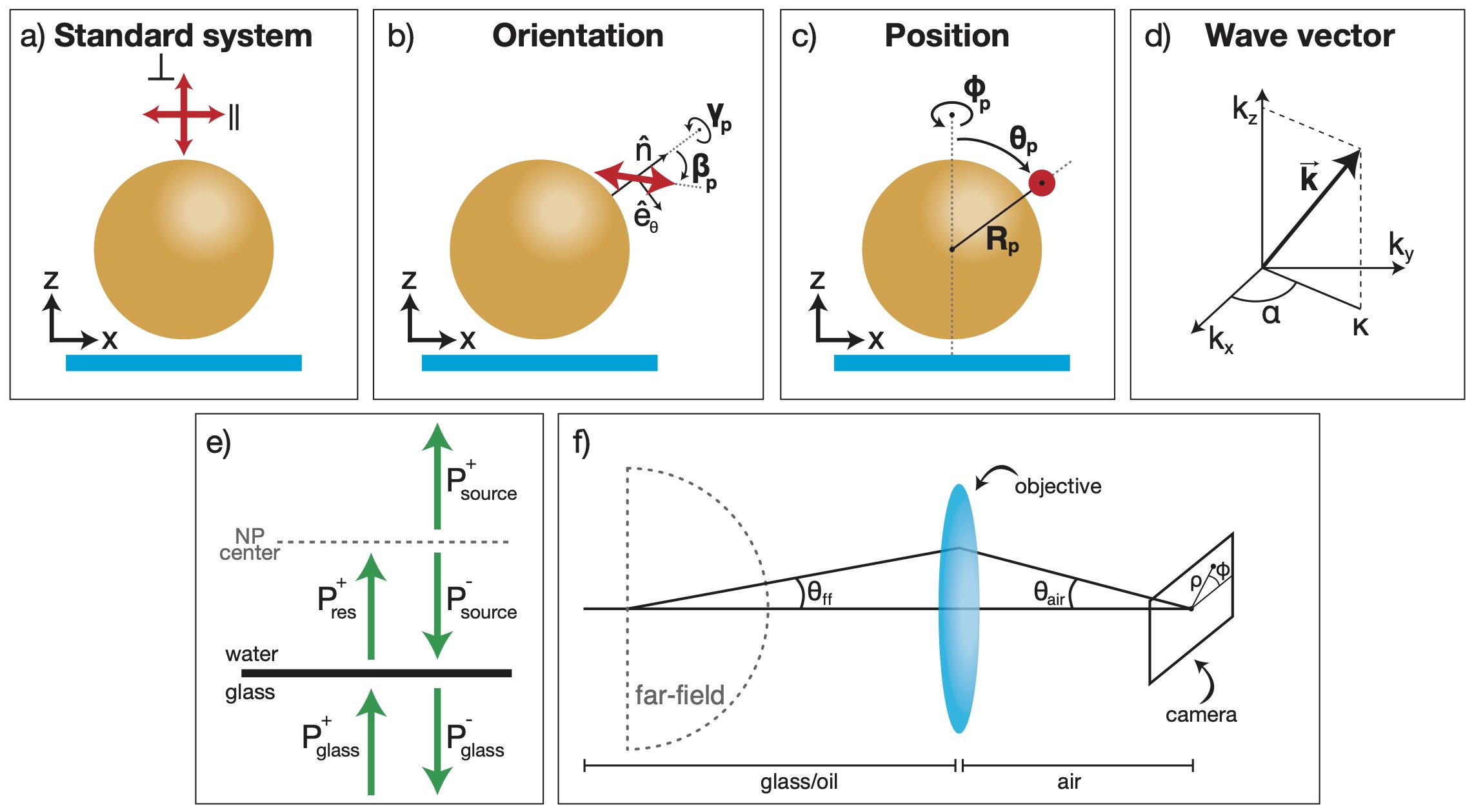}
    \caption{\textbf{Collection of figures for visual explanation of various topics covered in Supplementary Note 2.} \textbf{a} The standard Mie theory gives the solution for the perpendicular ($\perp$) and parallel ($\parallel$) dipole positioned on top of the nanoparticle. \textbf{b} The orientation of the dipole with respect to the nanoparticle is defined by the angles $\beta_p$ and $\gamma_p$, with $\beta_p$ the angle between the dipole moment (red arrow) and the surface normal $\hat{n}$, and $\gamma_p$ the rotation around the surface normal. \textbf{c} The position of the dipole with respect to the nanoparticle is defined by $R_p$ and the angles $\theta_p$ and $\phi_p$, with $R_p$ the distance from the nanopartile center, $\theta_p$ the polar position measured from the positive $z$-axis and $\phi_p$ the azimuthal position measured from $x$-axis. \textbf{d} Definition of the wave vector in cylindrical coordinates $(\kappa,\alpha,k_z)$, with $\kappa$ the length of wave vector $\vec{k}$ projected onto the $xy$-plane, $\alpha$ the angle between this projection and the positive $x$-axis, and $k_z$ the projection of $\vec{k}$ onto the $z$-axis. \textbf{e} Schematic representation of the relevant plane waves around the water-glass interface, with $P_{\mathrm{source}}^\pm$ the emission of the dipole in the water, $P_{\mathrm{res}}^+$ the response of the interface to source, and $P_{\mathrm{glass}}^\pm$ the plane waves in the glass medium (image adjusted from \cite{Egel2019}). \textbf{f} Schematic representation of the relevant angles in the focusing process, with $\theta_{\mathrm{ff}}$ the azimuthal angle of the far-field and $\theta_{\mathrm{air}}$ the azimuthal angle after diffraction in the objective lens. The image plane on the camera is described in polar coordinates, with $\phi$ the polar angle and $\rho$ the distance to the optical axis. }
    \label{fig:S25_help}
\end{figure}

\newpage
\begin{figure}[ht]
    \addcontentsline{toc}{section}{S26. Determining analytical parameters: far-field hemisphere discretization and radius}
    \centering
    \includegraphics[width=.8\linewidth]{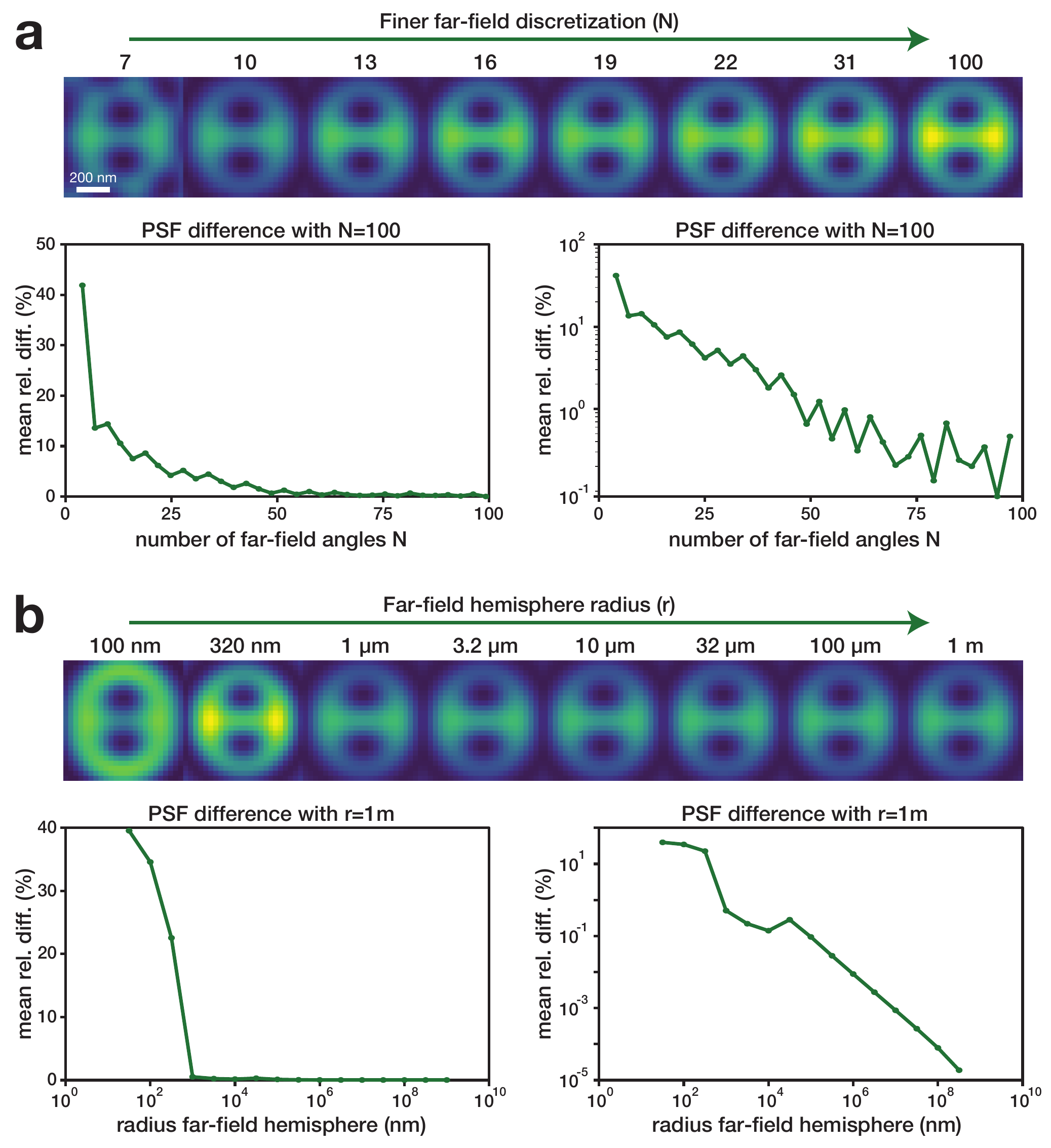}
    \caption{\textbf{Determining analytical parameters: far-field hemisphere discretization and radius}. \textbf{a} The analytical model is numerically evaluated since no closed-form solution exists to the integrals. Here we investigate how the numerical discretization of the far-field hemisphere (the number of far-field angles) affects the PSF. Multiple analytical calculations are performed with different numbers of far-field angles $N$ (ranging from 7 to 100), where the far-field hemisphere is discretized in $N\times N$ angles. Calculations are for a horizontally oriented dipole on top of a 200 nm gold spherical NP.
    Top row shows the analytical PSFs, where the numbers above each PSF indicates $N$. Bottom rows shows the difference in PSF for various $N$, with respect to the reference ($N=100$), on linear (left) and logarithmic (right) axes. The difference is defined as the mean relative difference with the reference image, calculated as $\textrm{mean}\big (|\textrm{PSF}-\textrm{ref}|/\textrm{max(ref)} \big )$. We can conclude that a bigger $N$ gives a better PSF, and that $N=100$ (the value we use in all calculations) is sufficient enough to calculate PSFs reliably, since the average pixel difference is less than 1$\%$. \textbf{b} Same as \textbf{a} but when varying the radius $r$ of the far-field hemisphere from 100 nm to 1 m. Based on these results, we conclude that $r=1$m is big enough to be in the far-field limit. Scale bar applies to all PSFs and full analytical parameters can be found in SI Tables 1 and 2.} 
    \label{fig:S26_NRsweep}
\end{figure}
\addtocontents{toc}{\protect\setcounter{tocdepth}{1}}

\chapter*{Supplementary Note 1: \\ Analytical PSF model - Textual}
\addcontentsline{toc}{chapter}{Supplementary Note 1: Analytical PSF model - Textual} 
\chaptermark{Supplementary Note 1: Analytical PSF model - Textual}

In this Note we will give an overview of how the analytical PSF model is constructed in a condensed and textual manner, without too much equations. For the full thorough mathematical derivation and explanation of the model, please consider Supplementary Note 2. The 6 chapters of this chapter corresponds to the same 6 chapters of Note 2.

In order to calculate the point-spread function (PSF) for a fixed dipole emitter near a spherical nanoparticle (NP), we start from the light emitted by the excited dipole emitter and trace these electro-magnetic (EM) fields throughout the described microscope system. The calculation consists of six steps, which we will go over one by one. The six steps are schematically outlined in Fig. \ref{fig:S2_overview}.

\subsection*{1. \quad Calculate the fields in the water by employing Mie scattering}
First, we compute the EM-fields in the water created by the excited dipole emitter and scattered by the NP. To calculate the fields in the water, we employ the theory of Mie scattering \cite{Mie1908,Bohren1983}, which describes the scattering properties of an EM plane wave by a homogeneous sphere. Mie scattering is a solution to Maxwell's equation that is exact regardless the particle size. The EM-fields $\mathbf{E}\left(\mathbf{r}\right) $ are expanded as an infinite series:

\begin{equation}
    \textbf{E}(\textbf{r}) = E_0 \sum_{\tau,n,m}\mathcal{S}_{\tau nm} \Psi_{\tau nm}^{(3)}(k,\mathbf{r}),
\end{equation}

where $\mathbf{\Psi}_{\tau nm}^{\left(3\right)}$ is a complete set of vector spherical harmonics (VSH) indexed by $\tau$, $n$ and $m$, $\mathcal{S}_{\tau nm}$ are the corresponding respective expansion coefficients, and $E_0$ the field intensity. Mie theory is generally defined with plane wave excitation. In our system, however, the NP scatters light that originates from a dipole emitter, the fluorophore. By extending the Mie theory to the case of dipole excitation \cite{LeRu2008}, we obtain the required coefficients $\mathcal{S}_{\tau nm}$ which allow us to calculate the EM-fields everywhere in the water (see Suppl. Note 2 for details). Since equation (1) is an infinite sum of vector spherical harmonics, we will further refer to it as a spherical wave expansion (SWE).

\subsection*{2. \quad Rotate SWE to obtain arbitrary dipole position and orientation}
For symmetry reasons, Mie theory is derived to calculate the solutions for a dipole positioned on top of the NP for two orientations, either perpendicular or parallel to the NP surface. By rotating the obtained SWE around the NP centre, and interpolating between the perpendicular and parallel solution, we can obtain the SWE for an arbitrary dipole position and orientation with respect to the NP and the glass surface (Fig. \ref{fig:S25_help}).

\subsection*{3. \quad Decompose the spherical wave expansion into plane wave expansion}
The thirds step is to calculate the fields in the glass, which are essentially the downward propagating EM-fields in the water after undergoing refraction in the water-glass interface. The process of refraction is well-defined for plane waves, where the direction of the refracted wave is given by Snell’s law, and amplitude by the polarization-dependent Fresnel transmission and reflection coefficients. Since refraction in a planar interface is naturally defined for plane waves, we choose to decompose every vector spherical harmonic $\mathbf{\Psi}_{\tau nm}^{\left(3\right)}$ of the SWE into plane waves using plane wave decomposition (PWD)\cite{Doicu2006}. Simplified, the PWD describes a VSH as an integral over all possible plane waves:

\begin{equation}
    \Psi_{\tau nm}^{(3)} = \sum_{j=1}^{2} \int_{\mathbb{R}^2} (B_{\tau nm,j} \mathbf{\Phi}_j) \mathrm{d}^2\mathbf{k}_{||},
\end{equation}

where $\mathbf{\Phi}_j$ represents a plane wave with $j$ as polarization, $B_{\tau nm,j}$ the set of coefficients indicating the contribution of every plane wave to this particular VSH, $\mathrm{d}^2\mathbf{k}_{||}$ the integral over all possible plane waves (See Chapter 3 of Supplementary Note 2 for details). Using this decomposition, we construct a plane wave expansion (PWE) by taking many plane wave directions and computing a coefficient for each direction. The expansion coefficient for each plane wave, $\mathcal{P}_j$, is calculated by multiplying the coefficient for a VSH $(\mathcal{S}_{\tau nm}$ in Eq. 1) with the coefficient of how strong this particular plane wave is present in this VSH $(B_{\tau nm,j}$ in Eq. 2), summed over all possible VSHs in the SWE:

\begin{equation}
    \mathcal{P}_j = \sum_{\tau,n,m} \mathcal{S}_{\tau nm} B_{\tau nm,j}.
\end{equation}

\subsection*{4. \quad Refract plane waves in water-glass interface}
In the fourth step, we compute the PWE in the glass by taking every downward propagating plane wave from the PWE in the previous step and refract it in the water-glass interface. We employ the transfer matrix formalism \cite{Dyakov2009} to relate the plane wave coefficients in the water to those in the glass:

\begin{equation}
    \binom{\mathcal{P}_{1,j}^+}{\mathcal{P}_{1,j}^-}=I_j^{1,2}\binom{\mathcal{P}_{2,j}^+}{\mathcal{P}_{2,j}^-},
\end{equation}

with the transfer matrix $I_j^{1,2}$ defined as:

\begin{equation}
    I_j^{1,2}=\ \frac{1}{t_j}\left[\begin{matrix}1&r_j\\r_j&1\\\end{matrix}\right].
\end{equation}

The factors $\mathcal{P}_{i,j}^\mp$ are the expansion coefficients for upward($+$)/downward($-$) propagating plane waves (see eq. 3), with polarization $j$, in medium $i$. The transfer matrix is constructed using the polarization-dependent Fresnel reflection and transmission coefficients $r_j$ and $t_j$. In this formalism, the water is medium $2$ and the glass medium $1$. The $2\times2$ transfer matrix $I_j^{1,2}$ with the Fresnel coefficients is only dependent on the polar angle of the plane waves since the azimuthal angle is unchanged by refraction. By using the transfer matrix formalism in combination with the PWE in the water that was determined before $(\mathcal{P}_{2,j}^\pm)$, we obtain the PWE coefficients in the glass medium $(\mathcal{P}_{1,j}^\pm)$.

\subsection*{5. \quad Far-field projection to obtain angular information}
After the plane waves have been propagated into the glass medium, we project them into the far-field. We essentially compute the angular information, which is used in the next step to focus onto the camera plane. To compute the EM-field at a given location, one has to include the contributions of all plane waves by integrating over every component of the PWE. However, the stationary phase approximation \cite{Chew1988} states that the EM-field at a certain location in the far-field is fully determined by the single plane wave propagating in that exact direction, since all other plane waves will destructively interfere. Since there is no need any more to integrate over all plane waves, we can compute the far-field directly and efficiently from the plane wave coefficients $\mathcal{P}$ in the glass, which were calculated above.

Our Python implementation of steps 2-5 makes use of existing functionalities from the SMUTHI software package designed to perform scattering calculations for multiple particles in thin-film systems, developed by Amos Egel \cite{Egel2019,egel2021smuthi}.

\subsection*{6. \quad Evaluate focusing integral to focus far-field information onto the camera}
After the light has been propagated into the far-field, the microscope system will focus the EM-fields onto the camera plane. The electric field for a certain location on the camera is affected by all far-field angles. This field is calculated by evaluating the focusing integral (Chapter 6 in Supplementary Note 2), which integrates the entire far-field while accounting for the limited collection angle (numerical aperture, NA) of the objective. The intensity of light on the camera is found by calculating the $z$-component of the Poynting vector. This gives a slightly different result than the commonly employed magnitude of the electric field, but the difference can be ignored for high magnification. This computation is repeated for every pixel in the image, resulting in the point-spread function. Ideally, one should integrate the light intensity over the area of the pixel to obtain the correct intensity. However, when the pixels are small enough, we can assume that the intensity profile is somewhat constant over one pixel, and we take the intensity at the center of the pixel multiplied by the pixel area. The required integrals in the analytical model are evaluated numerically (inspired by \cite{bloksma2021}), since no closed from solution exits, however, the discretization does not affect the outcome (Fig. \ref{fig:S26_NRsweep}).

\addtocontents{toc}{\protect\setcounter{tocdepth}{1}}

\chapter*{Supplementary Note 2: \\Analytical PSF model - Mathematical}
\addcontentsline{toc}{chapter}{Supplementary Note 2: Analytical PSF model - Mathematical} 
\chaptermark{Supplementary Note 2: Analytical PSF model - Mathematical}
\label{Note2:anMath}
\setcounter{section}{0}

The process of calculating the PSF of a dipole emitter next to a spherical nanoparticle is composed out of 6 steps (outlined textually in SI Note 1 and visually in SI Fig. \ref{fig:S2_overview}), which are subsequently: (1) calculation of the EM-fields in the water medium for the standard dipole position/orientation in terms of spherical waves, (2) rotation and interpolation of the fields to obtain desired dipole position/orientation relative to the nanoparticle and the glass, (3) decomposition of the fields into plane waves, (4) refraction of the plane waves in the water-glass interface, (5) projection of the fields into the far-field, and (6), focusing the fields onto the camera. In addition to the Methods section, each of the 6 steps is explained in detail in the chapters below. 

\section{Electromagnetic fields in water (Mie scattering)}
The first step towards calculating the electromagnetic (EM) fields on the camera is finding the fields in the water medium around the nanoparticle-dipole system (Supplementary Fig. \ref{fig:S2_overview}a). Here, a fluorescent molecule is present in close vicinity to a spherical nanoparticle. Fluorescent molecules in general contain a chemical group that has a dipole moment. When excited, the fluorophore goes into the excited state, with a probability proportional to the overlap of  the excitation dipole moment with the polarization of the excitation light. When transitioning back from the excited state to the ground state, the molecule emits light in the direction according to the emission dipole moment (which is not necessarily the same as excitation dipole moment). Due to these photo-physical effects, we model the fluorescent molecule as an excited dipole emitter, where we only consider the process of emission. 

The dipole emitter will emit light, which will be scattered by the nanoparticle. The fields in the water thus contain two contributions: the incident field, originating from the dipole emitter, and the scattered field, which is the effect of the nanoparticle on the incident field. For finding the scattered field we employ Mie scattering theory. 

\subsection{Mie scattering - overview}
\label{sec:Mie_overview}
Mie scattering is a solution to Maxwell's equation that describes the scattering properties of an electromagnetic plane wave by a homogeneous sphere \cite{Mie1908,Bohren1983} The solution is exact regardless of the nanoparticle size and describes the EM-fields as infinite series:

\begin{equation}
    \label{eq:Mie_E_standard}
    \mathbf{E}(\mathbf{r}) = E_0 \sum_{n=0}^{\infty} \sum_{m=-n}^{m=n} a_{nm}\mathbf{M}_{nm}^{(i)}(k,\mathbf{r}) + b_{nm}\mathbf{N}_{nm}^{(i)}(k,\mathbf{r}),
\end{equation}

where $E_0$ is the electric field amplitude and $\mathbf{r}$ the position vector in spherical coordinates, $k$ is the complex-valued wave number $k=2\pi \sqrt{\epsilon}/\lambda$, with $\epsilon$ the dielectric constant, normally referred to as the refractive index $n=\sqrt{\epsilon}$. The electric field is expressed as an infinite series of the vector spherical harmonics (VSHs) $\mathbf{M}_{nm}^{(i)}(k,\mathbf{r})$ and $\mathbf{N}_{nm}^{(i)}(k,\mathbf{r})$, each with their respective complex coefficients $a_{nm}$ and $b_{nm}$. The infinite sum runs over two indices, $n$ and $m$, with $n$ a positive integer representing the degree and integer $m$ representing the order, limited to $|m|\leq n$. There are multiple types of VSHs dependent on the nature of the electric field, which are represented by the superscript $(i)$ that can take values $i=1...4$. In our formulations we will only need regular ($i=1$) and outgoing ($i=3$) spherical waves, so we will not address the other options in this document. From the decomposition the electric field in VSHs, the magnetic field can be easily derived by substituting the coefficients: 

\begin{equation}
    \label{eq:Mie_H_standard}
    \mathbf{H}(\mathbf{r}) = H_0 \sum_{n=0}^{\infty} \sum_{m=-n}^{m=n} b_{nm}\mathbf{M}_{nm}^{(i)}(k,\mathbf{r}) + a_{nm}\mathbf{N}_{nm}^{(i)}(k,\mathbf{r}),
\end{equation}

with $H_0 = \frac{kE_0}{iw\mu_0}$.

In the Mie scattering theory, the field outside of the nanoparticle has two components: an incident field $\mathbf{E}_{\mathrm{inc}}$, which is applied externally (in our case by the dipole emitter), and a scattered field $\mathbf{E}_{\mathrm{scat}}$, which is the reaction of the nanoparticle to the incident field. The total field outside of the nanoparticle is the sum of the two: 

\begin{equation}
    \label{eq:E_total}
    \mathbf{E}_{\mathrm{out}} =
    \mathbf{E}_{\mathrm{inc}} +
    \mathbf{E}_{\mathrm{scat}}. 
\end{equation}

Both components of the electric field can be decomposed into VSHs, similar to Eq. \eqref{eq:Mie_E_standard}, with their own respective coefficients. For the incident field we have: 

\begin{equation}
    \label{eq:E_inc_near}
    \mathbf{E_\mathrm{inc}}(\mathbf{r}) = E_0 \sum_{n,m} a_{nm}\mathbf{M}_{nm}^{(1)}(k,\mathbf{r}) + b_{nm}\mathbf{N}_{nm}^{(1)}(k,\mathbf{r})
    \qquad (r<r_p),
\end{equation}

where the known coefficients $a_{nm}$ and $b_{nm}$ are determined fully by the type of incident field. Similarly, for the scattered field we have:

\begin{equation}
    \label{eq:E_scat}
    \mathbf{E_\mathrm{scat}}(\mathbf{r}) = E_0 \sum_{n,m} c_{nm}\mathbf{M}_{nm}^{(3)}(k,\mathbf{r}) + d_{nm}\mathbf{N}_{nm}^{(3)}(k,\mathbf{r}),
\end{equation}

where the unknown coefficients $c_{nm}$ and $d_{nm}$ depend on the scattering properties of the spherical nanoparticle and the incident field.  The corresponding magnetic fields can be derived using Eq. \eqref{eq:Mie_H_standard}. Note the different types of VSHs used for both electric fields, for which the difference will be explained later.

It is important to note that the expansion of the incident field $\mathbf{E}_{\mathrm{inc}}$ is only valid for $r < R_p$, where $R_p$ is the position of the dipole emitter. When interested in the field further away from the nanoparticle, an expansion in terms of outgoing spherical harmonics, instead of regular, is necessary:

\begin{equation}
    \label{eq:E_inc_far}
    \mathbf{E_\mathrm{inc}}(\mathbf{r}) = E_0 \sum_{n,m} e_{nm}\mathbf{M}_{nm}^{(3)}(k,\mathbf{r}) + f_{nm}\mathbf{N}_{nm}^{(3)}(k,\mathbf{r})
    \qquad (r>r_p).
\end{equation}

The expansion for the scattered field is valid everywhere outside of the nanoparticle. The coefficients $a_{nm}$, $b_{nm}$, $e_{nm}$ and $f_{nm}$ are known and depend on the type of incident field, which is a dipole field in the case of a fluorescent molecule. The unknown expansion coefficients $c_{nm}$ and $d_{nm}$ are proportional to the known coefficients and can be calculated from them:

\begin{equation}
    c_{nm}  = \Gamma_n a_{nm}
    \quad \mathrm{and} \quad
    d_{nm}  = \Delta_n b_{nm},
    \label{eq:cd_from_ab}
\end{equation}

where $\Gamma_n$ and $\Delta_n$ are called the susceptibilities of the spherical nanoparticle, and represent the optical response of the nanoparticle to the incident field. Effectively, all coefficients of the final solution are calculated from the coefficients of the incident field ($a_{nm}$ and $b_{nm}$). For the two most commonly used types of incident field, plane waves and dipole emission, these coefficients simplify significantly and $a_{nm}$ and $b_{nm}$ are only non-zero for $m=0$ and $|m|=1$. This property carries on to all other coefficients, so we only need to consider coefficients for these three values of $m$.

\subsection{Coordinate system}
\label{sec:coord_system}
All equations and definitions presented in this document assume spherical coordinates ($r$,$\theta$,$\phi$), according to the following convention: 

\begin{itemize}
    \item $r \geq 0$, the distance to the origin
    \item $0 \leq \theta \leq \pi$, the polar angle, measured from $\textbf{e}_z$ (positive $z$-axis) to the point of interest
    \item $0 \leq \phi \leq 2\pi$, the azimuthal angle, measured counter-clockwise from $\textbf{e}_x$ (positive $x$-axis) to the projection of the point of interest onto the $xy$-plane 
\end{itemize}

The conversion from spherical coordinates to Cartesian coordinates is therefore as follows: 
\begin{equation}
\label{eq:Cartesian}
\begin{cases}
x = r \sin \theta \cos \phi \\
y = r \sin \theta \sin \phi \\
z = r \cos \theta, \\
\end{cases}
\end{equation}

with the unit basis vectors: 

\begin{equation}
\label{eq:basis_vectors}
\left \{
\begin{array}{lllll}
\textbf{e}_r  \; &=
\sin \theta \cos \phi & \textbf{e}_x \; + 
\sin \theta \sin \phi & \textbf{e}_y \; + 
\cos \theta  &\textbf{e}_z\\
\textbf{e}_{\theta} \; &= 
\cos \theta \cos \phi & \textbf{e}_x \; + 
\cos \theta \sin \phi & \textbf{e}_y \; - 
\sin \theta  &\textbf{e}_z\\
\textbf{e}_{\phi} \; &= - \sin \phi           & \textbf{e}_x \; + 
\cos \phi & \textbf{e}_y. &\\
\end{array} \right.
\end{equation}

The origin of the coordinate system is placed at the center of the spherical nanoparticle.

\subsection{Definitions of vector spherical harmonics}

The vector spherical harmonics, in which the EM-fields are decomposed, are complicated functions of $r$, $\theta$ and $\phi$:

\begin{equation}
\label{eq:M_all}
\begin{cases}
\mathbf{M}_{nm}(k,\mathbf{r}) \cdot \mathbf{e}_r = 0\\
\mathbf{M}_{nm}(k,\mathbf{r}) \cdot \mathbf{e}_{\theta} = \mathrm{i} Z_n^0(kr)T_{nm}^1(\theta)e^{\mathrm{i}m\phi}\\
\mathbf{M}_{nm}(k,\mathbf{r}) \cdot \mathbf{e}_{\phi}= - Z_n^0(kr)T_{nm}^3(\theta)e^{\mathrm{i}m\phi},\\
\end{cases}
\end{equation}

\begin{equation}
\label{eq:N_all}
\begin{cases}
\mathbf{N}_{nm}(k,\mathbf{r}) \cdot \mathbf{e}_r = Z_n^1(kr)T_{nm}^2(\theta)e^{\mathrm{i}m\phi}\\
\mathbf{N}_{nm}(k,\mathbf{r}) \cdot \mathbf{e}_{\theta} = Z_n^2(kr)T_{nm}^3(\theta)e^{\mathrm{i}m\phi}\\
\mathbf{N}_{nm}(k,\mathbf{r}) \cdot \mathbf{e}_{\phi}= \mathrm{i} Z_n^2(kr)T_{nm}^1(\theta)e^{\mathrm{i}m\phi}.\\
\end{cases}
\end{equation}

They have the convenient characteristic that the $r-$, $\theta-$ and $\phi-$dependencies can be decoupled. The $r$-dependent auxiliary function $Z$ is defined as: 

\begin{align}
    & Z_n^0(x) = z_n(x), \nonumber \\
    & Z_n^1(x) = z_n(x)/x, \nonumber \\
    & Z_n^2(x) = [xz_n(x)]'/x,
    \label{eq:Z}
\end{align}

where $z_n$ is a type of spherical Bessel function, depending on the nature of the spherical waves (regular or outgoing). The two types of VSH (regular and outgoing) differ in their type of spherical Bessel function. Either the spherical Bessel function of the first kind ($j_n$) or the spherical Hankel function of the first kind ($h_n^{(1)}$) is used, the former in case of superscript $i=1$ (in $\mathbf{M}_{nm}^{(i)}$ and $\mathbf{N}_{nm}^{(i)}$) and the latter of superscript $i=3$. More details about the definitions and implementation of the spherical Bessel and Hankel functions can be found in Section \eqref{sec:Impl_details}

The $\theta$-dependent auxiliary functions are: 

\begin{align}
    & T_{nm}^1(\theta) = m \frac{T_{nm}(\theta)}{\sin\theta}, \nonumber \\
    & T_{nm}^2(\theta) = n(n+1)T_{nm}(\theta), \nonumber \\
    & T_{nm}^3(\theta) = \frac{\partial T_{nm}}{\partial \theta}(\theta). 
    \label{eq:T_all}
\end{align}

In these definitions, the $T$ functions are defined as: 

\begin{equation}
\label{eq:T_def}
    T_{nm}(\theta) = \frac{1}{\sqrt{n(n+1)}}Y_{nm}(\theta,\phi=0).
\end{equation}

$Y_{nm}(\theta,\phi)$ are the spherical harmonics:

\begin{equation}
\label{eq:Y_def}
   Y_{nm}(\theta,\phi) = \sqrt{\frac{2n+1}{4\pi}\frac{(n-m)!}{(n+m)!}}
   P_n^m(\cos(\theta))e^{\mathrm{i}m\phi},
\end{equation}

with $P_n^m(x)$ the associated Legendre functions, defined as: 

\begin{equation}
\label{eq:P_def}
   P_n^m(x) = \frac{(-1)^m}{2^nn!}(1-x^2)^{m/2}
   \frac{\mathrm{d}^{n+m}}{\mathrm{d}x^{n+m}}(x^2-1)^n \qquad (m\geq0).
\end{equation}

\subsubsection{Simplification of VSH for $|m|=1$}
We have seen in Section \ref{sec:Mie_overview} that for the dipole incident field, the coefficients are only non-zero for $m=0$ and $|m|=1$. With this knowledge, we can simplify the equations for the VSH (Eq. \eqref{eq:M_all}), starting for the case $|m|=1$. We simplify the $\theta$-dependent $T_{nm}$ functions, the only auxiliary functions that depends on $m$, by using recurrence relations for the associated Legendre functions:

\begin{equation}
    \label{eq:pi_tau}
    \pi_n(\theta) = -\frac{P_n^1(\cos\theta)}{\sin \theta}
    \quad \mathrm{and} \quad
    \tau_n(\theta) = - \frac{dP_n^1(\cos \theta)}{d\theta},
\end{equation}

where the recurrence relations are defined as: 

\begin{equation}
    \label{eq:pi_def}
    \pi_n(\theta) = \frac{2n-1}{n-1}\cos(\theta) \pi_{n-1}(\theta) - \frac{n}{n-1}\pi_{n-2}(\theta),
\end{equation}

\begin{equation}
    \label{eq:tau_def}
    \tau_n(\theta) = n\cos(\theta) \pi_{n}(\theta) - (n-1)\pi_{n-1}(\theta),
\end{equation}

with the initial conditions:

\begin{equation}
    \label{eq:pi_tau_init}
    \pi_0 = 0
    \quad \mathrm{and} \quad
    \pi_1 = 1.
\end{equation}

From this we can see that for $\theta=0$: 

\begin{equation}
    \label{eq:pi_tau_theta0}
    \pi_n(0) = \tau_n(0) = \frac{1}{2}n(n+1)
\end{equation}

The $\theta$-dependent $T_{nm}$ functions are, using these recurrence relations, defined as: 

\begin{align}
    T_{n,1}^1(\theta) = T_{n,-1}^1(\theta) & = -\mu_n \pi_n(\theta),\\
    T_{n,1}^2(\theta) = -T_{n,-1}^2(\theta) & = -\mu_n n(n+1) \sin (\theta) \pi_n(\theta), \label{eq:Tn12}\\
    T_{n,1}^3(\theta) = -T_{n,-1}^3(\theta) & = -\mu_n \tau_n(\theta), 
\end{align}

with:
\begin{equation}
    \label{eq:mu_n}
    \mu_n = \sqrt{\frac{2n+1}{4\pi}}\frac{1}{n(n+1)}.
\end{equation}

\subsubsection{Simplification of VSH for $m=0$}
In line with the previous section, the $\theta$-dependent $T_{nm}$ functions can also be simplified for the case of $m=0$. The recurrence relations for the associated Legendre functions: 

\begin{equation}
    \label{eq:p_t}
    p_n(\theta) = P_n^0(\cos \theta)
    \quad \mathrm{and} \quad
    t_n(\theta) = - \frac{dP_n^0(\cos \theta)}{d\theta},
\end{equation}

where the recurrence relations are defined as: 

\begin{equation}
    \label{eq:p_def}
    p_n = \frac{2n-1}{n}\cos(\theta) p_{n-1}(\theta) - \frac{n-1}{n}p_{n-2}(\theta),
\end{equation}

\begin{equation}
    \label{eq:t_def}
    t_n = \cos(\theta) t_{n-1}(\theta) - n\sin(\theta)p_{n-1}(\theta),
\end{equation}

with the initial conditions:

\begin{equation}
    \label{eq:p_t_init}
    p_0 = 1,  \quad
    p_1(\theta) = \cos\theta, \quad
    t_0 = 0, \quad
    \mathrm{and} \quad
    t_1(\theta) = -\sin\theta.
\end{equation}

From this we can see that for $\theta=0$: 

\begin{equation}
    \label{eq:p_t_theta0}
    p_n(0) = 1, \quad
    \mathrm{and} \quad
    t_n(0) = 0.
\end{equation}

The $\theta$-dependent $T_{nm}$ functions are, using these recurrence relations, defined as: 

\begin{align}
    T_{n,0}^1(\theta) & = 0 \label{eq:Tn01},\\
    T_{n,0}^2(\theta) & = \mu_{n,0} n(n+1) p_n(\theta), \label{eq:Tn02}\\
    T_{n,0}^3(\theta) & = \mu_{n,0} t_n(\theta), \label{eq:Tn03} 
\end{align}

with:
\begin{equation}
    \label{eq:mu_n0}
    \mu_{n,0} = \sqrt{\frac{2n+1}{4\pi n(n+1)}}.
\end{equation}

\subsection{Susceptibilities of nanoparticle}
We have seen in Section \ref{sec:Mie_overview} that the full solution of electromagnetic fields can be calculated from the expansion coefficients, where $a_{nm}$, $b_{nm}$, $e_{nm}$ and $f_{nm}$ are dictated by the incident field and $c_{nm}$ and $d_{nm}$ are calculated using the susceptibilities (Eq. \eqref{eq:cd_from_ab}). Before going into the exact definitions of the expansion coefficients of the incident field, we first look at the susceptibilities. The magnetic and electric susceptibilities, $\Gamma_n$ and $\Delta_n$, are the proportionality factors between the expansion coefficients for the incident field and the scattered field, and thereby represent the response of the spherical nanoparticle to the (known) incident field. 

We first define the following dimensionality conventions: 

\begin{equation}
    \label{eq:x_and_s}
    x = k_M a = \frac{2\pi a}{\lambda} \sqrt{\epsilon_M} , \quad
    \quad
    x_p = k_M R_p = \frac{2\pi R_p}{\lambda} \sqrt{\epsilon_M} , \quad
    \mathrm{and} \quad
    s = \frac{k_{in}}{k_M} = \frac{\sqrt{\epsilon_{in}}}{\sqrt{\epsilon_M}},
\end{equation}

with $a$ the radius of the nanoparticle and $R_p$ the distance from the dipole to the origin. The subscripts $in$ stand for internal, the properties of the nanoparticle, and the subscripts $M$ for medium, the water solution around the nanoparticle. Using these conventions, the magnetic susceptibility is:

\begin{equation}
    \label{eq:Gamma_n}
    \Gamma_n = \frac{s\psi_n(x)\psi_n'(sx) - \psi_n(sx)\psi_n'(x)}{\psi_n(sx)\xi_n'(x) - s\xi_n(x)\psi_n'(sx)},
\end{equation}

and electric susceptibility: 

\begin{equation}
    \label{eq:Delta_n}
    \Delta_n = \frac{\psi_n(x)\psi_n'(sx) - s\psi_n(sx)\psi_n'(x)}{s\psi_n(sx)\xi_n'(x) - \xi_n(x)\psi_n'(sx)}.
\end{equation}

$\psi_n$ and $\xi_n$ are the Riccati-Bessel function of the first and third kind, which are slight variations on the spherical Bessel and Hankel functions $j_n$ and $h_n^{(1)}$ (respectively, of the first and third kind): 

\begin{equation}
    \label{eq:Riccati_Bessel}
    \psi_n(\rho) = \rho j_n(\rho) , \quad
    \mathrm{and} \quad
    \xi_n(\rho) = \rho h_n^{(1)}(\rho) ,
\end{equation}

More details about the definition and implementation of the Riccati-Bessel functions and its derivatives, can be found in Section \eqref{sec:Impl_details}

\subsection{Mie coefficients}
\label{sec:mie_coeff}
Without going into the derivations, the expansion coefficients of the incident field (Eqs. \eqref{eq:E_inc_near} and \eqref{eq:E_inc_far}), valid for any arbitrary dipole position, are given by: 

\begin{equation}
\label{eq:coeffs}
\begin{cases}
a_{nm} = \sqrt{6\pi}(-1)^m \mathbf{e}_p \cdot \mathbf{M}_{n,-m}^{(3)}(k_M,\mathbf{R}_p) \\
b_{nm} = \sqrt{6\pi}(-1)^m \mathbf{e}_p \cdot \mathbf{N}_{n,-m}^{(3)}(k_M,\mathbf{R}_p) \\
e_{nm} = \sqrt{6\pi}(-1)^m \mathbf{e}_p \cdot \mathbf{M}_{n,-m}^{(1)}(k_M,\mathbf{R}_p) \\
f_{nm} = \sqrt{6\pi}(-1)^m \mathbf{e}_p \cdot \mathbf{N}_{n,-m}^{(1)}(k_M,\mathbf{R}_p), \\
\end{cases}    
\end{equation}

where $k_M$ is the wave number in the medium and $\mathbf{R}_p$ the position vector of the dipole emitter. These equations can be drastically simplified when we place the dipole emitter on top of the nanoparticle oriented within the (xOz) plane. This means that the dipole is positioned on the positive $z$-axis ($\theta=\phi=0$) outside of the nanoparticle ($\mathbf{R}_p>a$, with $a$ the radius of the nanoparticle). This is without any loss of generality, since the coordinate system can always be rotated to reach this situation, due to spherical symmetries.  

Furthermore, we only consider two dipole orientations, perpendicular and parallel to the nanoparticle surface. The dipole moment, $\mathbf{e}_p$, of the perpendicular dipole points into the positive $z$-direction ($\mathbf{e}_p=\mathbf{e}_z=\mathbf{e}_r$), and of the parallel dipole into the positive $x$-direction ($\mathbf{e}_p=\mathbf{e}_x=\mathbf{e}_\theta$). Using the definitions of the vector spherical harmonics \eqref{eq:M_all}, we can simplify the expressions for the coefficients of the \textit{perpendicular dipole}, where only $m=0$ coefficients are non-zero:  

\begin{align} 
\label{eq:simplify_coeffs_perp} 
a_{nm} & = 0  \quad \forall (n,m), \\
b_{nm} & = 0 \quad \forall m \neq 0, \\
b_{n,0} & = \sqrt{6\pi}(-1)^m \mathbf{e}_p \cdot \mathbf{N}_{n,-m}^{(3)}(k_M,\mathbf{R}_p) \nonumber \\
& = \sqrt{6\pi} \mathbf{e}_r \cdot \mathbf{N}_{n,0}^{(3)}(k_M,\mathbf{R}_p) \nonumber \\
& = \sqrt{6\pi} Z_n^1(x_p)T_{n,0}^2(\theta)e^{\mathrm{i}0\phi} \nonumber \\
& = \sqrt{6\pi} \frac{h_n^{(1)}(x_p)}{x_p} \mu_{n,0} n(n+1) p_n(\theta) \nonumber \\
& = \sqrt{\frac{3}{2}n(n+1)(2n+1)} \frac{\xi_n(x_p)}{x_p^2}. \label{eq:bn0}
\end{align}

We use Eq. \eqref{eq:M_all} to derive the first statement of \eqref{eq:simplify_coeffs_perp}, since $\mathbf{M}_{nm}$ has no component in the $r$-direction. For the second statement, we use Eq. \eqref{eq:N_all} in combination with \eqref{eq:Tn12} and $\theta=0$. The third statement, $b_{n,0}$, gives the only non-zero coefficients. In the 4-step derivation we subsequently use: the combination that $m=0$ and $\mathbf{e}_p = \mathbf{e}_r$, the definition of $\mathbf{N}_{n,0}$ (Eq. \eqref{eq:N_all}), the definitions of $Z_n^1$ and $T_{n,0}^2$ (Eqs. \eqref{eq:Z} and \eqref{eq:Tn02}), and lastly the definition of $\xi_n$ (Eq. \eqref{eq:Riccati_Bessel}) in combination with the insight that $p_n$ is always 1 for $\theta=0$ (Eq. \eqref{eq:p_t_theta0}). According to the same reasoning, with as only difference the type of spherical Bessel function: 

\begin{align} 
\label{eq:simplify_coeffs_perp_far} 
e_{nm} & = 0  \quad \forall (n,m), \\
f_{nm} & = 0 \quad \forall m \neq 0, \\
f_{n,0} & = \sqrt{\frac{3}{2}n(n+1)(2n+1)} \frac{\psi_n(x_p)}{x_p^2}. \label{eq:fn0}
\end{align}

We can also simplify the expressions for the expansions coefficients (Eq. \eqref{eq:coeffs}) for the case of a \textit{parallel dipole}, where only the $|m|=1$ expansion coefficients are non-zero: 

\begin{align} 
\label{eq:simplify_coeffs_para} 
a_{nm} & = 0  \quad \forall |m|\neq 1, \\
b_{nm} & = 0  \quad \forall |m|\neq 1, \\
a_{n,\pm 1} & = \sqrt{6\pi}(-1)^{\pm 1} \mathbf{e}_\theta \cdot \mathbf{M}_{n,\mp 1}^{(3)}(k_M,\mathbf{R}_p) \nonumber \\
& = \sqrt{6\pi}(-1) \mathrm{i} Z_n^0(x_p)T_{n,\mp 1}^1(\theta)e^0 \nonumber \\
& = \sqrt{6\pi} \mathrm{i} h_n^{(1)}(x_p) \mu_n \pi_n(\theta) \nonumber \\
& = \mathrm{i} \sqrt{\frac{3}{8}(2n+1)} \frac{\xi_n(x_p)}{x_p}, \label{eq:an1}\\
b_{n,\pm 1} & = \sqrt{6\pi}(-1)^{\pm 1} \mathbf{e}_\theta \cdot \mathbf{N}_{n,\mp 1}^{(3)}(k_M,\mathbf{R}_p) \nonumber \\
& = \sqrt{6\pi}(-1) Z_n^2(x_p)T_{n,\mp 1}^3(\theta)e^0 \nonumber \\
& = \mp \sqrt{6\pi} \frac{[x_p h_n^{(1)}(x_p)]'}{x_p} \mu_n \tau_n(\theta) \nonumber \\
& = \mp \sqrt{\frac{3}{8}(2n+1)} \frac{\xi_n'(x_p)}{x_p}. \label{eq:bn1}
\end{align}

The first statement is found by combining Eqs. \eqref{eq:M_all} with \eqref{eq:Tn01}, in combination with the knowledge that a dipole incident field only has non-zero coefficients for $m=0$ and $|m|=1$. For the second statement, we combine Eqs. \eqref{eq:N_all}, \eqref{eq:Tn03} and \eqref{eq:p_t_theta0}.In the 4-step derivation of the third statement ($a_{n,\pm1}$), we subsequently use: the definition of $\mathbf{M}_{nm}$ (Eq. \eqref{eq:M_all}), the definitions of $Z_n^0$ and $T_{n,0}^1$ (Eqs. \eqref{eq:Z} and \eqref{eq:Tn01}), and lastly the definitions of $\mu_n$, $\xi_n$ and $\pi_n$ (Eqs. \eqref{eq:mu_n}, \eqref{eq:Riccati_Bessel} and \eqref{eq:pi_tau_theta0}). The 4-step derivation of the fourth statement ($b_{n,\pm1}$) follows the same reasoning and equations, except with $\mathbf{N}_{nm}^{(3)}$ instead of $\mathbf{M}_{nm}^{(3)}$. According to the same reasoning, with as only difference the type of spherical Bessel function: 
\begin{align} 
\label{eq:simplify_coeffs_para_far} 
e_{nm} & = 0  \quad \forall |m|\neq 1, \\
f_{nm} & = 0  \quad \forall |m|\neq 1, \\
e_{n,\pm 1} & = \mathrm{i} \sqrt{\frac{3}{8}(2n+1)} \frac{\psi_n(x_p)}{x_p}, \label{eq:en1}\\
f_{n,\pm 1} & = \mp \sqrt{\frac{3}{8}(2n+1)} \frac{\psi_n'(x_p)}{x_p}.\label{eq:fn1}
\end{align}

More details about the definition and implementation of the Riccati-Bessel functions and its derivatives, can be found in Section \eqref{sec:Impl_details}

\subsection{Details and implementation - spherical Bessel functions}
\label{sec:Impl_details}

In the definitions of the VSHs and the expansion coefficients of the Mie scattering solution, we frequently encountered spherical Bessel function and its modifications and derivatives. This section will give a clear overview of the encountered functions, and how they are implemented in both the MATLAB and Python programming environments. 

The Mie scattering solution makes use of $j_n$ and $h_n^{(1)}$, and its modifications $\psi_n(x)$ and $\xi_n(x)$. Tabel \ref{tab:Bessel} gives an overview of how these functions relate to the original Bessel function, and how to compute their values in both MATLAB and Python. 

\begin{table}[ht]
  \begin{center}
    \caption{\textbf{Definition and implementation of Bessel functions.} (ss stands for scipy.special)}
    \label{tab:Bessel}
    \begin{tabular}{|c|c|c|c|c|}
      \toprule 
      \textbf{Symbol} & \textbf{Definition} & \textbf{Name} & \textbf{MATLAB} & \textbf{Python}\\
      \midrule 
      $J_n(x)$ & - & Bessel 1st kind & besselj$(n,x)$ & ss.jv\\
      $Y_n(x)$ & - & Bessel 2nd kind & bessely$(n,x)$ & ss.yv\\
      \midrule 
      $H_n^{(1)}(x)$ & $J_n + \mathrm{i}Y_n$ & Hankel 1st kind & besselh$(n,x)$ & ss.hankel1\\
      $H_n^{(2)}(x)$ & $J_n - \mathrm{i}Y_n$ & Hankel 2nd kind & besselh$(n,x)$ & ss.hankel2\\
      \midrule 
      $j_n(x)$ & $\sqrt{\frac{\pi}{2x}}J_{n+\frac{1}{2}}$ & Spherical Bessel 1st kind & $\sqrt{\frac{\pi}{2x}}\mathrm{besselj}(n+\frac{1}{2},x)$ & ss.spherical\_jn\\
      $y_n(x)$ & $\sqrt{\frac{\pi}{2x}}Y_{n+\frac{1}{2}}$ & Spherical Bessel 2nd kind & $\sqrt{\frac{\pi}{2x}}\mathrm{bessely}(n+\frac{1}{2},x)$ & ss.spherical\_yn\\
      \midrule 
      \multirow{2}{*}{$h_n^{(1)}(x)$} & \multirow{2}{*}{$j_n + \mathrm{i}y_n$} & \multirow{2}{*}{Spherical Hankel 1st kind} & \multirow{2}{*}{$\sqrt{\frac{\pi}{2x}}\mathrm{besselh}(n+\frac{1}{2},1,x)$} & ss.spherical\_jn \\
      &&&& $+\mathrm{i}$ ss.spherical\_yn \\
      \multirow{2}{*}{$h_n^{(2)}(x)$} & \multirow{2}{*}{$j_n - \mathrm{i}y_n$} & \multirow{2}{*}{Spherical Hankel 2nd kind} & \multirow{2}{*}{$\sqrt{\frac{\pi}{2x}}\mathrm{besselh}(n+\frac{1}{2},2,x)$} & ss.spherical\_jn \\
      &&&& $-\mathrm{i}$ ss.spherical\_yn \\
      \midrule 
      \multirow{2}{*}{$\psi_n(x)$} & \multirow{2}{*}{$xj_n(x)$} & \multirow{2}{*}{Riccati-Bessel 1st kind} & \multirow{2}{*}{$x\sqrt{\frac{\pi}{2x}}\mathrm{besselj}(n+\frac{1}{2},x)$}  & \multirow{2}{*}{$x$ $\cdot$ ss.spherical\_jn}\\
      &&&& \\
      \multirow{2}{*}{$\xi_n(x)$} & \multirow{2}{*}{$xh_n^{(1)}(x)$} & \multirow{2}{*}{Riccati-Bessel 2nd kind} & \multirow{2}{*}{$x\sqrt{\frac{\pi}{2x}}\mathrm{besselh}(n+\frac{1}{2},1,x)$} & $x$ $\cdot$ (ss.spherical\_jn\\
      &&&& $+\mathrm{i}$ ss.spherical\_yn) \\
      \bottomrule 
    \end{tabular}
  \end{center}
\end{table}

The derivatives of the Bessel functions are used in Eqs. \eqref{eq:Z}, \eqref{eq:Gamma_n} and \eqref{eq:Delta_n}. The following derivation shows how to compute the values of the derivative without having to perform a numerical derivative explicitly:
\begin{align} 
\label{eq:Bessel_der} 
\frac{[xz_n(x)]'}{x} & = \frac{1}{x} \frac{\mathrm{d} }{\mathrm{d} x} \big( xz_n(x) \big) \nonumber\\
& = \frac{1}{x} \bigg( z_n(x) + x\frac{\mathrm{d} }{\mathrm{d} x} \big(z_n(x) \big) \bigg) \nonumber\\
& = \frac{1}{x} \bigg( z_n(x) + x \bigg( z_{n-1}(x) - (n+1)\frac{z_n(x)}{x} \bigg) \bigg) \nonumber\\
& = \frac{1}{x} \bigg( z_n(x) + x z_{n-1}(x) - n z_n(x)  - z_n(x) \bigg) \nonumber\\
& = z_{n-1}(x) - n \frac{z_n(x)}{x},
\end{align}

where $z_n$ can be substituted by either $j_n$ or $h_n^{(1)}$. In this derivation we make use of the differentiation product rule and equation 10.1.1 of \cite{abramowitz1964} for the derivative of $z_n$. This simplification is used to evaluate derivatives when computing the VSHs and the susceptibilities.

\newpage
\section{Rotation of spherical wave expansion for arbitrary dipole position and orientation}

In the previous chapter, we expressed the EM-fields emitted by the nanoparticle-dipole system as a spherical wave expansion (SWE), which is linear combination of vector spherical harmonics (VSHs). We only considered the case where the dipole is positioned on top of the nanoparticle, for only two orientations: perpendicular and parallel to the nanoparticle surface (Supplementary Fig. \ref{fig:S25_help}a).

Due to spherical symmetries, any other position and orientation can be obtained by rotating the system around the center of the nanoparticle (Supplementary Fig. \ref{fig:S2_overview}b). In this chapter we will describe how VSHs are rotated and introduce a framework of 4 angles that together describe the position and orientation of the dipole with respect to the nanoparticle and the microscope (Supplementary Fig. \ref{fig:S25_help}b,c)

\subsection{Rotating VSHs}
Vector spherical harmonics can be rotated and have the convenient characteristic that a rotated VSH, with the Euler angles ($\alpha,\beta,\gamma$), can be expressed in linear combination of VSHs with the same degree:

\begin{equation}
    \label{eq:rot_VSH}
    \mathbf{\widetilde{M}}^{(1,3)}_{nm}(kr,\theta,\phi) = \sum_{m'=-n}^n D_{mm'}^n(\alpha,\beta,\gamma) \mathbf{M}^{(1,3)}_{nm'}(kr,\theta,\phi),
\end{equation}

where $\mathbf{\widetilde{M}}$ represents the rotated VSH \cite{Doicu2006,Mishchenko2017} and $D$ the Wigner D-functions \cite{wigner1959}. This relation holds for both $\mathbf{M}_{nm}$ and $\mathbf{N}_{nm}$, and for both the regular $(1)$ and outgoing $(3)$ type. The Wigner D-functions are defined as:

\begin{equation}
    \label{eq:wignerD}
    D_{mm'}^n(\alpha,\beta,\gamma) = (-1)^{m+m'}e^{\mathrm{i}m\alpha} e^{\mathrm{i}m'\gamma} \widetilde{d}_{mm'}^n(\beta).
\end{equation}

The function $\widetilde{d}_{mm'}^n(\beta)$ is given by: 

\begin{equation}
    \label{eq:d_tilde}
    \widetilde{d}_{mm'}^n(\beta) = \Delta_{mm'} d_{mm'}^n(\beta),
\end{equation}

with $d_{mm'}^n(\beta)$ the Wigner d-function (mind the case, no capital D here) and $\Delta_{mm'}$ defined as: 

\begin{equation}
    \label{eq:Delta_mm}
    \Delta_{mm'} =
    \left \{
    \begin{array}{lll}
        1  \quad \quad & \mathrm{for}    &\quad m\geq0, \quad m'\geq0 \\
        (-1)^{m'}      & &\quad m\geq0, \quad m'<0 \\
        (-1)^{m}       & &\quad m<0, \quad m'\geq0   \\
        (-1)^{m+m'}    & &\quad m<0, \quad m'<0.
    \end{array}
    \right.
\end{equation}

The Wigner d-function can be recursively computed \cite{Mishchenko2017}: 

\begin{multline}
    \label{eq:wigner_d}
    d_{mm'}^{n+1} = \frac{1}{n\sqrt{(n + 1)^2 - m^2}  \sqrt{(n + 1)^2 - m'^2}} \bigg( (2n+1)\big(n(n+1)x - mm'\big) d_{mm'}^n(\beta) \\
    - (n+1)\sqrt{n^2 - m^2}\sqrt{n^2 - m'^2} d_{mm'}^{n-1}(\beta) \bigg ) 
    \quad ,n\geq n_{\mathrm{min}},
\end{multline}

with $n_{\mathrm{min}} = \max{(|m|,|m'|)}$ and the initial conditions: 

\begin{equation}
    \label{eq:wigner_d_init1}
    d_{mm'}^{n_{\mathrm{min}}-1} = 0, 
\end{equation}
and
\begin{equation}
    \label{eq:wigner_d_init2}
    d_{mm'}^{n_{\mathrm{min}}} = \zeta_{mm'}2^{-n_{\mathrm{min}}} \sqrt{\frac{(2n_{\mathrm{min}})!}{(|m-m'|)!(|m+m'|)!}} (1-x)^{|m-m'|/2} (1+x)^{|m+m'|/2}
    , 
\end{equation}

with $x=\cos(\beta)$.

The rotation of the VSH is described in the Euler angles ($\alpha,\beta,\gamma$) that follow the ZY'Z'' convention for intrinsic rotations, which means that the rotations sequentially occur around the axes of the coordinate system of the rotating object. This means that the first rotation is $\alpha$ rad around the $z$-axis, changing the axis of the rotating object to x'y'z' (where z' is unchanged). The second rotation is $\beta$ rad around the new y'-axis changing the axis of the rotating body to x''y''z'' (where y'=y''). Finally, the third rotation is $\gamma$ rad around the z''-axis. 

Using this definition for rotating the VSHs, we will introduce a framework of 4 angles to describe the orientation and position of the dipole emitter.

\subsection{Orientation of dipole}
For the description of the orientation of a general 3D object, one requires 3 angles. However, the orientation of the dipole emitter can be described by two, due to the rotational symmetry around the dipole moment ($\mathbf{p}$). The orientation of the dipole can be described either relative to the global coordinate system (XYZ) or with respect to the nanoparticle surface. We choose the last option, because this will allow us to make use directly of the Euler angle convention mentioned above. 

The orientation of the dipole with respect to the nanoparticle surface is described using the following two angles: 

\begin{itemize}
    \item \makebox[2cm]{$\beta_p \in [0,\pi/2]$\hfill} =  smallest angle between the dipole moment $\mathbf{p}$ and the surface normal $\mathbf{\hat{n}}$
    \item \makebox[2cm]{$\gamma_p \in [0,2\pi]$\hfill} = angle between parallel component of dipole moment $\mathbf{p_\parallel}$ and polar unit vector $\mathbf{e_\theta}$,
\end{itemize}

where $\beta_p$ is the angle between the dipole moment and the surface normal, and $\gamma_p$ a rotation around the surface normal (see Supplementary Fig. \ref{fig:S25_help}b for a visual explanation of these definitions). The subscript $p$ indicates the relation to the dipole moment, to not confuse these angles with previous notations in this document. As repeated in the introduction of this chapter, the solution of the EM-field of the nanoparticle-dipole system was derived for two dipole orientations, perpendicular ($\perp$) and parallel ($\parallel$) to the nanoparticle surface. The angle $\beta_p$ describes how the dipole is orientation with respect to the surface normal, which is essentially an orientation somewhere between the perpendicular and the parallel case. Due to the linearity of the Maxwell equations, we can calculate the SWE for any intermediate orientation by interpolation between these two solutions: 

\begin{equation}
    \mathbf{p} = \cos(\beta_p) \mathbf{p_\perp} + \sin(\beta_p) \mathbf{p_\parallel},
\end{equation}

where $\mathbf{p_\perp}$ is the dipole moment pointing into the direction of the surface normal $\mathbf{\hat{n}}$ and $\mathbf{p_\parallel}$ pointing parallel to the surface pointing into the direction of $\mathbf{e_\theta}$, which is the polar unit vector (Eq. \eqref{eq:basis_vectors}) that is oriented parallel to the surface and points along the $\theta$-direction (along the surface towards the \textit{south-pole} of the nanoparticle). The expansion coefficients of the SWE for a dipole with orientation $\beta_p$ can be calculated using the same logic:

\begin{equation}
        \mathcal{S} = \cos(\beta_p) \mathcal{S}_\perp + \sin(\beta_p) \mathcal{S}_\parallel, 
\end{equation}

where $\mathcal{S_\perp}$ and $\mathcal{S_\parallel}$ represent symbolically the spherical wave expansion coefficients for the perpendicular and parallel dipole, as derived in Section \ref{sec:mie_coeff}.

The angle $\gamma_p$ describes the rotation of the dipole around the surface normal vector, which is exactly the third Euler angle in Eq. \eqref{eq:rot_VSH} for rotating VSHs. This angle is defined as the counter clock-wise rotation from $\mathbf{e_\theta}$ to the parallel component of the dipole moment (seen from the top of $\mathbf{\hat{n}}$).

In conclusion, the orientation of the dipole with respect to the nanoparticle normal is described by two angles, where $\beta_p$ is used to interpolate the expansion coefficients between a perpendicular and parallel dipole, and $\gamma_p$ as the third Euler angle in the rotation of the VSHs ($\gamma$ in Eq. \eqref{eq:rot_VSH}).

\subsection{Position of dipole}
The position of the dipole around the nanoparticle is described by three parameters: 

\begin{itemize}
    \item \makebox[2cm]{$R_p \in [0,\infty]$\hfill} = distance from dipole to nanoparticle center
    \item \makebox[2cm]{$\theta_p \in [0,\pi]$\hfill} = polar position of dipole on nanoparticle 
    \item \makebox[2cm]{$\phi_p \in [0,2\pi]$\hfill} = azimuthal position of dipole on nanoparticle,
\end{itemize}

where the subscript $p$ again indicates the relation to the dipole, to not confuse these angles with previous notations in this document. See Supplementary Fig. \ref{fig:S25_help}c for a visual explanation of these definitions.

The radial distance of the dipole to the nanoparticle center is defined as $R_p = a+d$, with $a$ the radius of the nanoparticle and $d$ the radial distance between the dipole and the nanoparticle surface. The two angles that together describe the position of the dipole on the nanoparticle surface, $\theta_p$ and $\phi_p$, are defined in the same way as in Section \ref{sec:coord_system}. 

When we compare these angles to the rotation of the VSH as described before, $\theta_p$ is equal to the second Euler angle ($\beta$ in Eq. \eqref{eq:rot_VSH}) and $\phi_p$ to the first Euler angle ($\gamma$ in Eq. \eqref{eq:rot_VSH}). 

In conclusion, using $\theta_p$, $\phi_p$, $\gamma_p$ and $R_p$, we can fully define the position/orientation of the dipole relative to the particle by interpolating between the solutions for $\mathbf{p_\perp}$ and $\mathbf{p_\parallel}$, and rotating the solution around the nanoparticle center (Supplementary Fig. \ref{fig:S25_help}b,c).
\newpage
\section{Plane wave decomposition}
\label{chap:PWD}

In the previous two chapters, we have derived expressions for the EM-fields around the nanoparticle-dipole system, for any arbitrary position and orientation of the dipole with respect to the nanoparticle. Until now, we decomposed the EM-fields into VSHs, which can be described as a spherical wave expansion (SWE). This was the obvious choice in Mie scattering theory, because the presence of spherical symmetries in the system heavily simplifies the mathematics of the derivations. The next step of the pipeline is the refraction of the EM-waves in the water-glass interface. Since refraction in a planar interface is only defined for plane waves, we choose to decompose every VSH of the SWE into plane waves using plane wave decomposition (PWD), see Supplementary Fig. \ref{fig:S2_overview}c. In this chapter we will cover the process of PWD and the necessary definitions and derivations.

\subsection{Adjusting coefficients to other convention}
The definition of the VSHs and the expansion coefficients in the previous chapters follows the convention of Bohren-Huffman and Le Ru \cite{Bohren1983,LeRu2008}. The PWD, however, is very clearly described by Amos Egel \cite{Egel2019,egel2021smuthi}, which follows the convention of Doicu-Wriedt-Eremin \cite{Doicu2006}. Even though the conventions are very similar to each other, there are is some minor differences. We therefore have one \textit{conversion rule} to convert the expansion coefficients we have defined before: all expansion coefficients regarding $m=1$ need to be multiplied by $-1$. Note that there is also a factor of $\sqrt{\pi}$ difference between the definitions of Bohren-Huffman and Doicu-Wriedt-Eremin (due to a difference in normalization of the Legendre functions), but we will not consider that for now, since we are not interested in the absolute value of the EM-fields. 

Table \ref{tab:coeffs} shows the bookkeeping of the expansion coefficients throughout the pipeline. In the Mie scattering theory (first row), the full solution requires 10 coefficients. In the implementation, we only need 6 coefficients (second row), since we know that the coefficients for $m=-1$ are the same as for $m=1$, except $d_{n,1}=-d_{n,-1}$ and $f_{n,1}=-f_{n,-1}$ (compare Eqs. \eqref{eq:bn0},\eqref{eq:fn0},\eqref{eq:an1},\eqref{eq:bn1},\eqref{eq:en1} and \eqref{eq:fn1}). The conversion of the expansion coefficients according the convention of Doicu-Wriedt-Eremin \cite{Doicu2006} requires that all coefficients for $m=1$ need an extra factor of $-1$ (third row).

\begin{table}[h]
\begin{center}
\caption{\textbf{Conversion and implementation of expansion coefficients.}}
\label{tab:coeffs}
\begin{tabular}{lrccccc}
\toprule
&& perpendicular ($\perp$) & \multicolumn{4}{c}{parallel ($\parallel$)} \\
\cmidrule(lr){3-3}\cmidrule(lr){4-7}
           &   & $\mathbf{N}_{n0} $ & \multicolumn{2}{c}{$\mathbf{M}_{n,\pm1}$}  & \multicolumn{2}{c}{$\mathbf{N}_{n,\pm1}$}  \\
           &   $m$ & $0$ & $1$ & $-1$ & $1$ & $-1$  \\
\midrule
    \multirow{2}{*}{Theory}  &scat & $d_{n0}$ & $c_{n,1}$ & $c_{n,-1}$ & $d_{n,1}$ & $d_{n,-1}$ \\
                             &init & $f_{n0}$ & $e_{n,1}$ & $e_{n,-1}$ & $f_{n,1}$ & $f_{n,-1}$\\
\midrule
    \multirow{2}{*}{Implem.} &scat & $d_{n0}$ & $c_{n,1}$ & $c_{n,1}$ & $d_{n,1}$ & $-d_{n,1}$\\
                             &init & $f_{n0}$ & $e_{n,1}$ & $e_{n,1}$ & $f_{n,1}$ & $-f_{n,1}$\\
\midrule
    \multirow{2}{*}{Doicu}   &scat & $d_{n0}$ & $-c_{n,1}$ & $c_{n,1}$ & $-d_{n,1}$ & $-d_{n,1}$\\
                             &init & $f_{n0}$ & $-e_{n,1}$ & $e_{n,1}$ & $-f_{n,1}$ & $-f_{n,1}$\\
\bottomrule
\end{tabular}
\end{center}
\end{table}

Furthermore, we will simplify the expressions of the SWE, to make the equations in this chapter more compact. In stead of writing the EM-fields in terms of $\mathbf{M}_{nm}^{(3)}$ and $\mathbf{N}_{nm}^{(3)}$:

\begin{align}
    \label{eq:swe_full}
    \mathbf{E_\mathrm{scat}}(\mathbf{r}) & = E_0 \sum_{n,m} c_{nm}\mathbf{M}_{nm}^{(3)}(k,\mathbf{r}) + d_{nm}\mathbf{N}_{nm}^{(3)}(k,\mathbf{r}),\\
    \mathbf{E_\mathrm{inc}}(\mathbf{r}) & = E_0 \sum_{n,m} e_{nm}\mathbf{M}_{nm}^{(3)}(k,\mathbf{r}) + f_{nm}\mathbf{N}_{nm}^{(3)}(k,\mathbf{r}),
\end{align}

we will combine them into one expression: 

\begin{equation}
    \label{eq:swe_compact}
    \mathbf{E_\mathrm{total}}(\mathbf{r}) = E_0 \sum_{\tau,n,m} \mathcal{S}_{\tau nm} \mathbf{\Psi}_{\tau nm}^{(3)}(k,\mathbf{r}),
\end{equation}

where $\mathbf{M}_{nm}^{(3)}$ and $\mathbf{N}_{nm}^{(3)}$ are combined into $\mathbf{\Psi}_{\tau nm}^{(3)}$, with the new parameter $\tau$ representing either $\mathbf{M}_{nm}^{(3)}$ $(\tau=0)$ or $\mathbf{N}_{nm}^{(3)}$ $(\tau=1)$. All expansion coefficients ($c_{nm}$, $d_{nm}$, $e_{nm}$ and $f_{nm}$) are aggregated into $\mathcal{S}_{\tau nm}$ and, for convenience, the scattered and incident field are added up into the total field.

\subsection{Plane wave expansion}
Just like for the complete set of VSHs, any EM-field can also be decomposed into plane waves, giving a plane wave expansion. A single plane wave is defined as: 

\begin{equation}
    \label{eq:pw}
    \Phi_j^\pm(\kappa,\alpha|\mathbf{r}) = e^{\mathrm{i}\mathbf{k}^\pm \cdot \mathbf{r}} \hat{\mathbf{e}}_j^\pm,
\end{equation}

where the subscript $j$ indicates the polarization, which is either in the azimuthal direction ($j=1$) or the polar direction ($j=2$). Accordingly, $\hat{\mathbf{e}}_j^\pm$ is the unit vector for both polarizations. The superscript $\pm$ indicates whether the plane wave is travelling in the positive ($+$) or negative ($-$) $z$-direction. The position vector $\mathbf{r}$, the position where the plane wave is evaluated, is relative to the reference point of the plane wave expansion. This reference point is chosen to be at the origin $(0,0,0)$, where the optical axis intersects with the water-glass interface. $\mathbf{k}^\pm$ is the wave vector in cylindrical coordinates $(\kappa,\alpha,\pm k_z)$, with $\kappa$ the length of $\mathbf{k}$ projected onto the $xy$-plane, $\alpha$ the angle between this projection and the positive $x$-axis, and $k_z$ the projection of $\mathbf{k}$ onto the $z$-axis (see Supplementary Fig. \ref{fig:S25_help}d). The value of $k_z$ is implicitly defined by  $k=|\mathbf{k}|$ and $k_z$:

\begin{equation}
    \label{eq:kz}
    k_z = \sqrt{k^2 - \kappa^2}.
\end{equation}

Due to the definition of the square root, $k_z$ can be either positive of negative, for which we use the following convention: 

\begin{equation}
\label{eq:kz_sign}
\begin{cases}
    \mathrm{Re}\{k_z\} \geq 0 \quad \mathrm{if} \; \mathrm{Im}\{k_z\}=0\\
    \mathrm{Im}\{k_z\} > 0 \quad \mathrm{else}. \\
\end{cases}    
\end{equation}

The unit vectors are defined as: 

\begin{equation}
    \label{eq:e1}
    \hat{\mathbf{e}}_1 = 
    \begin{pmatrix} x\\y\\ z
    \end{pmatrix}
    =\begin{pmatrix}
        -\sin(\alpha)\\
        \cos(\alpha)\\
        0
    \end{pmatrix},
\end{equation}
\begin{equation}
    \label{eq:e2}
    \hat{\mathbf{e}}_2 = 
    \begin{pmatrix} x\\y\\ z
    \end{pmatrix}
    =\begin{pmatrix}
        \cos(\alpha) \frac{k_z}{k}\\
        \sin(\alpha) \frac{k_z}{k}\\
        -\frac{\kappa}{k}       
    \end{pmatrix},
\end{equation}
         
which are the same for upward and downward plane waves, except $k_z$ will have either a positive or negative value, respectively. Since the plane waves form a complete set, any electric field can be decomposed into a set of upward/downward propagating plane waves, resulting in a plane wave expansion: 

\begin{equation}
    \label{eq:PWE}
    \mathbf{E}(\mathbf{r}) = \sum_{j=1}^2 \int_{\mathbb{R}^2} \big( g_j^+(\kappa,\alpha)\Phi_j^+ + g_j^-(\kappa,\alpha)\Phi_j^-\big) \mathrm{d}^2\mathbf{k}_{||},
\end{equation}

which is a combination of a sum over both polarizations and an integral in polar coordinates over all plane waves using the area element $\mathrm{d}^2\mathbf{k}_{||} = \kappa \mathrm{d}\kappa \mathrm{d}\alpha$. The coefficients $g_j^+$ and $g_j^-$ are the plane wave expansion coefficients for upward/downward propagating plane waves, respectively, in the direction defined by $\kappa$, $\alpha$ and $k$.

\subsection{SWE to PWE}
In order to translate our derived SWE into a PWE, we will decompose every VSH into plane waves. The following expression gives the decomposition of a particular outgoing VSH into plane waves \cite{Egel2019,egel2021smuthi,Bostrom1991}:

\begin{equation}
    \Psi_{\tau nm}^{(3)}(k,\mathbf{r}) = \frac{1}{2\pi} \sum_{j=1}^2 \int_{\mathbb{R}^2} \bigg( \frac{1}{k_zk} e^{\mathrm{i}m\alpha} B_{\tau nmj}(\pm k_z/k)\Phi_j^\pm(\kappa,\alpha|\mathrm{r}) \bigg )\mathrm{d}^2\mathbf{k}_{||} \quad \mathrm{for} \; z\gtrless0,
\end{equation}

with:
\begin{equation}
    B_{\tau nmj}(x) = \frac{-1}{\mathrm{i}^{n+1}}\frac{1}{\sqrt{2n(n+1)}} (\mathrm{i}\delta_{j1}+\delta_{j2})\bigg( \delta_{\tau j} \tilde{\tau}_n^{|m|}(x) + (1- \delta_{\tau j})m\pi_n^{|m|}(x)
    \bigg).
\end{equation}

Note the two different tau-parameters, where subscript $\tau$ represents the type of VSH and $\tilde{\tau}_n^{|m|}(x)$ the recurrently-determined Legendre polynomial that we used before (Eq. \eqref{eq:tau_def}). The reason for the tilde is that, compared to equation \eqref{eq:tau_def}, $\tilde{\tau}$ contains an extra factor $\sqrt{\frac{2n+1}{n(n+1)}}$. Filling the above plane wave decomposition into the SWE, we get: 

\begin{align}
    \label{eq:pwe_termsOfP}
    \mathbf{E_\mathrm{total}}(\mathbf{r}) & = E_0 \sum_{\tau,n,m} \mathcal{S}_{\tau nm} 
    \bigg[
    \frac{1}{2\pi} \sum_{j=1}^2 \int_{\mathbb{R}^2} \bigg( \frac{1}{k_zk} e^{\mathrm{i}m\alpha} B_{\tau nmj}(\pm k_z/k)\Phi_j^\pm(\kappa,\alpha|\mathrm{r}) \bigg )
    \mathrm{d}^2\mathbf{k}_{||}
    \bigg] \nonumber \\
    & = E_0 \int_{\mathbb{R}^2} \bigg( \sum_{\tau,n,m} \sum_{j=1}^2 \frac{1}{2\pi k_zk} e^{\mathrm{i}m\alpha} \mathcal{S}_{\tau nm} B_{\tau nmj}(\pm k_z/k)\Phi_j^\pm(\kappa,\alpha|\mathrm{r}) \bigg ) \mathrm{d}^2\mathbf{k}_{||} \nonumber \\
    & = E_0 \int_{\mathbb{R}^2} \bigg( \sum_{j=1}^2 \mathcal{P}_j^\pm(\kappa,\alpha) \Phi_j^\pm(\kappa,\alpha|\mathrm{r}) \bigg ) \mathrm{d}^2\mathbf{k}_{||},
\end{align}


with: 

\begin{equation}
    \label{eq:P_coeffs_def}
    \mathcal{P}_j^\pm(\kappa,\alpha) = \sum_{\tau,n,m} \frac{1}{2\pi k_zk} e^{\mathrm{i}m\alpha} \mathcal{S}_{\tau nm} B_{\tau nmj}(\pm k_z/k)
\end{equation}

For reasons that will become clear in a future chapter, we are only interested in particular plane waves and their respective coefficient, which is exactly the factor $\mathcal{P}^\pm_j(\kappa,\alpha)$ we obtained above. This factor $\mathcal{P}$ represents how strong every plane wave is present in the E-field emitted by the combined nanoparticle-dipole system.

\newpage
\section{Refraction in water-glass interface}
In the previous chapter we have derived the EM-fields in the water, as a plane wave expansion (PWE). The next step it to refract every plane wave in the water-glass interface (Supplementary Fig. \ref{fig:S2_overview}d). This process of refraction is well-defined for plane waves, where the direction of the refracted wave is given by Snell’s law and the polarization-dependent amplitude by Fresnel coefficients. In this chapter we will derive the expression for the PWE coefficients after refraction in the water-glass interface. 

\subsection{Fresnel coefficients}
When a plane wave is refracted in a planar surface, the intensities of the transmitted and reflected waves are given by the polarization-dependent Fresnel coefficients: 

\begin{equation}
    \label{eq:Fresnel}
    \begin{cases}
    r_{1}(\kappa) = \frac{k_{z,0}-k_{z,1}}{k_{z,0}+k_{z,1}} \\[5pt]
    r_{2}(\kappa) = \frac{n_1^2 k_{z,0} - n_0^2 k_{z,1}}{n_1^2 k_{z,0}+ n_0^2 k_{z,1}} \\[5pt]
    t_{1}(\kappa) = \frac{2k_{z,0}}{k_{z,0}+k_{z,1}} \\[5pt]
    t_{2}(\kappa) = \frac{2n_0n_1 k_{z,0}}{n_1^2 k_{z,0}+ n_0^2 k_{z,1}},\\
    \end{cases} 
    \Rightarrow \quad
    \begin{cases}
    r_{1} \\[5pt]
    r_{2}\\[5pt]
    t_{1} = 1+r_1\\[5pt]
    t_{2} = \frac{n_0}{n_1}(1+r_2),
    \end{cases}
\end{equation}

where $r_j$ and $t_j$ are the Fresnel reflection and transmission coefficients for either S- ($j=1$) or P-polarization ($j=2$), $n_i$ the refractive index of medium $i$ and $k_{z,i}$ the $\kappa$-dependent $z$-component of the wave vector in medium $i$ (Eq. \eqref{eq:kz}). The reflection coefficient $r$ describes how much of the incident wave is reflected back into the first medium ($i=0$), where the transmission coefficient $t$ describes how much of the incident wave is transmitted into the second medium ($i=1$). The reflection and transmission coefficients are all defined from medium 0 to medium 1, which means that the incident wave approaches the interface from medium $0$. For a full description of the plane waves on both sides of the interface, we also need the Fresnel coefficients in the other direction ($\tilde{r}$ and $\tilde{t}$), going from medium 1 to medium 0. These \textit{inverted} coefficients can be expressed in the \textit{normal} coefficients (Eq. \eqref{eq:Fresnel}) as follows:

\begin{equation}
    \label{eq:Fresnel_inv}
    \begin{cases}
    \tilde{r}_{1} = -r_1\\[5pt]
    \tilde{r}_{2} = -r_2\\[5pt]
    \tilde{t}_{1} = 1-r_1 = (1-r_1)\frac{1+r_1}{1+r_1} = \frac{1-r_1^2}{t_1}\\[5pt]
    \tilde{t}_{2} = \frac{n_1}{n_0}(1-r_2) = \frac{1-r_1^2}{t_1}
    \end{cases}
    \Rightarrow \quad
    \begin{cases}
    \\[5pt]
    \tilde{r}_{j} = -r_j \\[5pt]
    \tilde{t}_{j} = \frac{1-r_j^2}{t_j}.  \\[5pt]
    \\
    \end{cases}
\end{equation}

\subsection{Scattering matrix formalism}
The scattering matrix formalism describes how plane wave coefficients in two adjacent media (separated by a planar surface) are linked \cite{Egel2019,egel2021smuthi,ko1988}: 

\begin{equation}
    \label{eq:scat_mat}
    \begin{bmatrix}
    g_{i+1,j}^+(\kappa,\alpha) \\ g_{i,j}^-(\kappa,\alpha)
    \end{bmatrix}
    = S_j^{i,i+1}(\kappa)
    \begin{bmatrix}
    g_{i,j}^+(\kappa,\alpha) \\ g_{i+1,j}^-(\kappa,\alpha)
    \end{bmatrix},
\end{equation}

where $g_{i,j}^\pm$ are the plane wave coefficients for upward($+$)/downward($-$) propagating plane waves in medium $i$ with polarization $j$. The 2x2 scattering matrix $S_j^{i,i+1}(\kappa)$ links the plane waves that are approaching the interface ($g_{i,j}^+$ and $g_{i+1,j}^-$) to the plane waves propagating away from the interface ($g_{i,j}^-$ and $g_{i+1,j}^+$). Here we will derive the expression for the scattering matrix. 

When we consider a planar interface separating two media, there are four sets of PWE coefficients at play, upward and downward propagating waves both in medium $i$ and in medium $i+1$: $g_i^+$, $g_i^-$, $g_{i+1}^+$ and $g_{i+1}^-$. Using the Fresnel coefficients, we can define the relation between both types of plane wave in medium $i$ with both types in medium $i+1$. 

\begin{equation}
    \label{eq:scat_mat_def}
    \begin{cases}
    g_1^+ = t \cdot g_0^+ \\[5pt]
    g_0^- = r \cdot g_0^+ \\[5pt]
    g_1^+ = \tilde{r} \cdot g_1^- \\[5pt]
    g_0^- = \tilde{t} \cdot g_1^- \\
    \end{cases}
    \Rightarrow \quad
    \begin{cases}
    g_1^+ = t \cdot g_0^+ \\[5pt]
    g_0^- = r \cdot g_0^+ \\[5pt]
    g_1^+ = -r \cdot g_1^- \\[5pt]
    g_0^- = \frac{1-r^2}{t} \cdot g_1^- \\
    \end{cases}   
    \Rightarrow \quad
     S^{i,i+1} =
    \begin{bmatrix}
        t & -r   \\
        r & \frac{(1-r)^2}{t} \\
    \end{bmatrix},
\end{equation}

where Eq. \eqref{eq:Fresnel_inv} is used in the first step for the inverted Fresnel coefficients and $S^{i,i+1}$ is found in the second step by comparing the set of 4 equations to the scattering matrix formalism in Eq. \eqref{eq:scat_mat}. For clarity, $i=0$ is used and the polarization subscript $j$ and $\kappa$-dependence are left out.

\subsection{Transfer matrix formalism}
The obtained scattering matrix $S^{i,i+1}$ above links the plane waves that are approaching the interface ($g_i^+$ and $g_{i+1}^-$) to the plane waves propagating away from the interface ($g_i^-$ and $g_{i+1}^+$). It is more convenient to relate the plane waves in one medium to those in the other medium. The transfer matrix formalism does this \cite{Egel2019,egel2021smuthi,Dyakov2009}: 

\begin{equation}
    \label{eq:trans_mat}
    \begin{bmatrix}
    g_{i,j}^+(\kappa,\alpha) \\ g_{i,j}^-(\kappa,\alpha)
    \end{bmatrix}
    = I_j^{i,i+1}(\kappa)
    \begin{bmatrix}
    g_{i+1,j}^+(\kappa,\alpha) \\ g_{i+1,j}^-(\kappa,\alpha)
    \end{bmatrix}, 
\end{equation}

where $I_j^{i,i+1}(\kappa)$ is the 2x2 transfer matrix. Note the similarity with Eq. \eqref{eq:scat_mat}, with the only difference that $g_{i,j}^+$ and $g_{i+1,j}^+$ have been swapped. The expression of the transfer matrix $I^{i,i+1}_j(\kappa)$ can be derived using the definition of the scattering matrix $S^{i,i+1}_j(\kappa)$:

\begin{align}
    \label{eq:derive_I}
    &
    \begin{bmatrix}
        g_1^+  \\[5pt]
        g_0^-  \\
    \end{bmatrix}
    =   S^{0,1}
     \begin{bmatrix}
        g_0^+  \\[5pt]
        g_1^-  \\
    \end{bmatrix}
    \quad \Rightarrow \quad
    \begin{cases}
    g_1^+ = t \cdot g_0^+ - r \cdot g_1^- \\[5pt]
    g_0^- = r \cdot g_0^+ + \frac{1-r^2}{t} \cdot g_1^- \\
    \end{cases}
    \Rightarrow \quad
    \begin{cases}
    g_0^+ = \frac{1}{t} \cdot g_1^+ + \frac{r}{t} \cdot g_1^- \\[5pt]
    g_0^- = r \cdot g_0^+ + \frac{1-r^2}{t} \cdot g_1^- \\
    \end{cases} 
    \nonumber\\[10pt]
    &\Rightarrow \quad
    \begin{cases}
    g_0^+ = \frac{1}{t} \cdot g_1^+ + \frac{r}{t} \cdot g_1^- \\[5pt]
    g_0^- = \frac{r}{t} \cdot g_1^+ + \frac{1}{t} \cdot g_1^- \\
    \end{cases} 
    \Rightarrow \quad
    \begin{bmatrix}
        g_0^+  \\[5pt]
        g_0^-  \\
    \end{bmatrix}
    =\frac{1}{t}
    \begin{bmatrix}
        1&r \\[5pt]
        r&1 \\
    \end{bmatrix}
     \begin{bmatrix}
        g_1^+  \\[5pt]
        g_1^-  \\
    \end{bmatrix}
    \quad \Rightarrow \quad
    I^{0,1} = \frac{1}{t}
    \begin{bmatrix}
        1 & r   \\[5pt]
        r & 1  \\
    \end{bmatrix},
\end{align}

where we used $i=0$ for simplicity, but the found transfer matrix holds for any $i$. In the 5-step derivation of the transfer matrix above, we subsequently: use the definition of the scattering matrix (Eq. \eqref{eq:scat_mat_def}), rewrite the first equation in terms of $g_0^+$, substitute the first equation for $g_0^+$ into the second equation, convert the system of equation to matrix notation, and lastly, find the expression for the transfer matrix $I^{0,1}$ by comparing the matrix notation to Eq. \eqref{eq:trans_mat}. For clarity, the polarization subscript $j$ and $\kappa$-dependence of all variables are left out. Finally, following \cite{Dyakov2009}, we derive the expression for the transfer matrix: 

\begin{equation}
    \label{eq:trans_mat_def}
    I_j^{i,i+1}(\kappa) = \frac{1}{t_j(\kappa)}
    \begin{bmatrix}
    1 & r_j(\kappa)   \\[5pt]
    r_j(\kappa) & 1  \\
    \end{bmatrix}.
\end{equation}

\subsection{Downward propagating waves in our setup}
After properly formulating the process of refraction and deriving the transfer matrix, we can now use these tools to answer the question we started with: what are the PWE coefficients in the glass medium? In our system we have two media, water and glass. Since the water medium is on top of the glass medium, glass is the bottom medium ($i=0$) and water the top one ($i=1$). The only \textit{source} of light is the nanoparticle-dipole system, which is located at a certain height in the water, and the only interface is the water-glass interface in between the two media. Going further, we will consider the PWEs in both media one by one (Supplementary Fig. \ref{fig:S25_help}e). 

The water medium ($i=1$) contains the only \textit{source} of light, the nanoparticle-dipole system. This means that all EM-field in the water originate from the nanoparticle-dipole system or are responses of the interface. There will be no contributions originating from other media, since they have no sources of light. For convenience, we split up the PWE in the water in an source ($\mathcal{P}_{\mathrm{source}}^\pm$) and a response ($\mathcal{P}^\pm_{\mathrm{res}}$) component. The source component ($\mathcal{P}_{\mathrm{source}}^\pm$) is the light originating from the nanoparticle-dipole system, for which we derived the expression as a PWE in chapter \ref{chap:PWD}. The source PWE consists of upward and downward propagating plane waves and has the center of the nanoparticle as reference point. This means that the upwards propagating plane waves ($\mathcal{P}_{\mathrm{source}}^+$) are only present above the center of the nanoparticle ($z$-coordinates above the reference point of the PWE), and the downward propagating plane waves ($\mathcal{P}_{\mathrm{source}}^-$) only below the center of the nanoparticle. 

The response component ($\mathcal{P}_{\mathrm{res}}^\pm$) is the response of the source light to the interface. Since the interface is located below the nanoparticle, and can therefore only reflect $\mathcal{P}_{\mathrm{source}}^-$, the only existing response component is $\mathcal{P}_{\mathrm{res}}^+$, which is defined as: 

\begin{equation}
    \label{eq:Pres_plus}
    \mathcal{P}_{\mathrm{res}}^+ = \tilde{r} \cdot \mathcal{P}_{\mathrm{source}}^- = -r \cdot \mathcal{P}_{\mathrm{source}}^-,
\end{equation}

where we have to use the inverted reflection coefficient $\tilde{r}$ since the reflection happens in medium $i=1$ (water) and not $i=0$ (glass). Knowing all the components that are present in the water at the interface ($\mathcal{P}_{\mathrm{source}}^-$ and $\mathcal{P}_{\mathrm{res}}^+$), we can write down the process of refraction using the transfer matrix formalism (Eq. \eqref{eq:trans_mat}) to find the plane waves in the glass (Supplementary Fig. \ref{fig:S25_help}e):

\begin{align}
    \begin{bmatrix}
        \mathcal{P}_{\mathrm{glass}}^+  \\[5pt]
        \mathcal{P}_{\mathrm{glass}}^-  \\
    \end{bmatrix}
    &= I^{0,1}
    \begin{bmatrix}
        \mathcal{P}_{\mathrm{res}}^+  \\[5pt]
        \mathcal{P}_{\mathrm{source}}^-  \\
    \end{bmatrix} \nonumber \\
    & = I^{0,1}
    \begin{bmatrix}
        -r \cdot \mathcal{P}_{\mathrm{source}}^-  \\[5pt]
        \mathcal{P}_{\mathrm{source}}^-  \nonumber \\
    \end{bmatrix}\\
    & = I^{0,1}
    \begin{bmatrix}
        -r \\[5pt]
        1  \\
    \end{bmatrix}
    \mathcal{P}_{\mathrm{source}}^-    \nonumber\\
    & =
    \frac{1}{t}
    \begin{bmatrix}
        1 & r   \\[5pt]
        r & 1  \\
    \end{bmatrix}
    \begin{bmatrix}
        -r \\[5pt]
        1  \\
    \end{bmatrix}
    \mathcal{P}_{\mathrm{source}}^-   \nonumber\\
    & =
    \begin{bmatrix}
        0  \\[5pt]
        \frac{1-r^2}{t}  \\
    \end{bmatrix}
    \mathcal{P}_{\mathrm{source}}^-.
\end{align}

In the 4-step derivation we subsequently: use the definition for $\mathcal{P}_{\mathrm{res}}^+$ (Eq. \eqref{eq:Pres_plus}), pull $\mathcal{P}_{\mathrm{source}}^-$ outside of the vector, use the definition of $I^{0,1}$ (Eq. \eqref{eq:trans_mat_def}), and finally, perform the matrix-vector multiplication. As final result we obtain: 

\begin{equation}
    \label{eq:Pminus_final}
    \begin{cases}
        \mathcal{P}_{\mathrm{glass},j}^+(\kappa,\alpha)  &= 0 \\[5pt]
        \mathcal{P}_{\mathrm{glass},j}^-(\kappa,\alpha)  &= \frac{1-r_j^2(\kappa)}{t_j(\kappa)} \mathcal{P}_{\mathrm{source},j}^-(\kappa,\alpha)
    \end{cases},
\end{equation}

from which we can see that there are only downward propagating plane waves in the glass, with PWE coefficients proportional to the coefficients of the downward source field in the water ($\mathcal{P}_{\mathrm{source},j}^-(\kappa,\alpha)$ in Eq. \eqref{eq:Pminus_final} is the $\mathcal{P}^-_j(\kappa,\alpha)$ in Eq. \eqref{eq:P_coeffs_def}).

Note that, in this chapter, we only employed the Fresnel reflection and transmission coefficients to describe refraction and never considered Snell's law for the change in direction of the plane waves. This can be explained by the fact that all variables are a function of $\kappa$, which is not changed by refraction. Since both media have a different wave number $k_i$, the $z$-component of the wave vector $k_{z,i}$ differs in both media, from which one can derive Snell's law. This means that Snell's is not explicitly used, since it follows from the definition of $k_{z,i}$.

\newpage
\section{Far-field in the glass}

In the previous chapter we have derived the EM-fields in the glass medium as a plane wave expansion (PWE), which is a collection of plane waves with corresponding coefficients. This PWE allows one to evaluate the EM-fields at any point in space by integrating over all plane waves (Eq. \eqref{eq:pwe_termsOfP}).

After the light has propagated into the glass medium, the microscope system will image the EM-fields originating from the focal plane onto the camera plane (Supplementary Fig. \ref{fig:S2_overview}e). Mathematically, this process can be explained as that the the near-field information (all EM-fields in the focal plane, both plane waves and evanescent waves) is propagated into the far-field, from which they are focused onto the image plane by optics. The far-field is among microscopists better known as the \textit{back-focal plane}, or in the framework of Fourier optics as the \textit{angular information}. In this chapter we will discuss how the far-field can be determined from the PWE.

\subsection{Stationary phase approximation}
Before going into the derivation of the far-field, we first have to discuss the stationary phase approximation (SPA), which is a method that will allow us to simplify the far-field expression \cite{Novotny2006,Chew1988}.

When we have an Weyl-type integral of the following form: 

\begin{equation} 
    \label{eq:Weyl}
    I = \int_{\mathbb{R}^2} g(k_x,k_y)e^{\mathrm{i}(k_xx + k_yy + k_z|z|)} dk_xdk_y,
\end{equation}

with $k_z = \sqrt{k^2-k_x^2-k_y^2}$, the problem is that for large ($x,y,z$) the phase term oscillates rapidly. The trick is to split the integrand in two parts, one part that varies slowly and a part that varies rapidly with ($k_x,k_y$): 

\begin{equation} 
    I = \int_{\mathbb{R}^2} \bigg[ g(k_x,k_y)k_z \bigg] \bigg[ \frac{e^{\mathrm{i}(k_xx + k_yy + k_z|z|)}}{k_z} \bigg] dk_xdk_y,
\end{equation}

where $k_z$ is added to both terms for convenience later. The strength of the SPA lies in finding the stationary phase points ($k_{x0},k_{y0}$) where the phase of the rapidly oscillating part is constant. The stationary phase points are values of ($k_x,k_y$) for which holds:  

\begin{equation} 
    \frac{\partial }{\partial k_x} (k_xx + k_yy + k_z|z|)=0  \quad \mathrm{and} \quad \frac{\partial }{\partial k_y} (k_xx + k_yy + k_z|z|)=0,
\end{equation}

which can be shown to result in: 

\begin{equation} 
    k_{x0} = k\frac{x}{r}   \quad \mathrm{and} \quad k_{y0} = k\frac{y}{r},
\end{equation}

with $r = \sqrt{x^2+y^2+z^2}$. Since the phase is constant for the stationary phase points, we can remove the slowly varying part from the integral by evaluating it at the stationary phase points: 

\begin{equation}  
    \label{eq:SPA_intermediate1}
    I = g(k_{x0},k_{y0})k_{z0} \int_{\mathbb{R}^2}  \bigg[ \frac{e^{\mathrm{i}(k_xx + k_yy + k_z|z|)}}{k_z} \bigg] dk_xdk_y.
\end{equation}

Using the closed-form Weyl Identity \cite{Chew1988}: 

\begin{equation}  
    \label{eq:Weyl_id}
    \frac{e^{\mathrm{i}kr}}{r} = \frac{\mathrm{i}}{2\pi}  \int_{\mathbb{R}^2} e^{\mathrm{i}(k_xx + k_yy + k_z|z|)} dk_xdk_y,
\end{equation}

we see that Eq. \eqref{eq:Weyl} simplifies to: 
\begin{align} 
    \label{eq:SPA}
    I & = \int_{\mathbb{R}^2} g(k_x,k_y)e^{\mathrm{i}(k_xx + k_yy + k_z|z|)} dk_xdk_y \nonumber\\
    & = g(k_{x0},k_{y0})k_{z0} \frac{2\pi e^{\mathrm{i}kr}}{r\mathrm{i}} \nonumber\\
    & = -2\pi \mathrm{i} \; \frac{e^{\mathrm{i}kr}}{r} \; g(k_{x0},k_{y0})k_{z0},
\end{align}

which is valid for values of ($x,y,z$) far away from the focal plane.

\subsection{The far-field}
The far-field is typically determined by evaluating the EM-fields on a reference hemisphere very far away from the focal plane, centered around the focus. Every location on the hemisphere represents an angular direction, which essentially is identical to performing a Fourier transform of the fields in the focal plane. To evaluate the EM-fields on a particular location on the distant hemisphere, we have to consider the contribution of every plane wave, which is done by integrating over all plane waves. Since the far-field is far away in the $z$-direction, all evanescent waves have decayed in amplitude, and we only have to consider the plane wave contributions. Starting from Eq. \eqref{eq:pwe_termsOfP} and only considering downward propagating plane waves, we have:

\begin{align}
    \mathbf{E_\mathrm{total}}(\mathbf{r}) 
    & = E_0  \int_{\mathbb{R}^2} \bigg(\sum_{j=1}^2  \mathcal{P}_j^-(\kappa,\alpha) \Phi_j^-(\kappa,\alpha|\mathrm{r}) \bigg ) \mathrm{d}^2\mathbf{k}_{||} \nonumber\\
    & = E_0  \int_{\mathbb{R}^2} \bigg(\sum_{j=1}^2 \mathcal{P}_j^-(\kappa,\alpha) 
    \hat{\mathbf{e}}_j^- e^{\mathrm{i}\mathbf{k}^- \cdot \mathbf{r}}
    \bigg ) \mathrm{d}^2\mathbf{k}_{||}     \nonumber\\
    & = E_0  \int_{\mathbb{R}^2}\bigg( \sum_{j=1}^2 \mathcal{P}_j^-(\kappa,\alpha)  
    \hat{\mathbf{e}}_j^- e^{\mathrm{i} (k_xx + k_yy + k_z|z|)}
    \bigg ) \mathrm{d}^2\mathbf{k}_{||},
\end{align}

where we use the definition of the plane waves (Eq. \eqref{eq:pw}), with $k_x$ and $k_y$ as the in-plane components of the wave vector, over which we integrate, and $k_z$ defined as in Eq. \eqref{eq:kz}. Comparing this expression to the derived result of SPA (Eq. \eqref{eq:SPA}), we see that we can directly use the SPA with $ g(k_x,k_y) = \mathcal{P}^-(\kappa,\alpha)_j \hat{\mathbf{e}}_j^- $. Using the SPA result, we get the expression for the electric field on the far-field hemisphere: 

\begin{align}
    \label{eq:E_far}
    \mathbf{E_\mathrm{far-field}}(\mathbf{r_\infty}) =  -2\pi \mathrm{i} E_0 \frac{e^{\mathrm{i}kr}}{r} k_{z} \sum_{j=1}^2  \mathcal{P}_j^-(\kappa,\alpha)  
    \hat{\mathbf{e}}_j^-.  
\end{align}

What the SPA essentially gives us, is that the EM-field at a location in the far-field is fully determined by the single plane wave propagating in that exact direction. There is no need anymore to integrate over all possible plane waves, because they will all destructively interfere, except the plane wave in the direction of interest. This strong finding significantly reduces the computational speed and makes that we can compute the far-field very efficiently from the plane wave expansion coefficients in the glass. In the implementation of this pipeline, we take the far-field being a hemisphere with a radius of 1 meter.  

\newpage
\section{Focusing onto the camera}

In the previous section we have derived the EM-fields in the far-field. In the next step, the objective focuses these fields onto the camera (Supplemenetary Fig. \ref{fig:S2_overview}f). Since the coverglass, immersion oil and the glass objective are refractive index-matched, optically, the entire medium between the water before the coverglass and the air after the objective can be seen as one, with refractive index $n_{\mathrm{glass}}$ and wave vector $k_{\mathrm{glass}}$. 

The main assumption is that the objective is perfectly aplanatic. This means that the directions of the plane waves in the far-field linearly relate to the directions of the light after the objective. Due to this assumption, the term \textit{objective} in this document encompasses the complete optics that image the far-field onto the camera, and therefore  includes the tube lens, etc. \cite{Enderlein2003}.

Supplementary Fig. \ref{fig:S25_help}f shows schematically the imaging system, where the polar angle of each light ray changes going from the glass to the air medium, due to focusing. The far-field contains the angular information and can be expressed as $\mathbf{E}_{\mathrm{far-field}}(\theta_{\mathrm{ff}},\phi_{\mathrm{ff}})$, like in Eq. \eqref{eq:E_far}. Here, $\theta_{\mathrm{ff}}$ is the polar angle (defined as the angle between the plane wave in this particular direction and the negative $z$-axis) and $\phi_{\mathrm{ff}}$ the azimuthal angle, which are fully defined by the cylindrical coordinates $\kappa$ and $\alpha$ of the plane waves: 

\begin{equation}
    \begin{cases}
        \theta_{\mathrm{ff}}(\kappa,\alpha) = \arccos(k_z/k) =  \arccos \bigg( \sqrt{k_{\mathrm{glass}}^2-\kappa^2}/k_{\mathrm{glass}} \bigg) \\
        \phi_{\mathrm{ff}}(\kappa,\alpha) = \alpha.
    \end{cases}
\end{equation}

Remember that refraction and focusing do not affect the azimuthal angle $\phi_{\mathrm{ff}}$. In the process of focusing the polar angles before and after the objective are connected with the Abbe sine condition: 

\begin{equation}
    n_{\mathrm{glass}}\sin(\theta_{\mathrm{ff}}) = Mn_{\mathrm{air}}\sin(\theta_{\mathrm{air}}) , 
\end{equation}

with $M$ the magnification of the optical system, and $\phi_{\mathrm{ff}}$ unchanged.

\subsection{Focusing integral}
Using the terminology of the angles in the glass and air medium, we use the focusing integral to evaluate the field in the image plane. The image plane is described in cylindrical coordinates ($\rho_{\mathrm{im}},\phi_{\mathrm{im}},z_{\mathrm{im}}$):

\begin{itemize}
    \item \makebox[3cm]{$\rho_{\mathrm{im}} \in [0,\infty)$\hfill} = radial distance to the optical axis
    \item \makebox[3cm]{$\phi_{\mathrm{im}} \in [0,2\pi]$\hfill} = counter-clockwise angle with the $x$-axis of the image
    \item \makebox[3cm]{$z_{\mathrm{im}}$\hfill} = $z$-position of the image-plane, with $z_{\mathrm{im}}=0$ as perfect focus
\end{itemize}

See Supplementary Fig. \ref{fig:S25_help}f for further clarification of the cylindrical coordinates. The diffracted electric field at the image plane, the camera, can be found by evaluating the focusing integral. The electric field on one position in the image plane is affected by all angles in the far-field, hence the integral over the entire far-field \cite{Enderlein2000,Enderlein2003,Mortensen2010}: 


\begin{equation}
\begin{split}   
    \label{eq:focus_int}
    \mathbf{E}(\rho_{\mathrm{im}},\phi_{\mathrm{im}},z_{\mathrm{im}}) \propto \int\displaylimits_0^{\theta_\mathrm{ff,max}} \int\displaylimits_0^{2\pi} 
    &\sqrt{\frac{n_{\mathrm{air}}\cos(\theta_{\mathrm{air}})}{n_{\mathrm{glass}}\cos(\theta_{\mathrm{ff}})}} 
    \Bigl ( \mathbf{\hat{e}}_s \mathbf{E}_{\mathrm{p,ff}}( \theta_{\mathrm{ff}},\phi_{\mathrm{ff}})+\mathbf{\hat{e}}_p\mathbf{E}_{\mathrm{s,ff}}(\theta_{\mathrm{ff}},\phi_{\mathrm{ff}}) \Bigr) \\
    & e^{\mathrm{i}kz\cos(\theta_{\mathrm{ff}})}
    e^{\mathrm{i}k\rho\sin(\theta_{\mathrm{ff}}) \cos(\phi_{\mathrm{ff}}-\phi_{\mathrm{im}}) } \sin(\theta_{\mathrm{ff}})\mathrm{d}\phi_{\mathrm{ff}}\mathrm{d}\theta_{\mathrm{ff}},
\end{split}
\end{equation}

with $k$ as $k_{\mathrm{air}}$, $z$ as $z_{\mathrm{im}}$, and $\rho$ as $\rho_{\mathrm{im}}$, to avoid the amount of confusing subscripts. The electric field vector in the far-field (Eq. \ref{eq:E_far}) is split up in $S$- and $P$-polarization, indicated by the subscript $s$ or $p$ in $\mathbf{E}_{\mathrm{p,ff}}$, and projected onto the polarization unit vectors, which are defined as: 

\begin{equation}
    \label{eq:es}
    \hat{\mathbf{e}}_s
    =\begin{pmatrix}
        -\sin(\phi_{\mathrm{ff}})\\
        \cos(\phi_{\mathrm{ff}})\\
        0
    \end{pmatrix}
    \quad
    \text{and}
    \quad
    \hat{\mathbf{e}}_p
    =\begin{pmatrix}
        -\cos(\phi_{\mathrm{ff}})\cos(\theta_{\text{air}})\\
        -\sin(\phi_{\mathrm{ff}})\cos(\theta_{\text{air}})\\
        \sin(\theta_{\text{air}})
    \end{pmatrix}
\end{equation}

Due to the limited collection angle of the objective, not all far-field angles can be focus onto the camera, quantified by the numerical aperture (NA). The NA is defined as $\mathrm{NA}=n\sin\theta$, which leads by rearranging the terms to the highest far-field angle that can be focused:

\begin{equation}
    \theta_{\mathrm{max}} =  \theta_{\mathrm{ff,max}} = \arcsin(\mathrm{NA}/n_{\mathrm{glass}}),
\end{equation}

which is the limit for $\theta_{\mathrm{ff}}$ in the integral of Eq. \eqref{eq:focus_int}. The magnetic field at the image plane, $\mathbf{B}$, is calculated in the same way, but using the magnetic far-field and interchanged polarization unit vectors.

\subsection{Poynting vector}
To find the intensity of the light, one needs to calculate the Poynting vector, which is a vector that indicates the directional energy flux of an electromagnetic wave. Since we want to know the intensity on a camera that is oriented perpendicular to the optical axis, we are only interested in the negative $z$-component of the Poynting vector:

\begin{equation}
    \mathrm{PSF}(x,y) \propto -S_z = - \hat{e}_z \cdot (\mathbf{E} \times \mathbf{B}^*) = - (E_xB_y^* - E_yB_x^*), 
\end{equation}

where the $E$- and $B$-fields are found by evaluating the focusing integral (Eq. \eqref{eq:focus_int}), after converting the vectors from cylindrical to Cartesian coordinates. We take the negative $z$-component, because the camera detects the downward directed light intensity, since it is located beneath the sample. 

In conclusion, the electric and magnetic far-fields are focused onto the camera plane using the focusing integral. Subsequently, the light intensity is obtained by calculating the $z$-component of the Poynting vector. This process is repeated for every pixel in the image, resulting in the point-spread function. Ideally, one should integrate the above expression for the PSF over the area of the pixel to obtain the correct intensity. However, when the pixels are small enough, we can assume that the intensity profile is somewhat constant over one pixel, and we take the intensity at the center of the pixel multiplied by the pixel area. 
\chapter*{Supplementary Note 3: \\Proof that PSF for freely rotating dipole is identical to the average of the PSFs of three mutually-orthogonal fixed dipoles}
\addcontentsline{toc}{chapter}{Supplementary Note 3: Proof that PSF for freely rotating dipole is identical to the average of the PSFs of three mutually-orthogonal fixed dipoles}  
\label{chap:Note4}
\chaptermark{Supplementary Note 3: Proof for PSF of freely rotating dipole emitter}

\subsection*{Definition of fixed dipole emitter}
Assume we have a fixed dipole emitter $\vec{p}$ with a 3D orientation parametrized in spherical coordinates by azimuthal angle $\alpha$ and polar angle $\beta$. We can write the dipole vector $\vec{p}$ as a linear combination of three basis vectors. For convenience, we choose the Cartesian basis vectors, which are oriented along the $x$-, $y$- and $z$-axis and have unit length. The linear combination can be written as:

\begin{equation}
    \vec{p}_{\alpha\beta} = f(\alpha,\beta) \hat{x} + g(\alpha,\beta) \hat{y} + h(\alpha,\beta) \hat{z},
\end{equation}

with $f$, $g$ and $h$ orientation-dependent weights, defined as: 

\begin{equation}
    \begin{cases}
        f = \sin \beta \cos \alpha \\
        g = \sin \beta \sin \alpha \\
        h = \cos \beta. \\
    \end{cases}
\end{equation}

\subsection*{Electric field of fixed dipole emitter}
In our steps toward the PSF for a freely rotating dipole emitter, we first derive the electric field of a fixed dipole. Since Maxwell's equations are linear, we can decompose the electric field emitted by our fixed dipole $\vec{E}_p$ in the electric fields emitted by fixed dipole emitters oriented along the Cartesian axis: 

\begin{equation}
    \vec{E}_p = f(\alpha,\beta) \vec{E}_x + g(\alpha,\beta) \vec{E}_y + h(\alpha,\beta) \vec{E}_z,
\end{equation}

where $\vec{E}_x$ is the electric field emitted by a fixed dipole that is oriented along the $x$-axis. Following this logic, we find the expression of the squared magnitude of the electric field of the fixed dipole: 

\begin{equation}
    |\vec{E}_p|^2 = f^2 |\vec{E}_x|^2 + g^2 |\vec{E}_y|^2 + h^2 |\vec{E}_z|^2 
    + 2fg\Re(E_xE_y^*)
    + 2fh\Re(E_xE_z^*)
    + 2gh\Re(E_yE_z^*),
    \label{eq:proof2_fullE}
\end{equation}

where $\Re$ represents the real part and $^*$ the complex conjugate. We omit the orientational dependency on $\alpha$ and $\beta$ of the weights $f$, $g$, and $h$ for clarity. In addition, we used the following property of complex numbers: $|a+b|^2 = |a|^2 + |b|^2 + 2\Re(ab^*)$.

\subsection*{Electric field of freely rotating dipole emitter}
For high magnification, the intensity of the PSF is very well approximated by the squared magnitude of the electric field $|E|^2$. Since we assume that the dipole is freely rotating, and the dipole visits every possible orientation within the integration time of the camera, the PSF is represented by averaging $|E_p|^2$ over all possible orientations of $\vec{p}$, which we compute by integrating the expression of the electric field over the entire normalized orientation space $\Omega$: 

\begin{equation}
    |\vec{E}_{\textrm{free}}|^2  
    = \frac{1}{4\pi} \iint |\vec{E}_p|^2 \textrm{d}\Omega
    = \frac{1}{4\pi} \int_{\beta=0}^{\pi} \int_{\alpha=0}^{2\pi} |\vec{E}_p|^2 \sin(\beta)\textrm{d}\alpha \textrm{d}\beta.
\end{equation}

We evaluate this integral separately for all six combinations of weights in Eq. \ref{eq:proof2_fullE}: 

\begin{align}
    \iint f^2 \sin \beta \textrm{d}\alpha \textrm{d}\beta &= \iint \sin^3 \beta \cos^2 \alpha \textrm{d}\alpha \textrm{d}\beta = 4\pi/3 \\
    \iint g^2 \sin \beta \textrm{d}\alpha \textrm{d}\beta &= \iint \sin^2 \alpha \sin^3 \beta \textrm{d}\alpha \textrm{d}\beta = 4\pi/3 \\
    \iint h^2 \sin \beta \textrm{d}\alpha \textrm{d}\beta &= \iint \sin \beta \cos^2 \beta \textrm{d}\alpha \textrm{d}\beta = 4\pi/3 \\
    \iint fg \sin \beta \textrm{d}\alpha \textrm{d}\beta &= \iint \sin \alpha 
    \cos \alpha \sin^3 \beta \textrm{d}\alpha \textrm{d}\beta = 0 \\
    \iint fh \sin \beta \textrm{d}\alpha \textrm{d}\beta &= \iint \cos \alpha 
    \sin^2 \beta \cos \beta \textrm{d}\alpha \textrm{d}\beta = 0 \\
    \iint gh \sin \beta \textrm{d}\alpha \textrm{d}\beta &= \iint \sin \alpha 
    \cos \beta \sin^2 \beta \textrm{d}\alpha \textrm{d}\beta = 0,
\end{align}

which are evaluated analytically using Mathematica.

\subsection*{PSF of freely rotating dipole emitter}
Since the three cross-terms are cancelled (the integrals evaluate to $0$), the total expression for the PSF of a freely rotating dipole reduces to:

\begin{align}
    \textrm{PSF}_{\textrm{free}} \propto 
    |\vec{E}_{\textrm{free}}|^2  \nonumber
    &= \frac{1}{4\pi} \iint |\vec{E}_p|^2 \textrm{d}\Omega \\ \nonumber
    &= \frac{1}{3}|\vec{E}_x|^2 + \frac{1}{3}|\vec{E}_y|^2 + \frac{1}{3}|\vec{E}_z|^2 \\
    &\propto  \frac{1}{3}\textrm{PSF}_{\textrm{x}} +  \frac{1}{3} \textrm{PSF}_{\textrm{y}} +  \frac{1}{3} \textrm{PSF}_{\textrm{z}}.
\end{align}

This shows that when we want to calculate the PSF for a freely rotating dipole emitter, the PSF is identical to the average of three PSFs, each from a fixed dipole emitter along the $x$-, $y$-, and $z$-direction, which eliminates the need for any integration. 

Interestingly, the exact same reasoning holds when the fixed dipole emitter $\vec{p}$ is defined as a linear combination in any other mutually-orthogonal basis. This means that if we give our Cartesian basis vectors $\hat{x}$, $\hat{y}$ and $\hat{z}$ an arbitrary rotation, the freely-rotating PSF is still simply the average of the three PSFs along these new axes. This is true, because a rotated basis will only change the definitions of the angles $\alpha$ and $\beta$, but integrating over the full rotation domain $\Omega$ will always cancel out the cross-terms from Eq. \ref{eq:proof2_fullE}. In the context of a fluorophore near a spherical nanoparticle, this proof shows that the PSF for a freely rotating dipole emitter is identical to the average of three PSFs, each oriented along the $x$-, $y$-, and $z$-axis. Or, identically, three dipoles oriented along the radial-, azimuthal- and polar- direction, which is convenient, since the dipole oriented radially with respect to the nanoparticle will have the strongest coupling and dominate the average.

\chapter*{Supplementary Table 1: \\Analytical PSF parameters}
\addcontentsline{toc}{chapter}{Supplementary Table 1: Analytical PSF parameters} 
\chaptermark{Supplementary Table 1: Analytical PSF parameters}

\vspace{-1cm}
Table S1 below shows the analytical parameters used for all the analytically calculated PSFs throughout the paper. The analytical parameters that are the same for all calculations are not mentioned in the table, and are: the refractive index of water (1.333), refractive index of glass (1.52) and the numerical aperture of the microscope (1.49). All parameters are explained in Supplementary Notes 1 and 2.

\begin{table}[ht!]
\begin{tabular}{ccccccccccc}
\toprule
                & \multicolumn{3}{c}{NP}                  & \multicolumn{6}{c}{Dipole emitter}      
                \\
\cmidrule(rl){2-4} \cmidrule(rl){5-10}
\textbf{Figure}      & \textbf{radius}        & \textbf{mode}        & \textbf{refr idx}         & \textbf{d}             & \textbf{lam}           & \textbf{angle}    & \textbf{alpha}          & \textbf{beta}           & \textbf{gamma}        \\
\multicolumn{1}{c}{} & \multicolumn{1}{c}{[nm]} & \multicolumn{1}{c}{} & \multicolumn{1}{c}{}      & \multicolumn{1}{c}{[nm]} & \multicolumn{1}{c}{[nm]} & \multicolumn{1}{c}{[rad]} & \multicolumn{1}{c}{[rad]} & \multicolumn{1}{c}{[rad]} & \multicolumn{1}{c}{[rad]}  \\
\midrule
2d(1)/S7a            & 100                    & solid                & gold                      & 10                     & 675                    & 0.5pi                   & 0              & 0                       & 0                        \\
2d(2)/S7b            & 100                    & solid                & 1.63   & 10                     & 400                    & 0                       & 0              & 0.25pi                  & 0                            \\
2d(3)/S7c            & 50                     & core-shell           & 1.4559 & 10                     & 675                    & 0                       & 0              & 0.5pi                   & 0                          \\
S3b                  & 100                    & solid                & gold                      & 10                     & 675           & 0.5pi                   & 0              & 0                       & 0                         \\
S3c                  & 100                    & solid                & gold                      & 10                     & 675           & 0                       & 0              & 0.5pi                   & 0                       \\
S4a-f                  & 100                    & solid                & gold                      & 10                     & 675           & 0.5pi          & 0              & 0                       & 0                          \\
S4g-l                  & 100                    & solid                & gold                      & 10                     & 675           & 0.5pi          & 0              & 0.5pi                   & 0                       \\
S5                   & 100                    & solid                & gold                      & 10                     & 675                    & 0                   & 0                       & 0.25pi                       & 0                              \\
S6                   & 100                    & solid                & gold                      & 10                     & 675                    & 0.5pi                   & 0                       & 0                       & 0                              \\
S14a                 & 50                     & solid                & gold                      & 5                      & 680                    & 0                       & 0                       & 0.25pi                  & 0                                 \\
                     & 50                     & solid                & gold                      & 5                      & 680                    & 0                       & pi                      & 0.75pi                  & 0                                  \\
S17                  & 50                    & solid                & gold                      & 5                     & 680                    & 0                       & 0                       & 0.25pi                  & 0                        \\
S19                  & 100                    & solid                & gold                      & 10                     & 675                    & 0                       & 0                       & 0.25pi                  & 0                        \\
S21                  & 50                    & solid                & gold                      & 5                     & 680                    & 0:0.5pi                      & 0                       & 0:pi                  & 0:0.5pi                        \\
S26a/b                 & 100                    & solid                & gold                      & 10                     & 675                    & 0.5pi                   & 0                       & 0                       & 0                                \\
\bottomrule
\end{tabular}
\end{table}

\setcounter{table}{0}

\begin{table}[ht!]
\begin{tabular}{cccccccccc}
\toprule
                & \multicolumn{9}{c}{microscope}              
                \\
\cmidrule(rl){2-9} 
\textbf{Figure} & \textbf{focus} & \textbf{dist\_glass} & \textbf{l\_max} & \textbf{m\_max} & \textbf{N} & \textbf{r} & \textbf{M} & \textbf{num\_px} & \textbf{px\_size} \\
                & [nm]            & [nm]                   &                 &                 &            & [m]          &            &                  & [nm]                \\
\midrule
2d(1)/S7a       & 120            & 20                   & 10              & 3               & 100        & 1          & 363.63     & 50               & 15                \\
2d(2)/S7b       & 120            & 20                   & 10              & 3               & 100        & 1          & 363.63     & 50               & 15                \\
2d(3)/S7c       & 70             & 20                   & 10              & 3               & 100        & 1          & 363.63     & 50               & 15                \\
S3b/c             & 120            & 20                   & 10              & 3               & 100        & 1          & 363.63     & 50               & 15                \\
S4a-l             & 120            & 20                   & 1:20            & l\_max          & 100        & 1          & 363.63     & 50               & 15                \\
S5              & 120            & 20                   & 2:30              & 3               & 4:100        & 1          & 363.63     & 2:32               & 30                \\
S6              & 120            & 20                   & 10              & 3               & 100        & 1          & 363.63     & 50               & 15                \\
S14a            & 56             & 6                    & 10              & 3               & 100        & 1          & 100     & 65               & 10                \\
S17              & 56            & 6                   & 10              & 3               & 100        & 1          & 100     & 50               & 15                \\
S19             & 120            & 20                   & 10              & 3               & 100        & 1          & 363.63     & 20               & 30     \\
S21              & 56            & 6                   & 10              & 3               & 100        & 1          & 100     & 50               & 15                \\
S26a            & 120            & 20                   & 10              & 1               & 1:100      & 1          & 363.63     & 24               & 30                \\
S26b            & 120            & 20                   & 10              & 1               & 100        & 1e-9:1     & 363.63     & 24               & 30                \\
\bottomrule
\end{tabular}
\caption{\textbf{Parameters for Analytically calculated PSFs.}}
\label{tab:params_analytical}
\end{table}

\chapter*{Supplementary Table 2: \\Numerical PSF parameters}
\addcontentsline{toc}{chapter}{Supplementary Table 2: Numerical PSF parameters} 
\chaptermark{Supplementary Table 2: Numerical PSF parameters}

Table S2 below shows the numerical parameters used for all the numerically calculated PSFs throughout the paper. The numerical parameters that are the same for all calculations are not mentioned in the table, and are: 

\begin{itemize}
    \item gap between NP and glass = 20 nm
    \item gap between NP and dipole = 10 nm
    \item dipole orientation azimuthal angle  = 0 deg
    \item FDTD simulation time = 40 fs
    \item detector z-position = -20 nm (20 nm below water-glass interface)
    \item refractive index water = 1.333
    \item refractive index glass = 1.52
    \item refractive index gold = taken from Johnson and Christy \cite{johnson1972}
\end{itemize}

\hfill
\hfill
\hfill

\begin{table}[ht!]
\begin{tabular}{cccccccc}
\toprule
                & NP              & \multicolumn{3}{c}{Dipole emitter}                                                                                                                  & \multicolumn{3}{c}{FDTD}     
                \\
\cmidrule(rl){2-2} \cmidrule(rl){3-5} \cmidrule(rl){6-8}
\textbf{Figure} & \textbf{radius} & \textbf{lambda} & \textbf{orientation} & \textbf{positon}                                                                  & \textbf{size} & \textbf{accuracy} & \textbf{fine mesh} \\
                & {[}nm{]}        & {[}nm{]}        & polar [deg]      &                     & {[}$\mu$m{]}  &                   & nm                 \\
\midrule
2c              & 100             & 675             & 90             & top                                                                                                  & 8             & 1-8               & 3                  \\
2d (1)/S7a      & 100             & 675             & 90             & top                          & 8             & 6                 & 3                  \\
2d (2)/S7b      & 100             & 400             & 45             & \begin{tabular}[c]{@{}c@{}}45 deg \end{tabular}  & 8             & 6                 & 5                  \\
2d (3)/S7c      & 50              & 675             & 90             & side                                                                                                 & 8             & 6                 & 0.6                \\
S3b             & 100             & 675             & 90             & top                                                                                                  & 8             & 6                 & 3                  \\
S3c             & 100             & 675             & 90             & side                                                                                                 & 8             & 6                 & 3                  \\
S6a             & 100             & 675             & 90             & top                                                                                                  & 0.25-16       & 6                 & 3                  \\
S6b             & 100             & 675             & 90             & top                                                                                                  & 8             & 1-8               & 3      \\
\bottomrule
\end{tabular}
\caption{\textbf{Parameters for numerically calculated PSFs.} This able shows the numerical parameters used for all numerically calculated PSFs throughout this paper. Explanation of the parameters can be found in the Methods section of the main text.}
\label{tab:params_numerical}
\end{table}

\addcontentsline{toc}{chapter}{Bibliography for SI} 
\bibliographystyle{apalike}   
\bibliography{z_biblio}

\end{document}